\documentclass[a4paper,fleqn]{cas-sc}

\usepackage[authoryear]{natbib}
\usepackage{xcolor}
\usepackage{graphicx}
\usepackage{subfigure}
\usepackage{multirow}
\usepackage{amsmath}
\usepackage{amssymb}
\usepackage{amsbsy}
\usepackage{bm}
\usepackage{hyperref}
\usepackage{booktabs}
\hypersetup{
	colorlinks = true,
	urlcolor   = red,
	citecolor  = blue,
}

\def\tsc#1{\csdef{#1}{\textsc{\lowercase{#1}}\xspace}}
\tsc{WGM}
\tsc{QE}
\tsc{EP}
\tsc{PMS}
\tsc{BEC}
\tsc{DE}

\begin{document}
\let\WriteBookmarks\relax
\def\floatpagepagefraction{1}
\def\textpagefraction{.001}
\shorttitle{Stochastic force model for a finite-size spherical particle in turbulence}
\shortauthors{Wang \& Hu}

\title [mode = title]{A stochastic force model for a finite-size spherical particle in turbulence}                      

\author[1]{Yuqi Wang}[type=editor,
                        auid=000,bioid=1,
                        orcid=0000-0001-7105-2703,
                        style=chinese]
\credit{Methodology, Software, Analysis and Writing}

\author[1]{Ruifeng Hu}[type=editor,
					   auid=002,bioid=2,
					   orcid=0000-0002-2505-5433,
					   style=chinese]
\cormark[1]
\ead{hurf@lzu.edu.cn}
\credit{Conceptualization, Methodology, Software, Analysis and Writing}

\affiliation[1]{organization={Center for Particle-Laden Turbulence, Key Laboratory of Mechanics on Disaster and Environment in Western China, Ministry of Education, College of Civil Engineering and Mechanics},
                addressline={Lanzhou University}, 
                city={Lanzhou},
                postcode={730000}, 
                state={Gansu},
                country={China}}

\cortext[cor1]{Corresponding author}

\begin{abstract}
Predicting particle-laden flows requires accurate fluid force models. However, a reliable particle force model for finite-size particles in turbulent flows remains lacking.
In the present work, a fluid force model for a finite-size spherical particle in turbulence is developed by simulating turbulent flow past a fixed spherical particle using particle-resolved direct numerical simulation (PRDNS).  
Our simulation demonstrates that turbulence increases the mean drag force of the particle, which is consistent with previous studies. By correlating the DNS data as functions of the Reynolds number of particles, the ratio of the particle-to-turbulence scale, and the intensity ratio of the turbulence, an empirical correlation for the mean drag force is obtained.
Furthermore, we find that the fluctuations of both the drag and lateral forces follow the Gaussian distribution. Consequently, the temporal variations of the fluctuating drag and lateral forces are modeled using a stochastic Langevin equation.
Empirical correlations of the fluctuation intensities and time scales involved in the stochastic model are also determined from the DNS data. 
Finally, we simulate the movement of a finite-size particle in turbulence and the dispersion of particles in a turbulent channel flow to validate the proposed model. 
The proposed fluid force model requires the mean flow velocity, the kinetic energy of the turbulence, and the dissipation rate of the turbulence as inputs, which makes it well suited for combination with the Reynolds-averaged Navier-Stokes (RANS) approach.
\end{abstract}




\maketitle

\section{Introduction}

Particle-laden flows are ubiquitous in nature and engineering applications, such as sediment transport, dust storms, fluidized bed reactor flow, and the spread of haze pollutants in cities. 
For decades, numerical methods for particle-laden flows have been greatly developed, and numerical simulation has become a powerful tool in both fundamental research and practical processes. 
Depending on the different levels of assumptions, the numerical methods for particle-laden flows can be roughly classified as the Eulerian two-fluid method, the Lagrangian particle method, and the fully resolved method \citep{Drew1983,Elghobashi1994,Crowe1996,Loth2000,Eaton2009,Wang2009,BalaEaton2010,Yu2010,Fox2012,Subramaniam2013,Tenneti2014,Kuerten2016,Maxey2017,Elghobashi2019,Brandt2022,Fox2024,Schneiderbauer2024}.
The prediction accuracy of particle-laden flows is critically dependent on the fluid force models used in numerical simulations \citep{Michaelides1997,Michaelides2003,Loth2009}.

The point-particle method is a widely used approach based on the Lagrangian framework, which has a long history in particle-laden flow simulations \citep{Mclaughlin1994,Balachandar2009,Eaton2009,Soldati2009,Kuerten2016}.
In this method, particles are modeled as infinitesimal points with finite mass and momentum, and there exists an exact theory for particle motion in the Stokes flow regime at very small particle Reynolds numbers \citep{Tsai1957Sedimentation,gatignol1983faxen,Maxey1983,Tsai2022Sedimentation}. 
The fluid forces exerted on a spherical particle include the Stokes drag force, the Basset history force, the added mass force, and the pressure gradient force.
With a large particle-to-fluid density ratio, it is usually assumed that the drag force is the most dominant and other force terms can be neglected \citep{Armenio2001}. 
At a finite particle Reynolds number,
empirical correlation for the drag force is commonly leveraged, \emph{e.g.} the Schiller-Naumann (S-N) drag correlation \citep{Schiller&Neumann1933}. 
The point-particle method requires that the particle size is generally much smaller than the smallest scale of turbulence, \emph{i.e.}, the Kolmogorov length scale. 
However, for a finite-size particle whose size is larger than the Kolmogorov length scale, the effect of small-scale turbulent motions around the particle may lead to an unreasonable prediction of the particle force with the point-particle method.

In the last three decades, particle-resolved or fully resolved direct numerical simulation has become a vital approach to resolving the flow around particles directly and accurately predicting particle-laden flows. Several different particle-resolved direct numerical simulation (PRDNS) approaches have been developed, such as the immersed boundary (IB) method \citep{mittal2005immersed,Verzicco2022immersed,mittal2023origin}, the distributed Lagrange multiplier (DLM)/fictitious domain approach \cite[]{glowinski1999distributed,glowinski2001fictitious,yu2007direct,Yu2010}, and others.
PRDNS have been widely used to obtain high-fidelity simulation data, which is greatly helpful in studying particle transport in turbulent flows \citep{uhlmann2008interface,Shao2012fully,picano2015turbulent,vowinckel2019settling,costa2020interface,peng2024preferential}, sediment transport \citep{jiDirectNumericalSimulation2013,kidanemariam2014direct,kidanemariam2014interface,vowinckel2016entrainment,kidanemariam2017formation,mazzuoli2020interface,schererRoleTurbulentLargescale2022,zhu2022particle,schwarzmeier2023particle}, turbulence modulation by particles \citep{Pan1996,vowinckel2014fluid,peng2019direct,peng2020flow,Costa2021near,Yu_Xia_Guo_Lin_2021,balachandar2024turbulence}, and developing particle force models \citep{vanderhoef2005,cate2006,beetstra2007,zeng2009,tenneti2011,tang2015, Zhou2015direct,Akiki2017pairwise,Seyed2020microstructure,seyed-ahmadi2022,Xia2022,Peng2023,vanwachem2024,xia2024}.
Despite its high accuracy, conducting large-scale simulations of particle-laden flows with PRDNS remains prohibitively costly. Therefore, it is necessary to develop reliable particle force models that can be integrated with cost-effective Reynolds-averaged Navier-Stokes (RANS) or large-eddy simulation (LES) techniques for turbulent flows. Various studies have been conducted on particle force models utilizing PRDNS data. 
However, a considerable proportion of these models focus on laminar flows rather than turbulent flows, which substantially limits their applicability in numerous practical situations. Studies have been conducted on the impact of turbulence on the mean particle drag force, while models for the instantaneous drag force are significantly deficient. 
In the following, we will review the current progress of the turbulence effects on the drag force of a finite-size particle and the prediction models.

\subsection{Effects of turbulence on the drag force of a finite-size particle and models}

Some studies investigated the influence of turbulence on particle drag. 
\citet{bagchi2003effect} investigated the influence of freestream isotropic turbulent flow on the drag and lift forces exerted on a spherical particle by PRDNS. Their findings indicated that freestream turbulence does not have a significant or systematic influence on the time-averaged mean drag, and the standard empirical drag formula, based on the instantaneous or mean relative velocity, may accurately predict the mean drag with reasonable accuracy. The prediction accuracy of instantaneous drag reduces as the particle size increases. For smaller particles, the standard drag law effectively captures low-frequency oscillations of particle drag; however, for larger particles, a notable discrepancy arises even in the low-frequency component.
\cite{BURTON_2005} studied the interaction between a fixed particle and the decaying homogeneous isotropic turbulence by PRDNS. They found that the Lagrangian point-particle equation of motion underpredicts the particle root mean square (RMS) force, and the RMS force errors between the PRDNS results and those predicted by the Lagrangian particle equation of motion are between 15\% and 30\%.
\cite{ZENG_2008} conducted a PRDNS investigation on the force of a single fixed finite-size particle in turbulent channel flow, with the particle positioned at various wall-normal locations. It was found that when the particle is away from the viscous sublayer, the instantaneous drag is accurately captured by the standard drag law; in contrast, the particle drag is underestimated when the particle is in the buffer layer.
\cite{kim_2012} performed PRDNS of a particle in freestream turbulence. They stated that the standard drag law defined with uniform flow simulation can accurately predict particle drag when turbulent intensity is sufficiently low (below 5\%), while flow non-uniformity substantially influences the forces acting on a finite-size particle as turbulent intensity increases, a factor often overlooked in the context of small particles. They suggested that both deterministic and stochastic components must be incorporated into the particle force model to accurately predict the motion of finite-size particles in turbulent flows.

In addition, several studies proposed empirical formulas for particle drag considering the turbulence effect. 
\cite{BRUCATO1998} conducted experiments to quantify the drag coefficient of particles in the Couette-Taylor turbulence. They determined that particle drag depends on particle size and turbulence intensity. Turbulence pulsation can markedly increase particle drag for larger particles, although it exerts no impact on the drag force for smaller particles. They proposed a correlation for predicting the effect of freestream turbulence on the mean drag coefficients of intermediate-sized particles,
\begin{equation}
	\Delta C_{D}= A \left({D_p}/{\eta}\right)^3,
	\label{cd1}
\end{equation}
where $\Delta C_{D}=\left( C_D-C_{D0} \right)/C_{D0}$ is the relative drag coefficient increment, $C_D$ is the mean particle drag coefficient in a turbulent flow, $C_{D0}$ is the particle drag coefficient from the standard S-N drag law \citep{Schiller&Neumann1933}
\begin{equation}
	C_{D0}=C_D^{SN} = \frac{24}{Re_p} \left( 1 + 0.15Re_p^{0.687}  \right),
	\label{CDSN}
\end{equation}
$D_p/\eta$ is the ratio of particle diameter $D_p$ to the Kolmogorov length scale $\eta$, and the constant $A=8.76 \times 10^{-4}$.

\cite{Homann_Bec_Grauer_2013} examined the influence of turbulence on the drag force of a fixed finite-size particle through PRDNS. Their findings indicate that the mean particle drag increases with increasing turbulence intensity, and this enhancement is significantly greater than the predictions provided by the standard drag law. They proposed two correlations:
\begin{equation}
	\Delta C_{D}=0.18 I^2 Re_p Re_L^{1/2},
	\label{cd2}
\end{equation}
and
\begin{equation}
	\Delta C_{D}=\frac{0.45Re_p^{0.687}}{1+0.15Re_p^{0.687}}I^2,
	\label{cd3}
\end{equation}
in which $I=u_{rms}/U_s$ is the relative turbulent intensity ($u_{rms}$ is turbulent fluctuation velocity, $U_s$ is the magnitude of mean particle-to-fluid slip velocity), $Re_p=U_s D_p/\nu$ is the particle Reynolds number ($\nu$ is fluid kinematic viscosity), and $Re_L=u_{rms}L/\nu$ is the integral scale Reynolds number of turbulence ($L=u_{rms}^3/\varepsilon$, $\varepsilon$ is the mean kinetic energy dissipation rate). 

\cite{Peng2023} performed PRDNS to investigate the mechanism and model of drag enhancement in a finite-size particle in an anisotropic turbulent flow. They found that large-scale turbulence increases particle drag by enhancing the pressure drop on the particle surface, whereas small-scale turbulence improves the mixing of high- and low-speed flow at the particle boundary layer, consequently intensifying the viscous stresses on the particle surface. They additionally offered a model for the enhancement of particle drag in turbulent flow, taking into account the anisotropy of the flow, as
\begin{equation}
	\Delta C_{D}=0.06\beta^{1.4}I^{0.5}Re_p^{-0.1}(D_p/\eta)^{0.9},
	\label{cd4}
\end{equation}
where $\beta=\langle u_1^{\prime} \rangle /( \langle u_2^{\prime} \rangle \langle u_3^{\prime} \rangle )^{0.5}$ is the anisotropy parameter of turbulence ($\langle u_1^{\prime} \rangle$, $\langle u_2^{\prime} \rangle$ and  $\langle u_3^{\prime} \rangle$ are the component-wise RMS velocities in three directions). 

\citet{Wang_Lei_Zhu_Zheng_2023} explored the drag force on the saltating particles over an erodible particle bed in a turbulent open channel flow using PRDNS. Their findings indicate that the mean drag force on the saltating particles is related to the square of the vertical particle velocity, and a drag force correlation based on the S-N drag law was established using data fitting, which is
\begin{equation}
	\Delta C_{D}=\frac{40(u_{p,y}/u_{p,0})^2+27}{Re_p}-1,
	\label{cd5}
\end{equation}
where $u_{p,y}$ is the vertical particle velocity at the height of $y_p$ and $u_{p,0}$ is the initial vertical particle velocity. 

\begin{table}
	\centering
	\def~{\hphantom{0}}
	\caption{Empirical correlations for the mean particle drag coefficient in turbulence.} 
	\begin{tabular}{ccccc}
		\toprule
		& Drag model   & $Re_p$ & $I$    & $D_p/\eta$ \\[5pt]
		\midrule
		\cite{BRUCATO1998} & Eqn.~(\ref{cd1}) & $<40$ & $0.05, 0.5$  & $2.5\sim30$\\[10pt]
		\cite{Homann_Bec_Grauer_2013} & Eqns.~(\ref{cd2}) and (\ref{cd3}) & $20\sim400$ & $0.05\sim0.6$ & $1.1\sim44$\\[10pt]
		\cite{Peng2023} & Eqn.~(\ref{cd4}) & $100\sim300$ & $0.17\sim0.27$  & $6.89\sim22.1$ \\[10pt]
		\cite{Wang_Lei_Zhu_Zheng_2023} & Eqn.~(\ref{cd5}) & $<80$ & /  & / \\[10pt]
		\cite{Xia2022} & Eqns.~(\ref{cd6}), (\ref{cd7}) and (\ref{cd8}) & $<227$ & $0.1\sim30$ & / \\[10pt]
		\cite{Xia2023} & Eqn.~(\ref{cd9}) & $<227$ & $0.1\sim30$  & / \\[10pt]
		\bottomrule
	\end{tabular}
	\label{tab:dragmodel}
\end{table}

\cite{Xia2022} proposed several mean drag correlations for particles in upward particle-laden turbulent channel flows from the PRDNS data of \cite{Yu_Xia_Guo_Lin_2021}. They are,
\begin{equation}
	\Delta C_{D}=0.448\left[ k/U_s^2-(k/U_s^2)_0 \right]^{0.525},
	\label{cd6}
\end{equation}
\begin{equation}
	\Delta C_{D}=\left\{ 
	\begin{array}{ll}
		\displaystyle
		1.25 \frac{0.15Re_m^{0.687}}{1+0.15Re_m^{0.687}} I^2, & I < 1, \\[15pt]
		\displaystyle
		1.64 \frac{0.15Re_m^{0.687}}{1+0.15Re_m^{0.687}} I^{0.687}, & I > 50, \\[1pt]
	\end{array} \right.
	\label{cd7}
\end{equation}
and
\begin{equation}
	\Delta C_{D}=1.32 \frac{0.15Re_m^{0.687}}{1+0.15Re_m^{0.687}} (I-I_0)^{1.1},
	\label{cd8}
\end{equation}
where $k$ is the turbulence kinetic energy (TKE), $(k/u_s^2)_0=0.0125+2.15\phi_s-5.83\phi^2_s$ is the particle-induced TKE ($\phi_s$ is the particle volume fraction), $Re_m=\phi_fRe_p$ is the particle Reynolds number considering the effect of particle volume fraction ($\phi_f$ is the fluid volume fraction), and $I_0=[2(k/u_s^2)_0/3]^{1/2}$ is the particle-induced turbulent intensity. 

Also based on the PRNDS data of \cite{Yu_Xia_Guo_Lin_2021}, \cite{Xia2023} developed another particle drag correlation, as
\begin{equation}
	\Delta C_{D}=0.125\beta^{0.365}(I-I_0)^{0.574}Re_m^{-0.647}(D_p/\eta)^{1.581},
	\label{cd9}
\end{equation}
where $\beta$ is the anisotropy parameter of the turbulence, which is the same as that in equation \ref{cd4}.

Table \ref{tab:dragmodel} summarizes the range of parameters of the studies on the particle drag force models mentioned above.

\subsection{Stochastic models in particle-laden flow}
Employing PRDNS to deal with practical particle-laden flows remains impossible due to the enormous computational cost. 
RANS or LES are more practical alternatives for computing fluid flows in actual applications. 
Nevertheless, throughout the averaging and filtering procedures in RANS and LES, subgrid-scale (SGS) flow information is filtered out and absent, necessitating modeling its effects on the particle.
The existing particle SGS models can be classified into two groups, \emph{i.e.} structural models and stochastic models \citep{marchioli2017,Pozorski2017}. 
The structural models aim to reconstruct the entire SGS flow velocity field \citep{kuerten2006,shotorban2006,marchioli2008,marchioli2008issues,ray2014,park2017,zhou2018,bassenne2019,zhou2020,hausmann2023large,hausmann2023wavelet}. The stochastic models try to retrieve some stochastic characteristics of the SGS flow field using stochastic methods \citep{WangQunzhen1996,fede2006,shotorban2006stochastic,berrouk2007,pozorski2009,jin2010,cernick2015,minier2015,innocenti2016,taniere2016a,Tenneti2016,innocenti2019,sabelnikov2019a,Lattanzi2020,knorps2021,lo2022,rousta2024}.
In this work, we focus on the stochastic method due to its capability in particle force modeling.

Two distinct stochastic modeling approaches have emerged to achieve this purpose. The first approach used the Lagrange stochastic technique to simulate the velocity of the unresolved flow experienced by particles, grounded in the point-particle framework and referred to as the flow Lagrangian stochastic model. 
This method utilizes statistical flow characteristics obtained from RANS or LES to generate instantaneous flow velocities encountered by particles via a stochastic model in the Lagrangian framework. 
The flow Lagrange stochastic methods can be classified into discontinuous random walk models (DRW) and continuous random walk models (CRW).
The DRW model posits that when a particle traverses an eddy during a specific duration, the fluctuating velocity of the flow within that eddy remains constant. Upon moving to the subsequent eddy, a new fluctuating velocity is established, independent of the prior period \citep{Gosman1983AspectsOC,Bocksell_Loth2001,Dehbi2008,Mofakham2020}.
The CRW model connects the current velocity fluctuation of the flow field to that of the preceding time step through a Markov chain constructed from the Langevin equation \citep{wang1992,Bocksell_Loth2006}, offering a continuous variation of the fluid velocity fluctuation and thus deemed more realistic than the DRW model. 
The CRW model has been widely used in simulating particle dispersion in turbulent flows \citep{Bocksell_Loth2006,Dehbi2008,Dehbi2010,Mofakham2019,Mofakham2020,Jaiswal2022,Li2023}.

The other approach for modeling the impacts of the unresolved flow originates at the particle level and incorporates stochastic models for particle position, velocity, or acceleration based on the Langevin equation, referred to as the particle Lagrangian stochastic model.
For example, \cite{Garz2012} proposed a model for particle acceleration by modeling particle velocity fluctuation as a Langevin model to account for the stochastic nature of neighbor particle effects in the Stokes flow regime. 
\cite{Tenneti2016} extended the particle acceleration Langevin model proposed by \cite{Garz2012} to moderate particle Reynolds numbers by PRDNS.
\cite{Esteghamatian2018} proposed a stochastic formulation for the drag of particles considering the effects of the filtered microstructure of turbulence through PRDNS. 
\cite{Lattanzi2020} introduced a detailed description of stochastic methods that can be used in Eulerian-Lagrangian simulations to account for fluctuations in the neighbor-induced drag force of the particles, including the Langevin equation for the position, velocity, and fluctuating drag force of the particles. 
\cite{Lattanzi2022} developed a Langevin drag force model that treats neighbor-induced drag fluctuations as a stochastic force within the Eulerian-Lagrangian framework. They closed the model with the integral timescale for the fluctuating hydrodynamic force and the standard deviation in drag.

\subsection{The present work}
The studies mentioned above indicate that while there are some correlations for the mean particle drag considering the effects of turbulence, no model for instantaneous particle force is available for direct application to simulate the motion of finite-size particles in turbulence. 
To fill this gap, we perform PRDNS of a finite-size particle in turbulence across a wide range of turbulence and particle parameters and offer a stochastic model for the instantaneous particle drag and lateral force, accounting for the effects of turbulence.

The rest of the paper is organized as follows: \S \ref{sec:methods} introduces the simulation methods employed in this work, including PRDNS, turbulent forcing, and particle wake elimination methods. In \S \ref{sec:results}, the new correlations for the mean particle drag and the stochastic models for the fluctuating particle drag and the lateral forces are given. 
In \S \ref{sec:validation}, the proposed model is validated by simulating the motion of a finite-size particle in a turbulent flow and the dispersion of particles in a turbulent channel flow, respectively. Finally, the main conclusions are drawn in \S \ref{sec:conclusion}.

\section{Simulation methods}\label{sec:methods}

\subsection{Particle-resolved direct numerical simulation (PRDNS)}

This study employs PRDNS to evaluate the particle-fluid interaction to quantify the force exerted by the fluid on the particle. The governing equations of fluid flow are the three-dimensional Navier-Stokes equations for incompressible flow
\begin{equation}
	\frac{\partial \boldsymbol u}{\partial t} + \bm{\nabla}\bm{\cdot}\left(\boldsymbol {uu}\right)= {-\bm{\nabla} p}+\frac{1}{Re}\nabla^2 \boldsymbol u+\boldsymbol f_{IB} + \boldsymbol f_{TF},
	\label{eq1}
\end{equation}
and the continuity equation,
\begin{equation}
	\bm{\nabla} \bm{\cdot} \boldsymbol u=0.
	\label{eq2}
\end{equation}
where $\boldsymbol u$ is the fluid velocity, $p$ is the fluid pressure, $Re$ is the Reynolds number, $\boldsymbol f_{IB}$ is the immersed-boundary volumetric force of the particle on the flow, and $\boldsymbol f_{TF}$ is the turbulence forcing term. The numerical solver used in the PRDNS study is the opensource code CP3d \citep{Gong2023}. 
The fractional step approach from \cite{kim1985} is utilized for equations~(\ref{eq1}) and (\ref{eq2}) to ensure that the flow velocity field remains divergence-free. The second-order Adams-Bashforth scheme is employed for the convection term, while the Crank-Nicolson scheme is adopted for the viscous term in the time advancement. The second-order central difference scheme is used for spatial discretization. The efficacy of the PRDNS code has been thoroughly validated \citep{Gong2023}.

The interaction between a particle and fluid flow is computed using the direct-forcing immersed boundary (IB) approach developed by \cite{uhlmann2005immersed}, where a Cartesian grid is employed in the flow solver, and the particle is represented by Lagrangian surface points. 
The IB method enforces the non-slip boundary condition on the particle surface by including $\boldsymbol f_{IB}$ on the right-hand side of equation~(\ref{eq1}), which is determined by the difference in velocity between the Lagrangian point and the interpolated flow velocity.
The interpolation between the Lagrangian points and the Eulerian flow field is performed using a regularized Dirac delta function \citep{roma1999adaptive}.
More specifically, the IB force $\boldsymbol f_{IB}$ is calculated using a multidirect forcing scheme \citep{luo2007full,Breugem2012second}. 
This method incorporates an external iterative process to determine the interaction force $\boldsymbol f$ between the particle and the flow with the direct forcing IB method and can reduce the error associated with the diffused interaction between Lagrangian points and the flow field, thus improving the accuracy of the particle-flow interaction computation.

\subsection{Turbulence forcing method}

\begin{figure}
	\centering
	\subfigure{\includegraphics[width=0.45\textwidth]{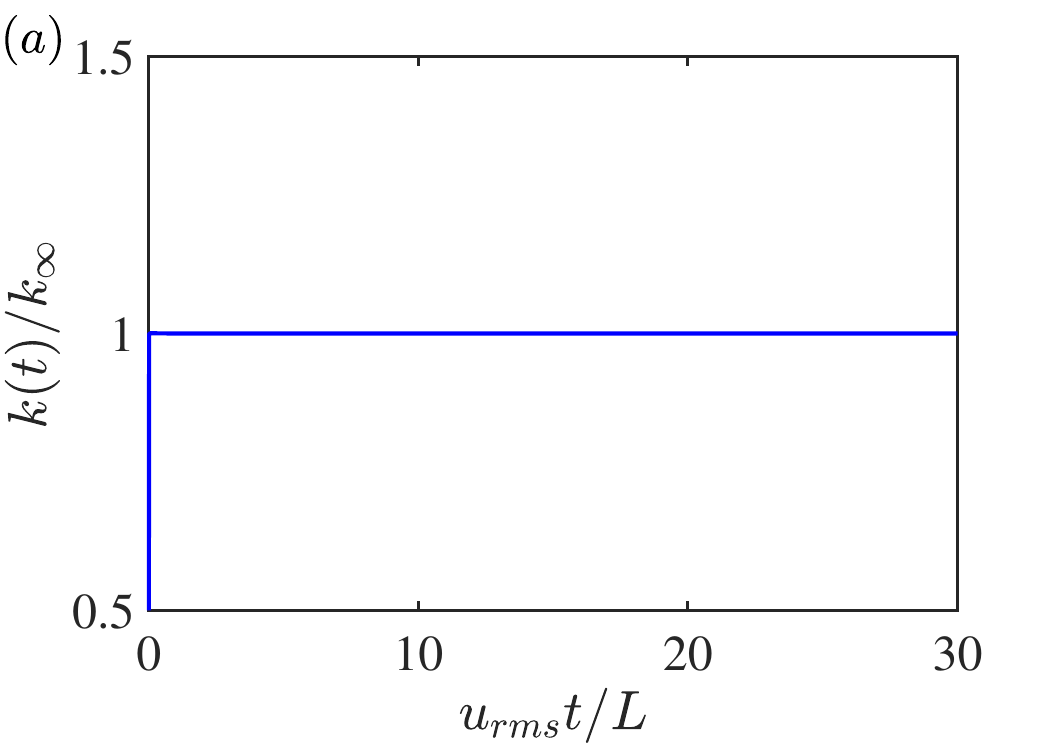}}
	\subfigure{\includegraphics[width=0.45\textwidth]{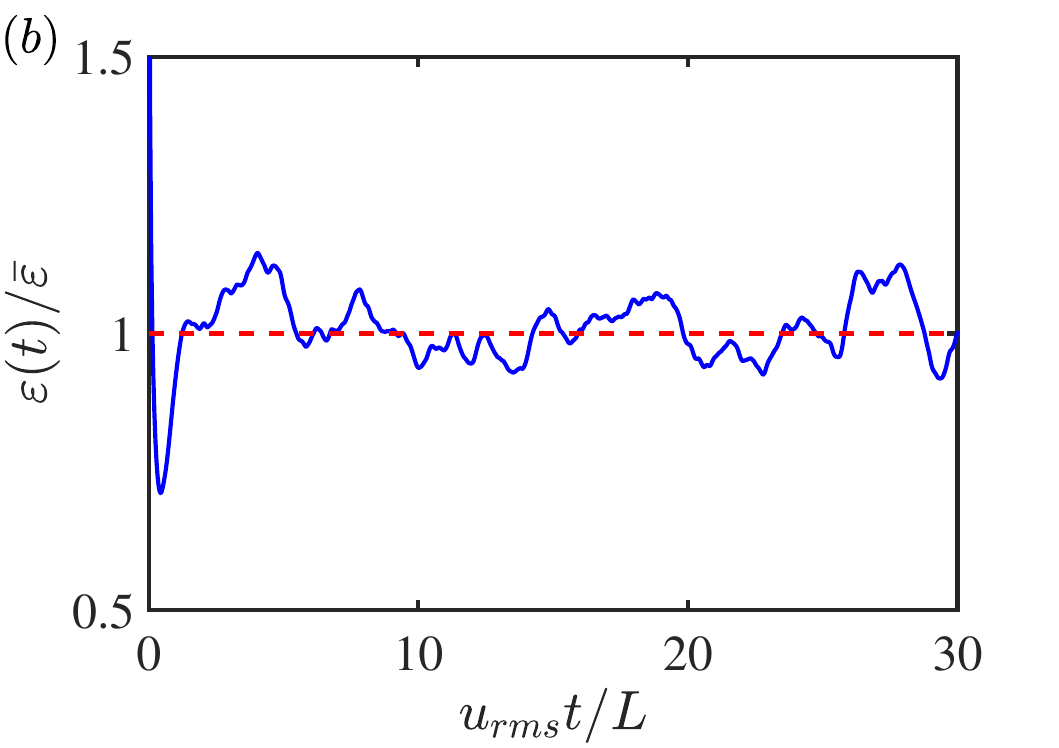}}
	\caption{Time variations of the spatially averaged TKE (a) and TDR (b) of the turbulent flow using the turbulence forcing method. Here 
		$k_\infty=(3/2)u_{rms}^2$ is the target TKE, $u_{rms}$ is the target rms value of turbulent velocity, and $L=u_{rms}^3/\bar{\varepsilon}$ is the integral length scale. Here $\overline{(\cdot)}$ represents temporal averaging.}
	\label{figtke}
\end{figure}

To accurately determine the relationship between the forces of the particle and the turbulence, a statistically stationary background turbulent-flow field must be generated and evolved.
However, a turbulent flow without mean shear, such as homogeneous and isotropic turbulence (HIT), will decay over time in the absence of external energy input. 
Therefore, a forcing method should be employed to maintain the spatially averaged TKE and the turbulence dissipation rate (TDR) to be statistically stationary in the simulation.
The main idea behind the turbulent forcing method is to add a forcing term $\boldsymbol{f}_{TF}$ in the right-hand side of the momentum equation~(\ref{eq1}), where
\begin{equation}
	\boldsymbol f_{TF}=A \widetilde {\boldsymbol u}^{(2)}.
	\label{eq3}
\end{equation}
Here $\widetilde {\boldsymbol u}^{(2)}$ represents the large-scale velocity fluctuation obtained by a low-pass filter that preserves the two lowest Fourier modes of the turbulent flow. 
The time-varying prefactor $A$ is determined from the TKE balance according to \citet{Bassenne2016Constant} 
\begin{equation}
	A(t)=\frac{\varepsilon(t)-G[k(t)-k_\infty]/t_{l,\infty}}{2k(t)},
	\label{eq4}
\end{equation}
where {$\varepsilon(t)=\langle{\nu\left( \partial u_i/\partial x_j \right) \left( \partial u_i/\partial x_j \right)}\rangle$ is the time-varying spatially averaged TDR, the angle bracket $\langle \cdot \rangle$ denotes spatial averaging, $k(t)=\langle{u_i^\prime u_i^\prime}\rangle/2$ is the time-varying spatially averaged TKE, $u_i^\prime$ is the flow velocity fluctuation, $k_\infty$ is the target TKE, and $G/t_{l,\infty}$ is set to a constant of 50, which can ensure a fast development of background turbulence to a statistically stationary state. 
Figure~\ref{figtke} shows the time variations of $k(t)$ and $\varepsilon(t)$ of a turbulent flow using the turbulence forcing method at $Re_\lambda=82$. It is seen that both $k(t)$ and $\varepsilon(t)$ reach a statistically stationary state very quickly once the turbulent forcing is applied. 

{Furthermore, to demonstrate the accuracy of the present turbulent forcing method, we conduct a HIT DNS with the same setting as that on a $1024^3$ grid in JHTDB (Johns Hopkins Turbulence Database) \citep{JHTDB-HIT}. Once the simulation reaches a statistically stationary state, the turbulent kinetic energy is determined to be $k = 0.706$, the dissipation rate $\varepsilon = 0.099$, and the Taylor microscale Reynolds number $Re_\lambda = 427$. They are in close agreement with those reported in JHTDB, where $k = 0.705$, $\varepsilon = 0.103$, and $Re_\lambda = 433$. Figure \ref{HIT1024} compares the turbulent kinetic energy spectrum of the HIT between the JHTDB and the present DNS. It can be seen that the results are in excellent coincidence, indicating that the present turbulent forcing method is feasible.}

\begin{figure}
\centering
\includegraphics[width=0.8\textwidth]{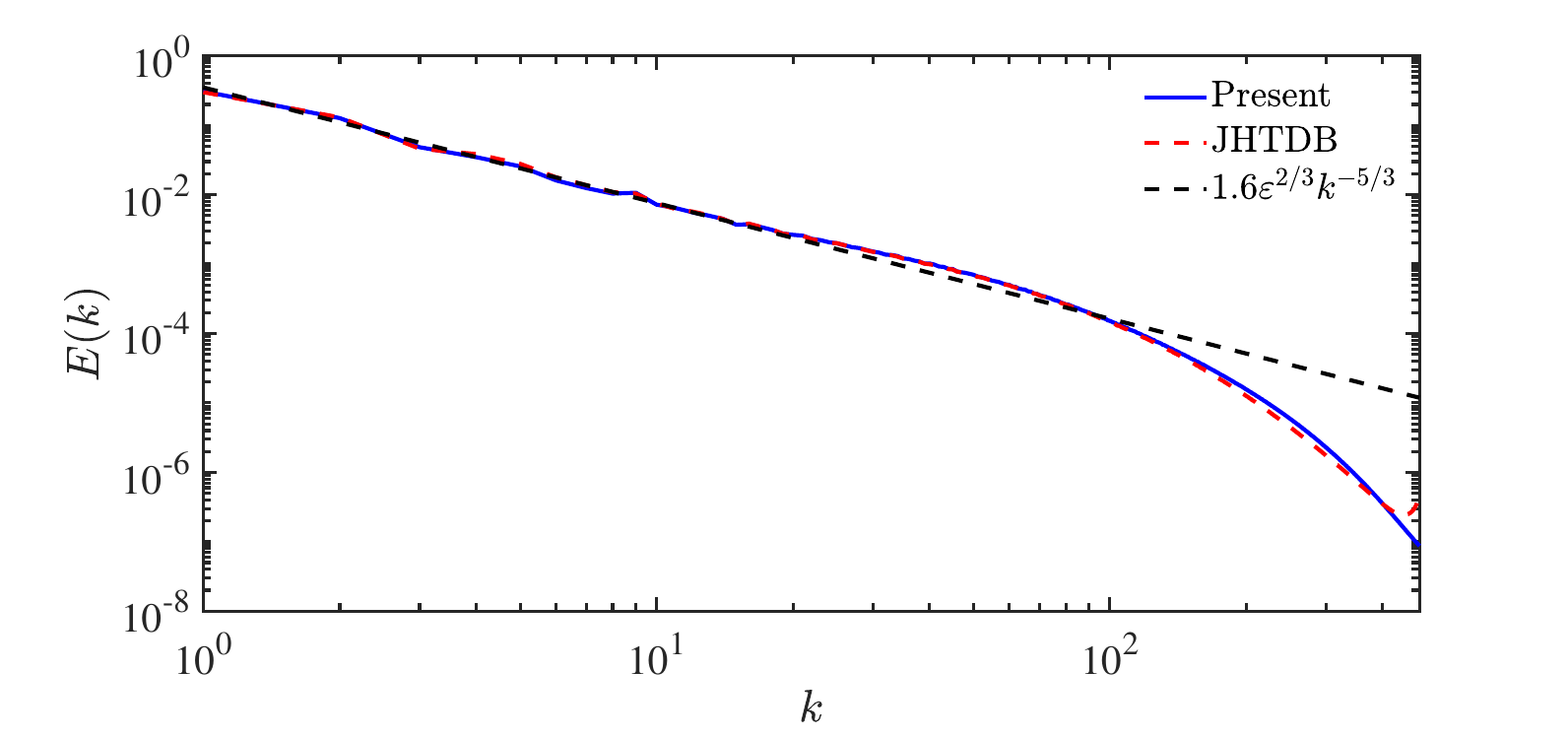}
\caption{Comparison of the turbulent kinetic energy spectrum of HIT between JHTDB and the present DNS.}
\label{HIT1024}
\end{figure}

\subsection{Removing the influence of particle wake}

\begin{figure}
\centering
\subfigure{\includegraphics[width=0.45\textwidth]{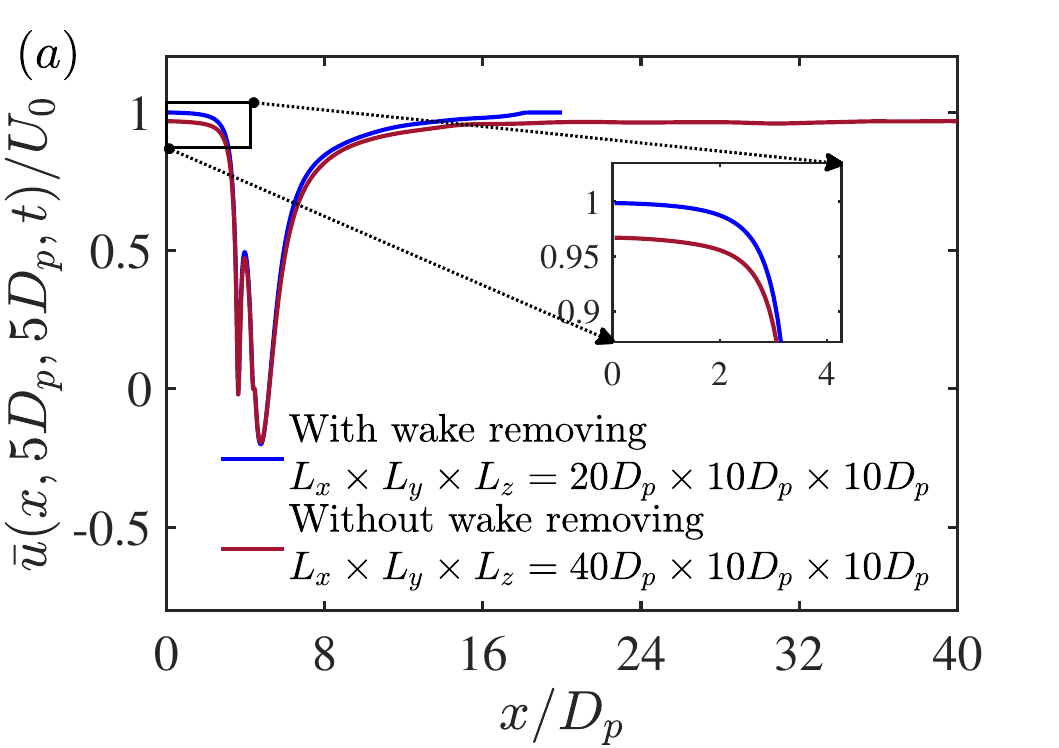}\label{U_DEFICIT}}
\subfigure{\includegraphics[width=0.45\textwidth]{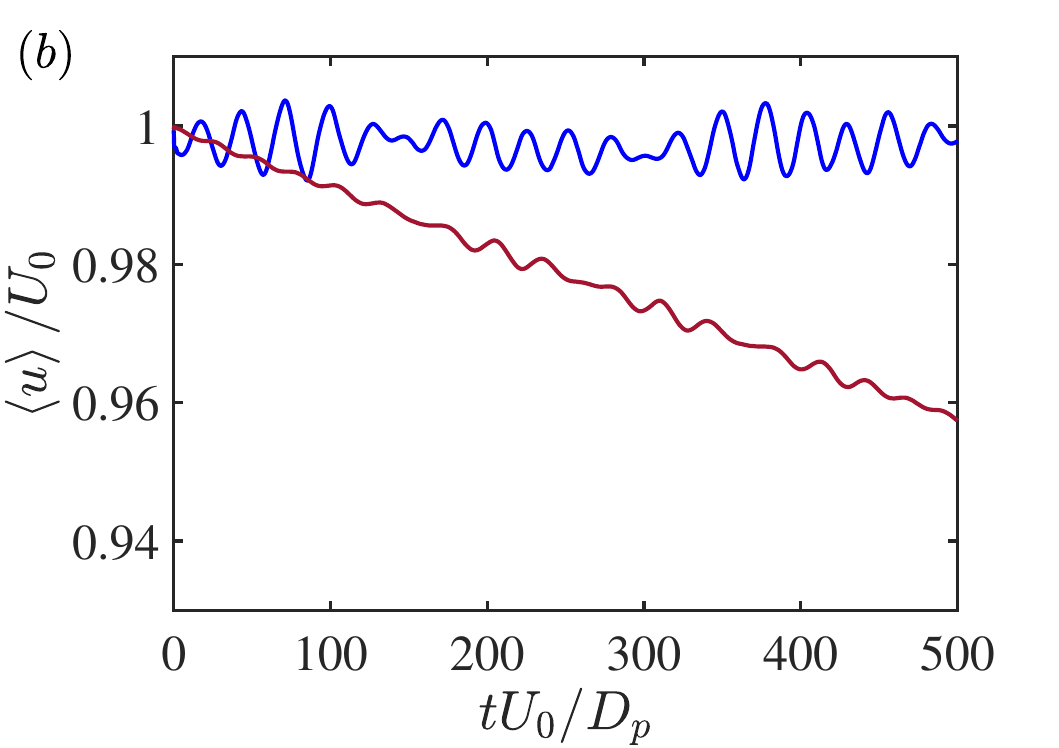}\label{umean}}
\caption{(a) The temporal mean streamwise flow velocity at the central line along the streamwise direction. (b) The time variation of the spatially averaged streamwise velocity.}
\label{figUdeficit}
\end{figure}

\begin{figure}
\centering
\includegraphics[width=0.8\textwidth]{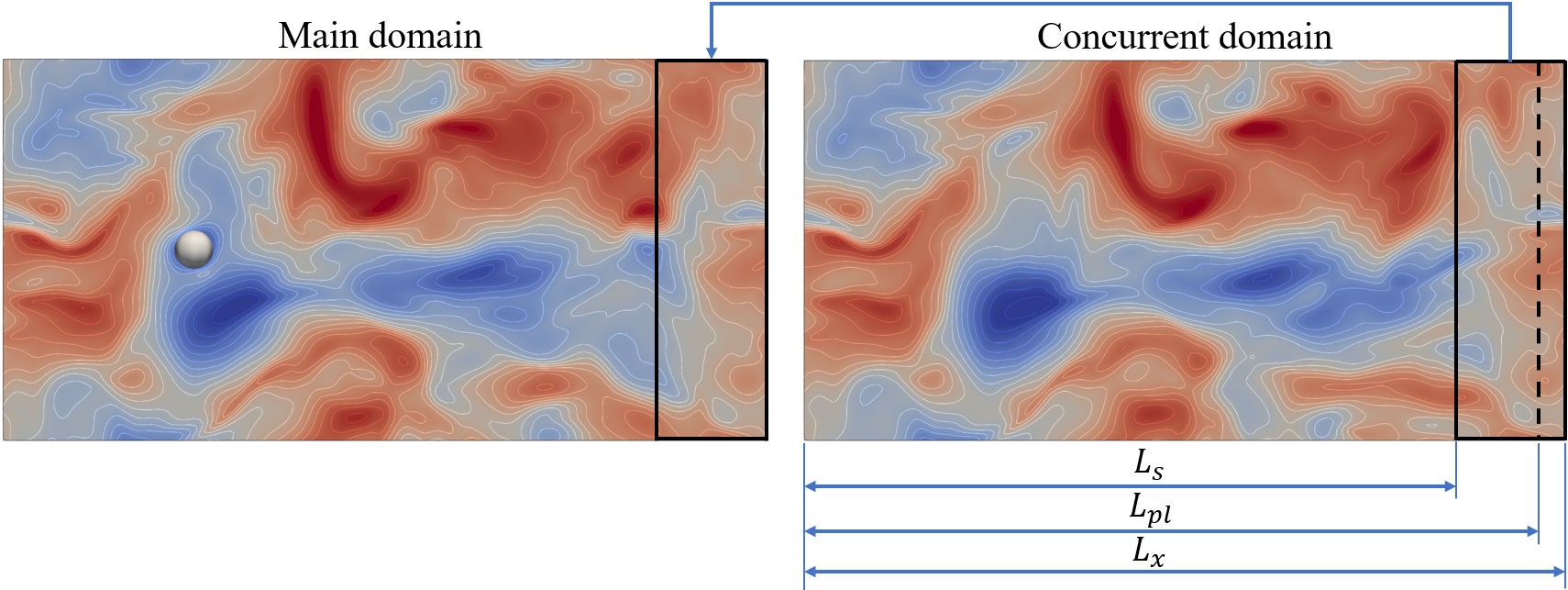}
\caption{Schematic diagram of the concurrent precursor inflow method.}
\label{concurrent_precursor}
\end{figure}

To calculate the fluid force of a particle in a turbulent environment, it is necessary to superimpose a mean flow on the turbulence to ensure the statistical steady state. If traditional inlet and outlet boundary conditions are applied in the streamwise direction, the turbulent flow will decay along the streamwise direction, leading to a nonuniform turbulent flow field around the particle. To address this issue, we employ periodic boundary conditions in the streamwise direction. However, the limited extent of the computational domain will inevitably lead to a reduction in the upstream flow velocity experienced by the particle due to the particle wake.
Figure~\ref{figUdeficit} illustrates that the mean streamwise flow velocity at the inlet is below $U_0$ (where $U_0$ is the desired mean inflow velocity), and the mean streamwise flow velocity all over the domain decreases over time if the influence of the particle wake is not adequately removed. 

This work employs a concurrent precursor method, as proposed by \cite{Stevens2014Concurrent}, to eliminate the impact of particle wake on the upstream flow of the particle, similar to its wide applications in wind turbine/farm flow simulations.
We note that both flows are bluff-body flows, and wakes are generated after the obstacles; therefore, the concurrent precursor method can be quite useful in their simulations. 
This method involves simultaneously simulating a separate flow field with the same configuration free of a particle alongside the flow field that includes a particle, as illustrated in figure~\ref{concurrent_precursor}. 
{At each time step, the flow velocity of the last $0.09L_x$ ($L_x$ is the streamwise length of the simulation domain) of the concurrent precursor simulation domain in the streamwise direction is interpolated into the last $0.09L_x$ of the main simulation domain using the interpolation function as follows:}
\begin{equation}
 \boldsymbol{u}_{m}=w(x)\cdot \boldsymbol{u}_{c}+\left( 1-w(x) \right) \cdot \boldsymbol{u}_{m},
\end{equation}
\begin{equation}
  w(x) = \left\{
    \begin{array}{ll}
    \displaystyle
     \frac{1}{2}\left[1-\cos \left( \pi\frac{x-L_s}{L_{pl}-L_s} \right) \right], & L_s < x \leq L_{pl}, \\[2pt]
    \displaystyle
      1,         & L_{pl} < x \leq L_x,
    \end{array} \right.
\end{equation}
where $\boldsymbol{u}_{m}$ and $\boldsymbol{u}_{c}$ are the flow velocities in the last $0.09L_x$ ($x > 0.91L_x$) of the main and concurrent precursor domains, respectively. $w(x)$ is the interpolation function, with $L_s = 0.91L_x$ and {$L_{pl}=L_s+0.75(L_x-L_s)=0.9775L_x$}, in which $x$ is the streamwise coordinate.
Figure~\ref{figUdeficit} shows that the mean streamwise velocity at the inlet is the desired $U_0$ and the mean bulk streamwise velocity remains constant over time once the influence of the particle wake is removed by the concurrent precursor simulation method. The influence of the concurrent precursor method on the prediction accuracy of particle drag in uniform flows is evaluated in the Appendix \ref{Appendix B}.

\subsection{Simulation setup}\label{Simulation_setup}

The simulation domain in this work spans $ L_x \times L_y \times L_z=20D_p \times 10D_p \times 10D_p$ ($D_p$ represents particle diameter) in the $x$ (streamwise), $y$ (vertical), and $z$ (spanwise) directions, respectively, as seen in figure~\ref{computational_domain}. 
The size of the simulation domain is close to the suggestion of \citet{elmestikawyInfluencePeriodicBoundary2025}.
The periodic boundary conditions are applied in all three directions for the fluid.
The finite-size spherical particle is fixed at $\boldsymbol{x}_p=(5D_p,5D_p,5D_p)$, namely in the center of the cross-plane and 5$D_p$ from the inlet. 

\begin{figure}
\centering
\includegraphics[width=0.6\textwidth]{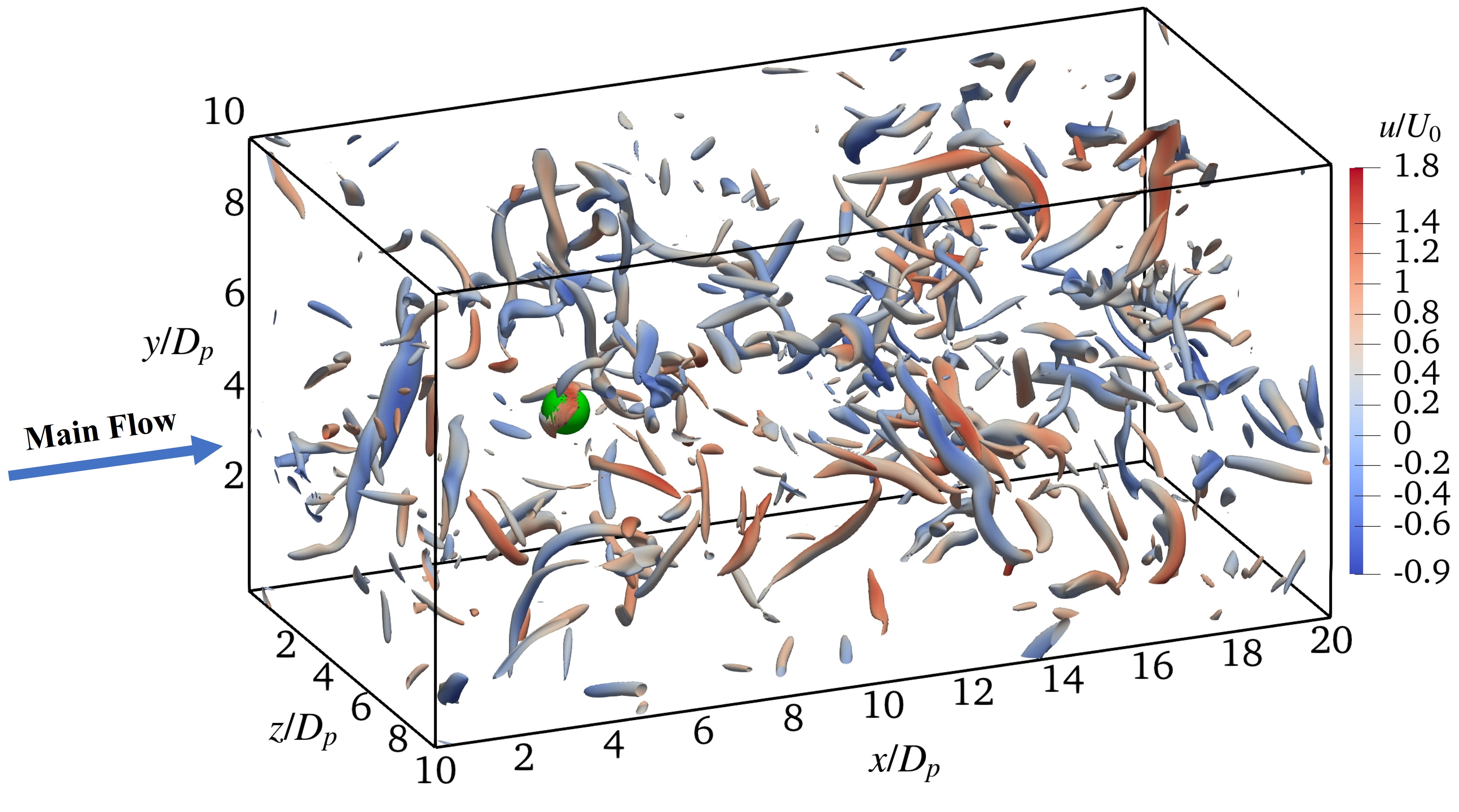}
\caption{Sketch of the simulation domain.}
\label{computational_domain}
\end{figure}

The kinematic simulation method \citep{Kraichnan1970,Saad2017} is used to generate the initial velocity field. Then, the turbulent forcing and concurrent precursor inflow methods previously introduced are adopted to maintain statistically stationary turbulence and eliminate the influence of particle wake, respectively. 
We employ a uniform mesh in all three dimensions to fulfill the prerequisites of the direct-forcing immersed boundary method \citep{uhlmann2005immersed,Breugem2012second}. 
Moreover, to accurately resolve the smallest turbulent motions within the flow field, the grid size $h$ must meet the criteria $h/\eta<2.1$ \citep{Pope_2000}. Consequently, a grid size of $h=0.05$ or $h=0.025$ is selected depending on the Kolmogorov scale of turbulence. Meanwhile, the ratio of particle size to grid size $D_p/h$ ($D_p=1$) is 20 or 40 in this work, which can sufficiently resolve the flow around the particle. 

There are five independent dimensional parameters involved in this problem, \emph{i.e.} the mean inflow velocity (mean slip velocity between particle and flow) $U_0$, the particle diameter $D_p$, the fluid kinematic viscosity $\nu$, the turbulent velocity intensity $u_{rms}$, and the mean turbulence dissipation rate $\varepsilon$. 
These parameters can constitute a group of three independent dimensionless variables, like the slip-velocity-based particle Reynolds number $Re_p=U_0D_p/\nu$, the turbulence intensity ratio $I=u_{rms}/U_0$, and the particle-to-turbulence scale ratio $D_p/\eta$, or the turbulence-velocity-based particle Reynolds number $Re_p^\prime=u_{rms}D_p/\nu$, $I=u_{rms}/U_0$, and $D_p/\eta$, or the Taylor-scale-based turbulence Reynolds number $Re_\lambda=u_{rms}\lambda/\nu$, $I=u_{rms}/U_0$, and $D_p/\eta$, where the Kolmogorov-scale size $\eta=(\nu^3/\varepsilon)^{1/4}$ and the Taylor-microscale size $\lambda={(15\nu u_{rms}^2/\varepsilon)}^{1/2}$.
Different combinations of the dimensionless parameters are, in fact, equivalent to one another, \emph{e.g.} $Re^\prime_p=IRe_p$, $Re_\lambda=\sqrt{15}\left( D_p/\eta \right)^{-2} Re^{\prime2}_p$.
There are totally 28 cases in the PRDNS, covering {$Re_p=1\sim300$, $D_p/\eta=1.6\sim76.9$, and $I=0.1\sim20$.}  
Table~\ref{tab:parameters} in the Appendix \ref{Appendix A} shows the corresponding values of the 
parameters in the PRDNS in the upper part of $D_p/\eta$ greater than 1. 
Furthermore, to cover a larger range of $D_p/\eta$, point-particle direct numerical simulation (PPDNS) using the Schiller-Neumann (S-N) drag law \citep{Schiller&Neumann1933} is also conducted with 15 cases, covering {$Re_p=0.1\sim3.2$, $D_p/\eta=0.22\sim0.99$, and $I=0.5\sim20$}, as shown in the bottom part of table~\ref{tab:parameters} in the Appendix \ref{Appendix A}. 

\subsection{Flow characteristics}

\begin{figure}
\centering
\subfigure{\includegraphics[width=0.32\textwidth]{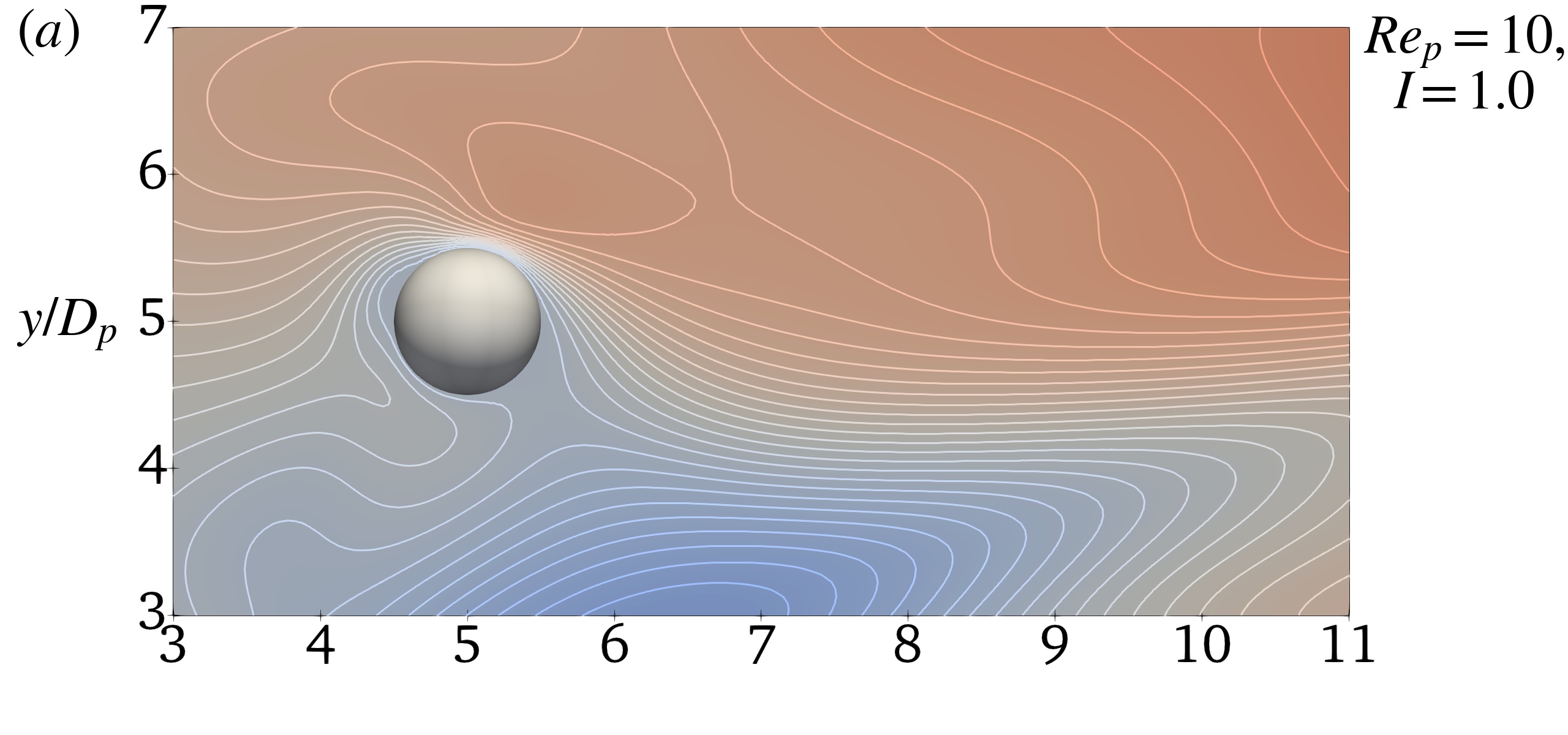}}
\subfigure{\includegraphics[width=0.32\textwidth]{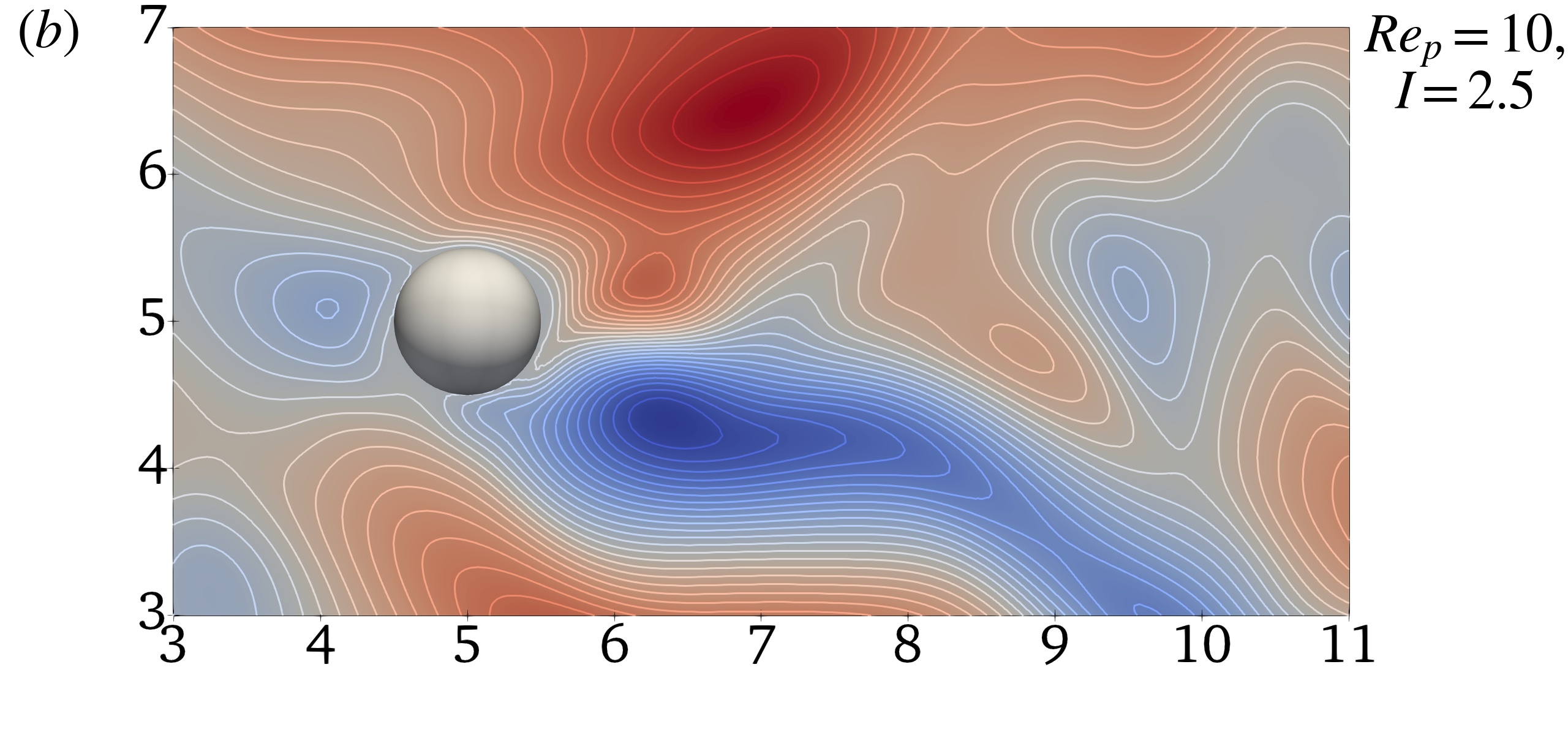}}
\subfigure{\includegraphics[width=0.32\textwidth]{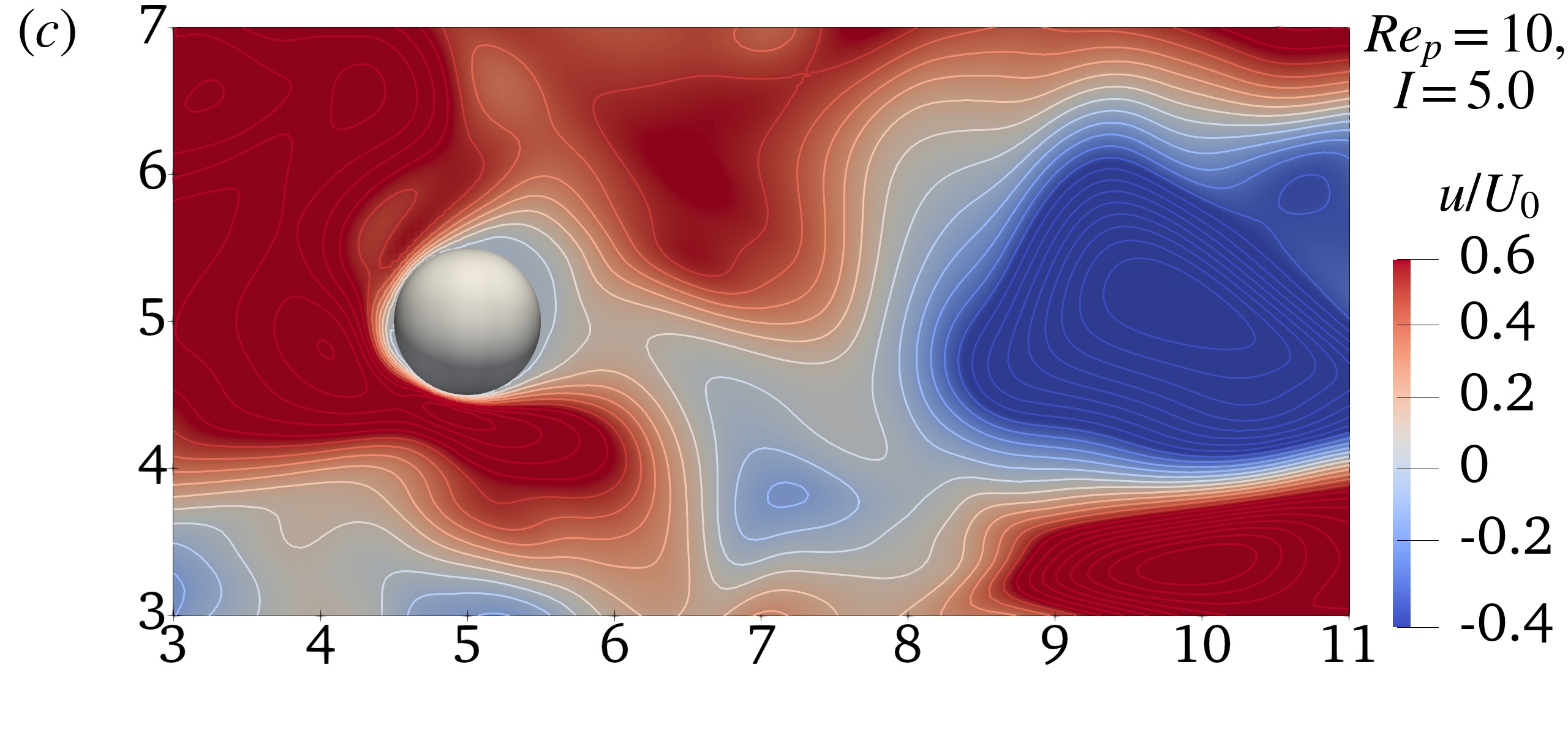}}
\subfigure{\includegraphics[width=0.32\textwidth]{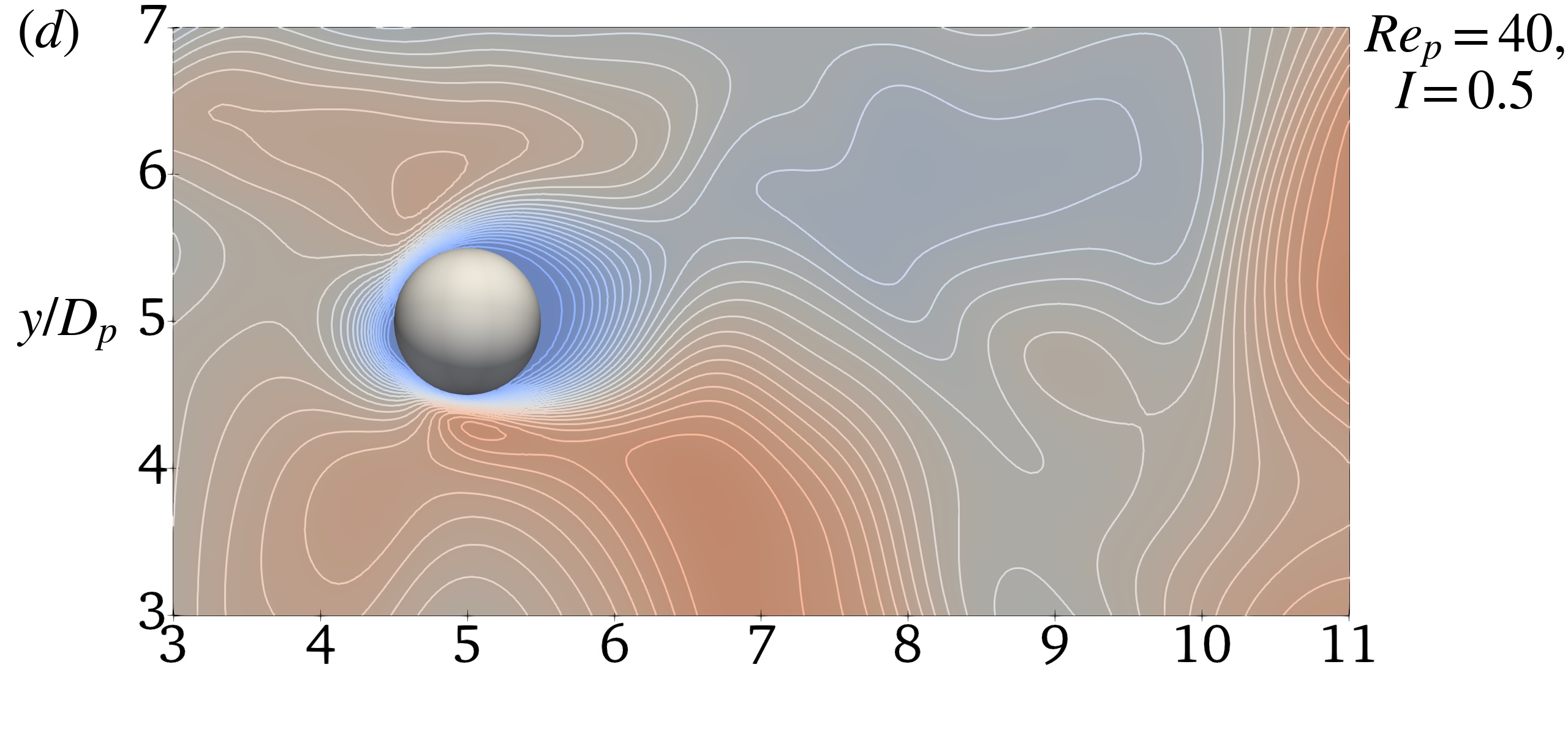}}
\subfigure{\includegraphics[width=0.32\textwidth]{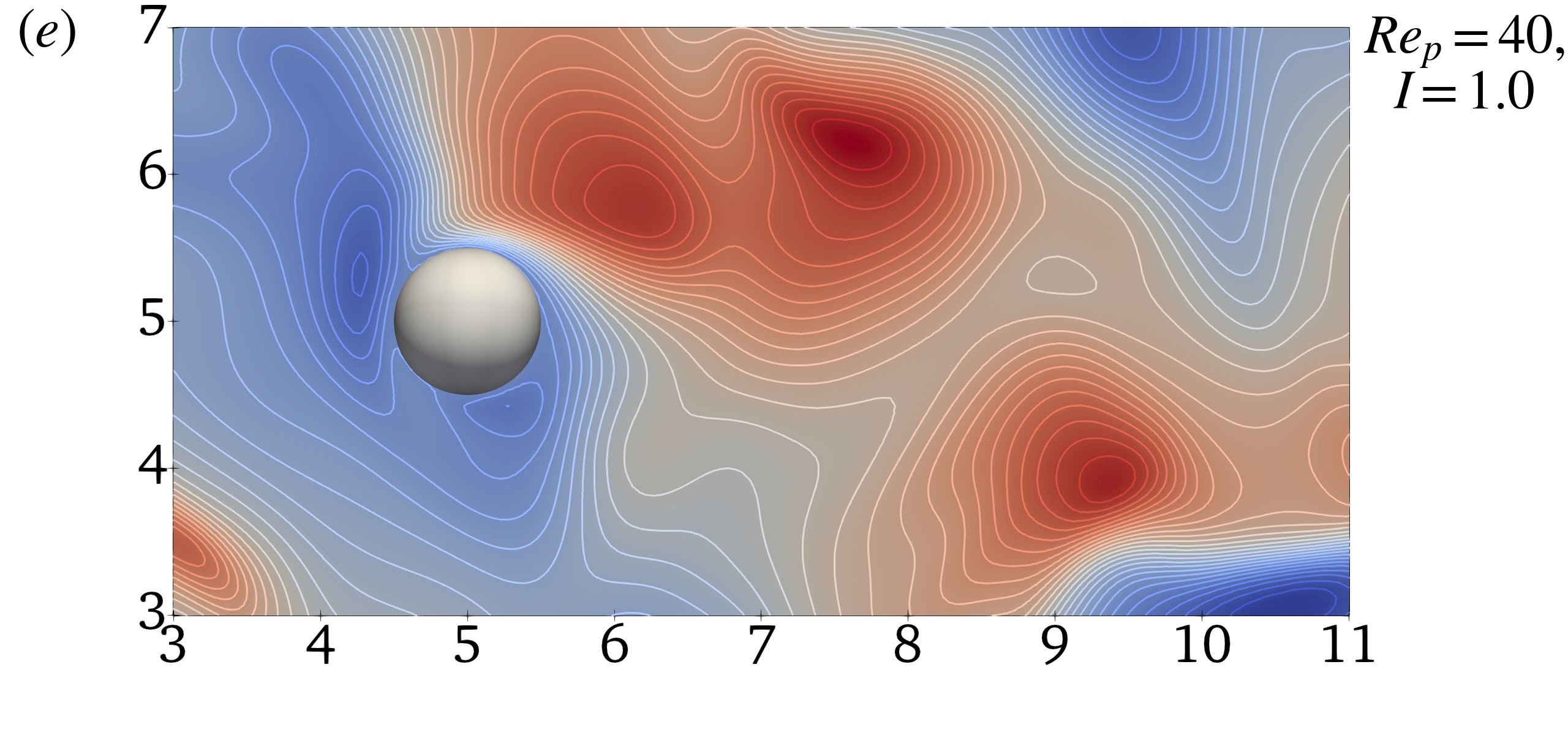}}
\subfigure{\includegraphics[width=0.32\textwidth]{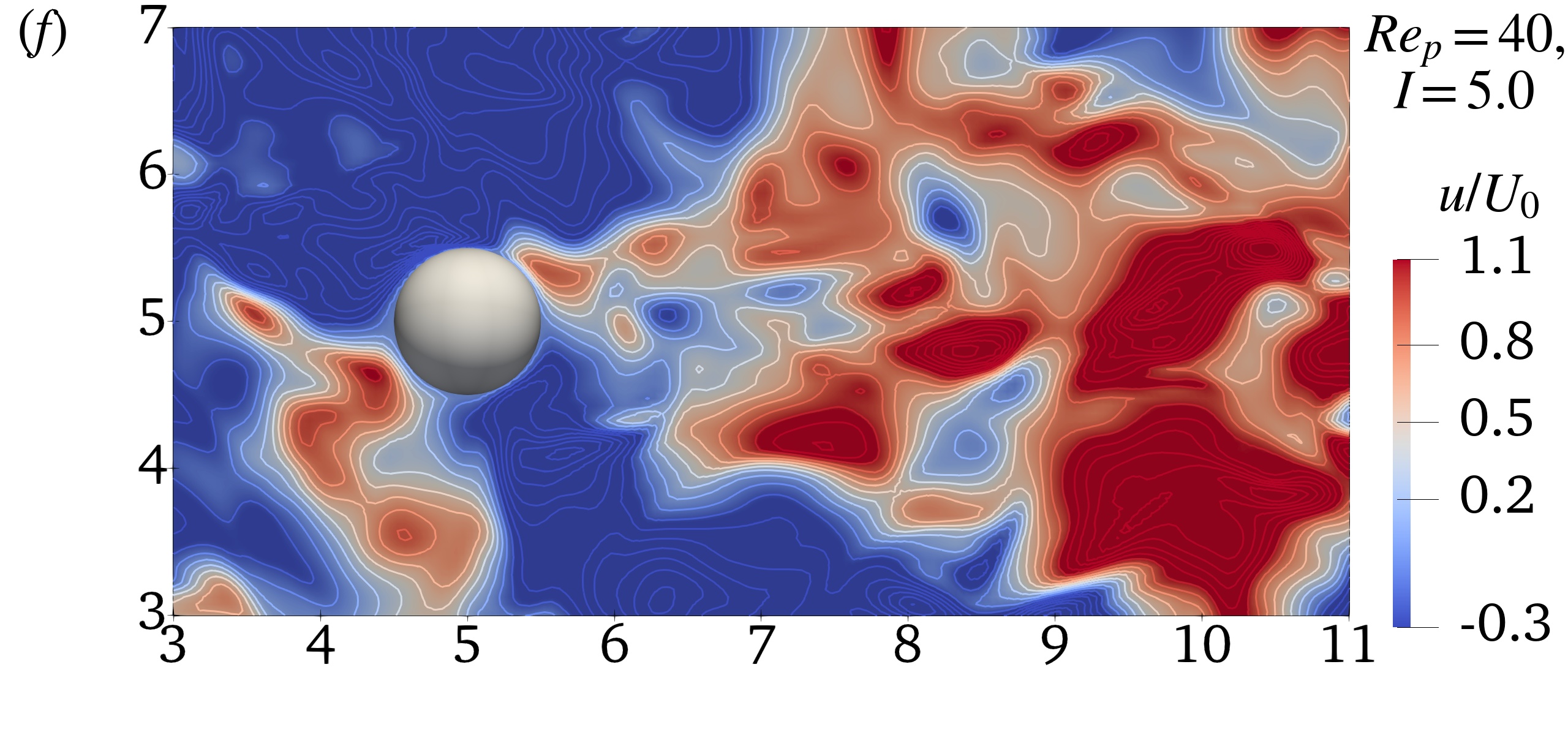}}
\subfigure{\includegraphics[width=0.32\textwidth]{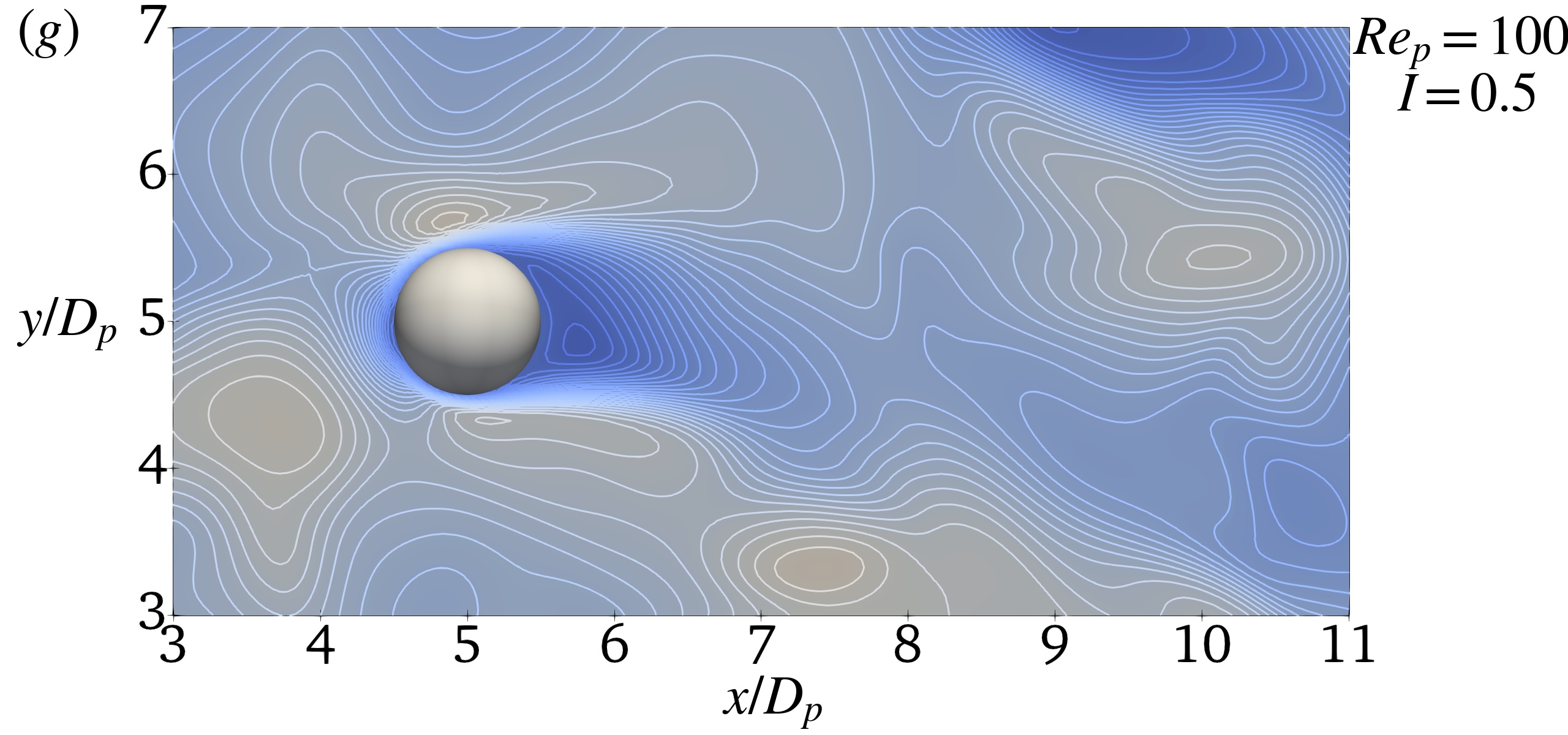}}
\subfigure{\includegraphics[width=0.32\textwidth]{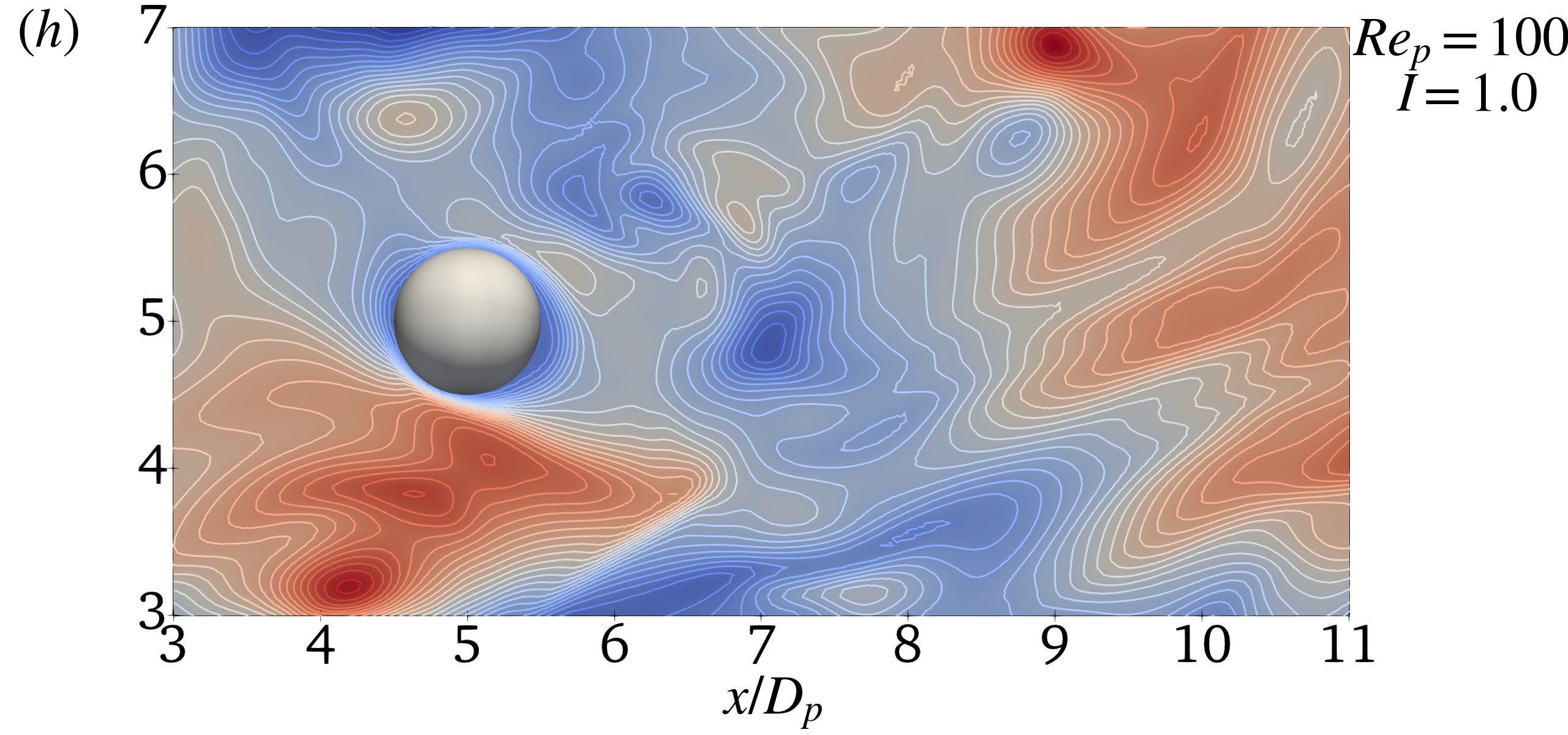}}
\subfigure{\includegraphics[width=0.32\textwidth]{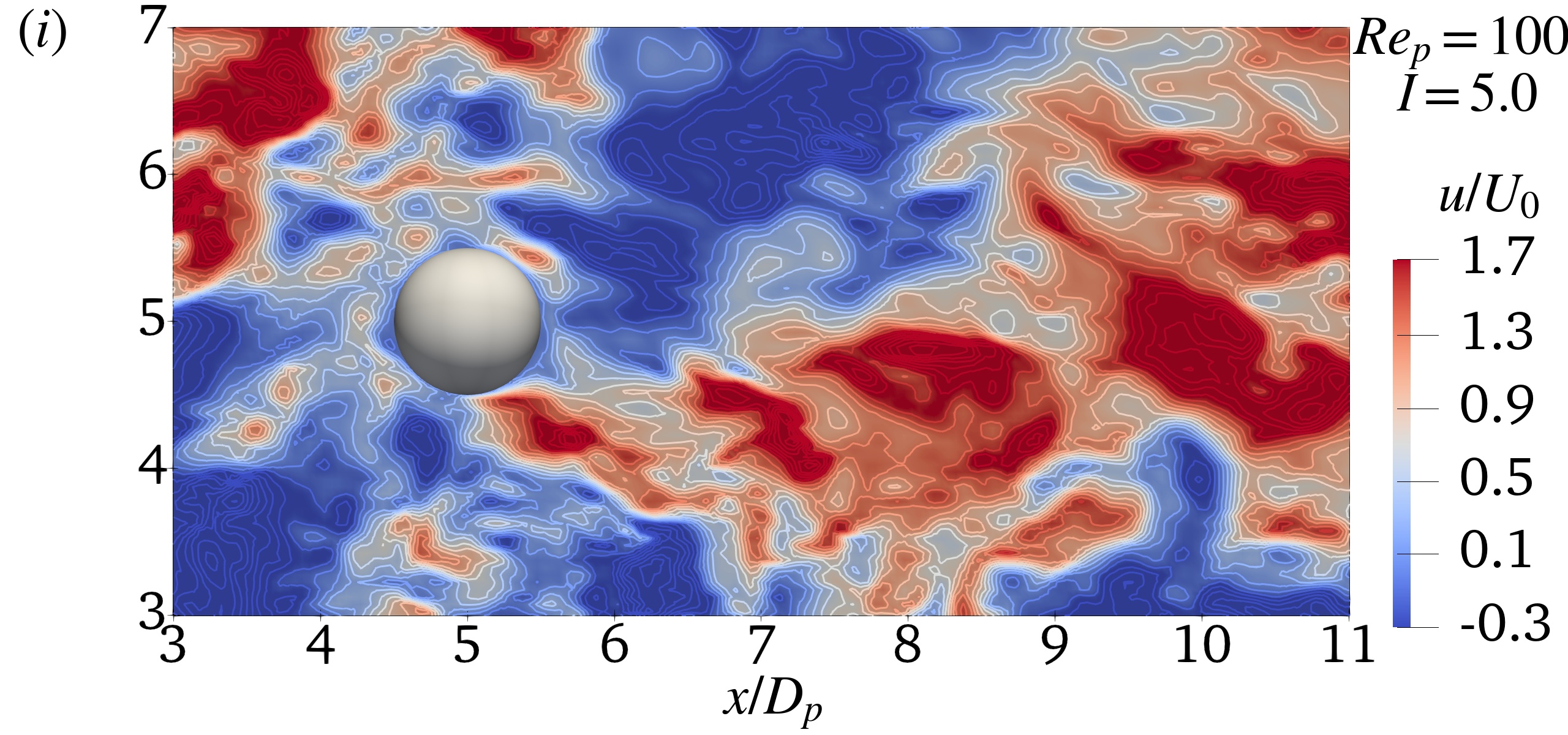}}
\caption{Streamwise velocity contours of the flow field around the particle under different conditions.}
\label{flow_characteristic}
\end{figure}

Figure~\ref{flow_characteristic} illustrates the streamwise velocity contours of the flow field around the particle under different conditions.
In general, the particle is fully immersed in the turbulence, and a very thin boundary layer can be observed adjacent to the particle.
The particle may interact with and modify the local flow, although only one particle is considered and the disperse phase is quite dilute. 
The modulation effect of the finite-size particle on the evolution of the turbulence is left for future studies.
Furthermore, it can be seen that as the turbulence intensity progressively increases, the flow structures surrounding the particles become more complex and multiscale due to the increase in the flow Reynolds number, and the magnitude of the flow velocity gradually intensifies. Furthermore, comparing the second row ($Re_p = 40$) with the third row ($Re_p = 100$), it is evident that an increase in the particle Reynolds number leads to a greater abundance of small-scale structures within the flow field, which can also be attributed to the increase of the flow Reynolds number with the same turbulence intensity ratio $I$ (see table~\ref{tab:parameters}).

\section{Fluid force models}\label{sec:results}

\subsection{Mean drag force model}

This section presents new correlations for the mean particle drag coefficient that consider turbulence effects. 
For each condition shown in table~\ref{tab:parameters} of Appendix \ref{Appendix A}, a time series of fluid force can be calculated from PRDNS or PPDNS. The drag force is the part that aligns with the mean slip velocity direction, which is the direction in the mean flow.
Figure~\ref{deltacd_I} illustrates the relative increments of the mean particle drag coefficient across various turbulence intensities and particle Reynolds numbers, showing only the PRDNS data. The relative increments in the mean particle drag coefficient are always positive, suggesting that turbulence enhances the particle drag force compared to the uniform flow condition. The relative increase $\Delta C_D$ also increases with the turbulence intensity ratio $I$ with the same particle Reynolds number. These observations are consistent with previous studies \citep{BRUCATO1998,ZENG_2008,Homann_Bec_Grauer_2013,Wang_Lei_Zhu_Zheng_2023,Peng2023,Xia2022,Xia2023}. 

\begin{figure}
\centering
\includegraphics[width=0.5\textwidth]{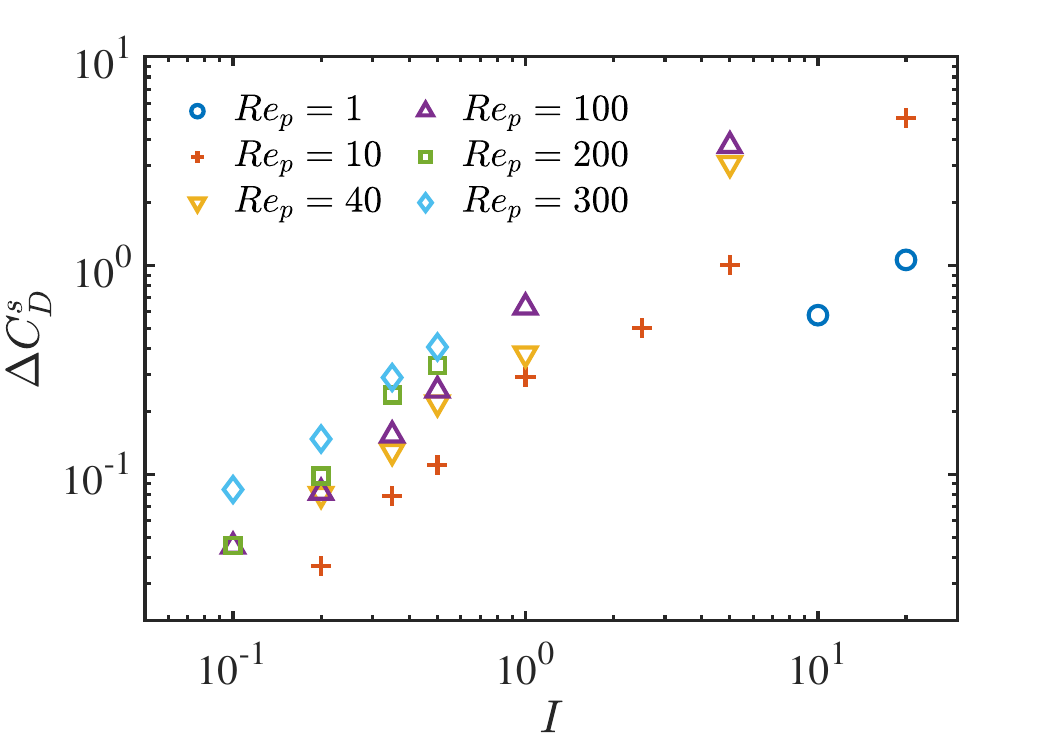}
\caption{Relative increments of the mean particle drag coefficient at different turbulent intensities and particle Reynolds numbers.}
\label{deltacd_I}
\end{figure}

The correlation~(\ref{cd3}) proposed by \cite{Homann_Bec_Grauer_2013} for the mean particle drag coefficient in turbulence was based on the following two assumptions: (1) The instantaneous particle drag is determined by the S-N drag law using the instantaneous slip velocity; (2) the instantaneous slip velocity follows a Gaussian distribution. 
With similar assumptions, \cite{Xia2022} proposed the correlation (\ref{cd7}). In the following, we briefly present the general derivation of the correlations~(\ref{cd3}) and (\ref{cd7}).
Commencing with the instantaneous slip velocity $\boldsymbol{U}_s=(u_s, v_s, w_s)$, the instantaneous particle drag force can be written as
\begin{equation}
  {F}_D(t) = \frac{24}{\Tilde{Re}_p(t)} \left[ 1 + 0.15\Tilde{Re}_p^{0.687}(t) \right] \cdot \frac{1}{2}\rho_f |\boldsymbol{U}_s(t)| {u}_s(t)A,
\end{equation}
where $\Tilde{Re}_p(t)=|\boldsymbol{U}_s(t)|D_p/\nu$ is the particle Reynolds number based on the instantaneous slip velocity, $\rho_f$ is the fluid density, and $A=\pi D_p^2/4$ is the particle sectional area. 

\begin{figure}
\centering
\subfigure{\includegraphics[width=0.45\textwidth]{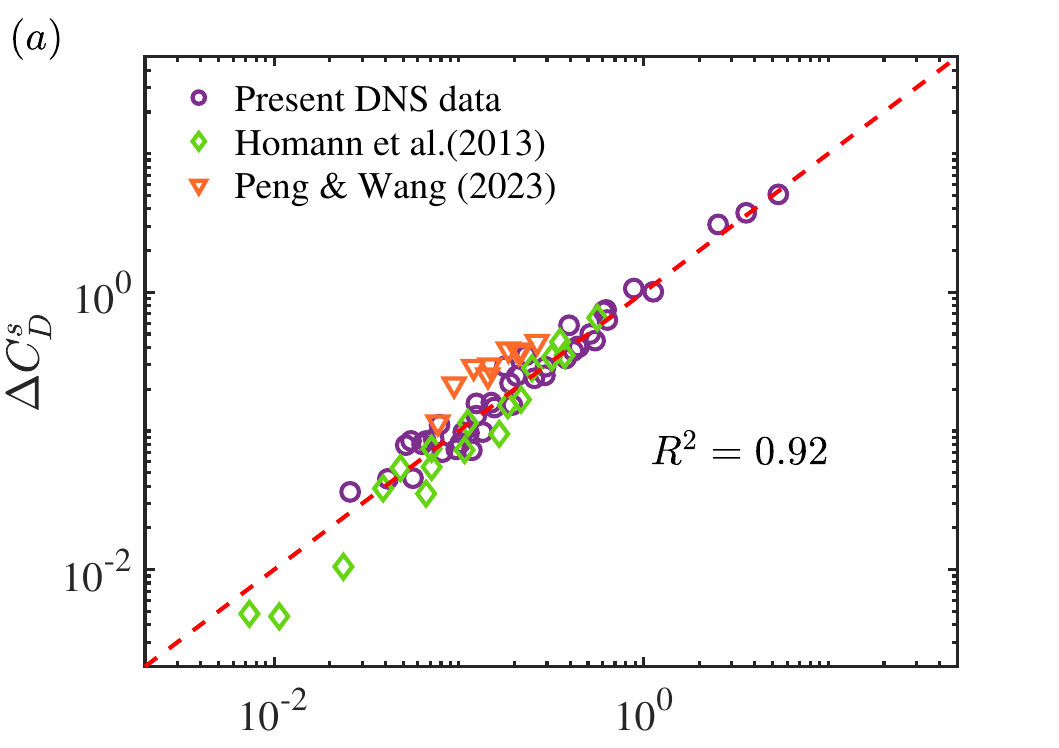} \label{deltacd_WangHomann_WangPre}}
\subfigure{\includegraphics[width=0.45\textwidth]{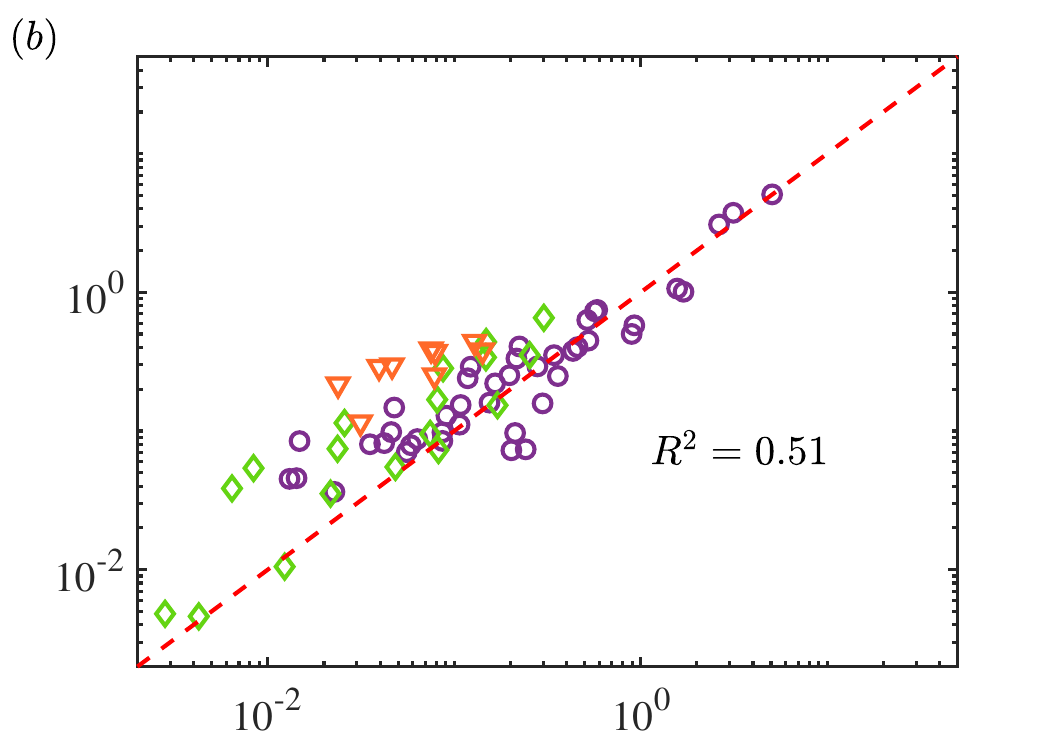} \label{deltacd_WangHomann_WangPre1}}
\subfigure{\includegraphics[width=0.45\textwidth]{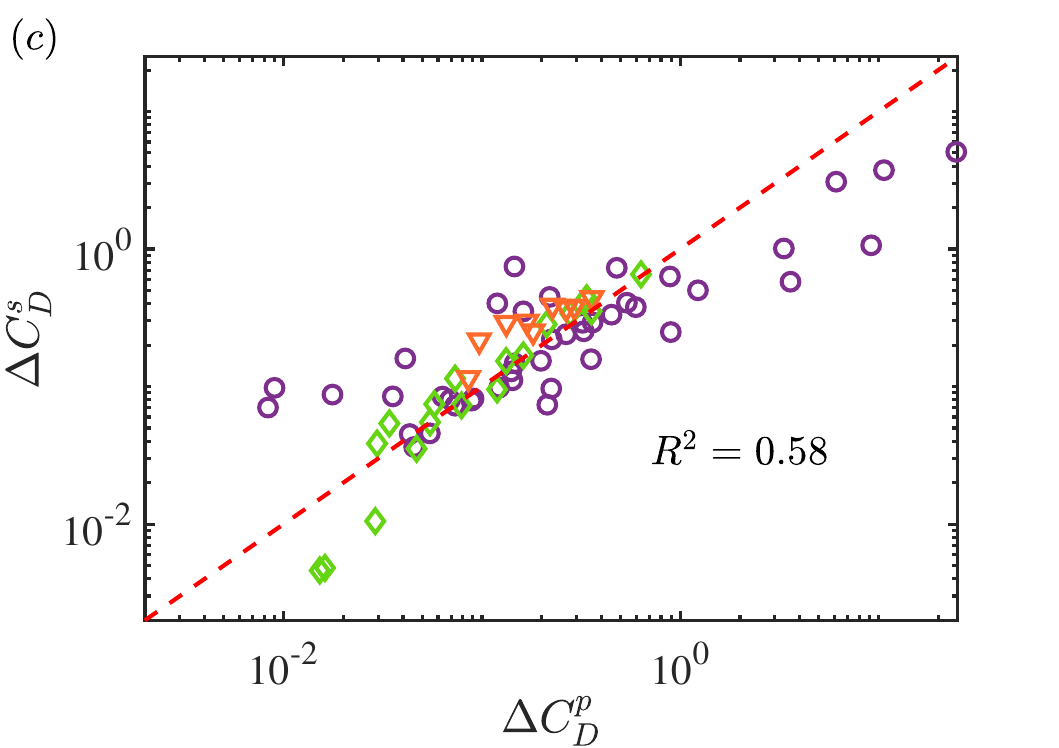} \label{deltacd_WangHomann_HomannPre}}
\subfigure{\includegraphics[width=0.45\textwidth]{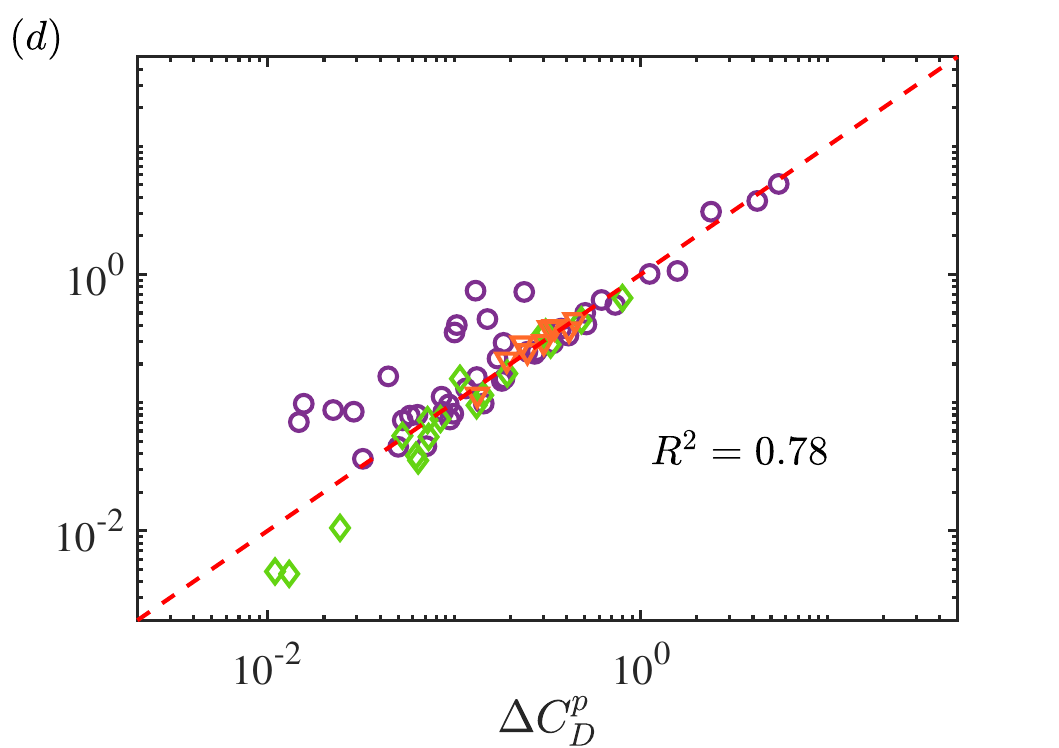} \label{deltacd_WangHomann_PengPre}}
\caption{Comparisons of the relative increment of the mean particle drag coefficient between the PRDNS data and the predictions by (a) the correlation (\ref{deltacdp2}), (b) the correlation (\ref{deltacdp1}), (c) the correlation (\ref{cd2}), and (d) the correlation (\ref{cd4}). Here, $R^2$ represents the coefficient of determination; the closer $R^2$ is to one, the better the agreement between the prediction and the DNS data.}
\label{deltacd_comp}
\end{figure}

Therefore, the mean particle drag coefficient can be obtained as follows:
\begin{eqnarray}
  {C}_D = \frac{\overline{F_{D}(t)}}{(1/2) \rho_f \overline{|\boldsymbol{U}_s(t)|}^2 A} &=& \frac{24}{Re_p} \frac{ \overline{u_s(t)\left[ 1+0.15\Tilde{Re}_p^{0.687}(t) \right]}} {\overline{|\boldsymbol{U}_s(t)|}} \nonumber\\
  & = & 
  \frac{24}{Re_p} \left[1+ \frac{\overline{0.15\Tilde{Re}_p^{0.687}(t)u_s(t)}} {\overline{|\boldsymbol{U}_s(t)|}} \right] \nonumber\\
  & = & 
  \frac{24}{Re_p} \left(1+ 0.15Re_p^{0.687} f_1\right),
  \label{cdmean}
\end{eqnarray}
where $Re_p=\overline{|\boldsymbol{U}_s(t)|} D_p/\nu$ ( here $\overline{|\boldsymbol{U}_s(t)|}=U_0$ is the mean slip velocity) and
\begin{equation}    
    f_1=\frac{\overline{|\boldsymbol{U}_s(t)|^{0.687}u_s(t)}}{\overline{|\boldsymbol{U}_s(t)|}^{1.687}}.
    \label{f_1}
\end{equation}
Therefore,
\begin{equation*}
    \Delta C_D = \frac{0.15Re_p^{0.687}}{1+0.15Re_p^{0.687}} (f_1-1).
\end{equation*}

To close (\ref{f_1}), although it can be assumed that the slip velocity $\boldsymbol{U}_s$ follows a Gaussian distribution with a mean of $(\overline{u}_s,0,0)$ and a standard variation of $(u_{rms},u_{rms},u_{rms})$, it is very difficult to get an explicit expression of $f_1$ \citep{Homann_Bec_Grauer_2013}. Therefore, the numerical approach is usually used. For example, \citet{Xia2022} employed numerical integration to calculate the increase in mean drag as a function of $I$. Here we stochastically generate the $\boldsymbol{U}_s$ data and determine the relationship between $f_1$ and $I$, \emph{i.e.}  
$f_1=1+0.815I^{1.687}$ when $I\leq1$ and $f_1=1.662I^{0.687}$ when $I>1$. Consequently, the relative increment of the mean particle drag coefficient can be modeled as follows:
\begin{equation}
  \Delta C_D =  \left\{
    \begin{array}{ll}
    \displaystyle
     \frac{0.122Re_p^{0.687}}{1+0.15Re_p^{0.687}}I^{1.687}, & I \leq 1 \\[15pt]
    \displaystyle
     \frac{0.15Re_p^{0.687}}{1+0.15Re_p^{0.687}}\left(1.662I^{0.687}-1\right), & I > 1. \\[4pt]
    \end{array} \right.
    \label{deltacdp1}
\end{equation}

In the derivation of the correlation
(\ref{deltacdp1}), $f_1$ is assumed to be a function of $I$ only, as it depends only on $\boldsymbol{U}_s$. 
However, \citet{Homann_Bec_Grauer_2013} found that their correlation (\ref{cd3}) does not match the PRDNS data very well at high $I$, so they proposed (\ref{cd2}) by introducing the flow Reynolds number as an additional parameter. This is because the physical problem is controlled by three independent dimensionless parameters, while only two are invoked in (\ref{deltacdp1}). Following a similar idea, we introduce the Taylor-microscale-based Reynolds number $Re_\lambda$ in the modeling $f_1$. Therefore, the second correlation is obtained by fitting the present PRDNS data with a form of $f_1=aI^bRe_\lambda^c+d$, which is
\begin{equation}
 \displaystyle
  \Delta C_D = \frac{0.015Re_p^{0.687}}{1+0.15Re_p^{0.687}}I^{0.858}Re_\lambda^{0.474}.
    \label{deltacdp2}
\end{equation}
The fitting is performed with the curve fitting toolbox of MATLAB by prescribing the functional form.
The correlations (\ref{deltacdp2}) and (\ref{deltacdp1}) are compared to the PRDNS data of the present study, \cite{Homann_Bec_Grauer_2013} and \citet{Peng2023} in figure \ref{deltacd_comp}, as well as the correlations (\ref{cd2}) and (\ref{cd4}). The correlation (\ref{deltacdp2}) exhibits the best concordance with the PRDNS data. It is also noted that $\Delta C_D \rightarrow 0$ as $I \rightarrow 0$ or $Re_\lambda \rightarrow 0$, which means that the mean particle drag force precisely adheres to the S-N drag law with a negligible turbulence effect.

\subsection{Fluctuating drag force model}

In this section, a stochastic model based on the Langevin equation is proposed for the fluctuating particle drag force. The fluctuating drag force denotes the component of the drag force obtained by subtracting its mean value, \emph{i.e.} $F_D^\prime=F_D-\overline{F}_D$. The fluctuating particle drag force in turbulence is intrinsically connected with turbulence fluctuations; hence, $u_{rms}^2$ is utilized as the reference velocity for normalizing the fluctuating particle drag force, which is
\begin{equation}
\displaystyle
  C_D^\prime = \frac{F_{D}^\prime}{(1/2) \rho_f u_{rms}^2A}. \\
\end{equation}

\begin{figure}
\centering
\subfigure{\includegraphics[width=0.45\textwidth]{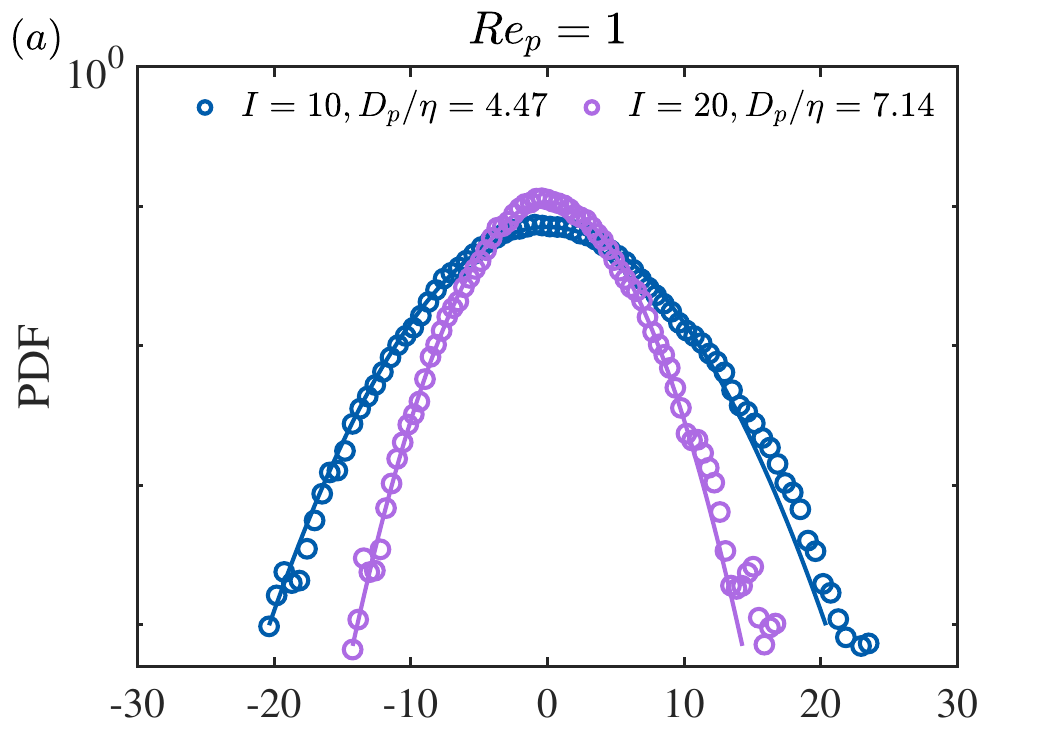}}
\subfigure{\includegraphics[width=0.45\textwidth]{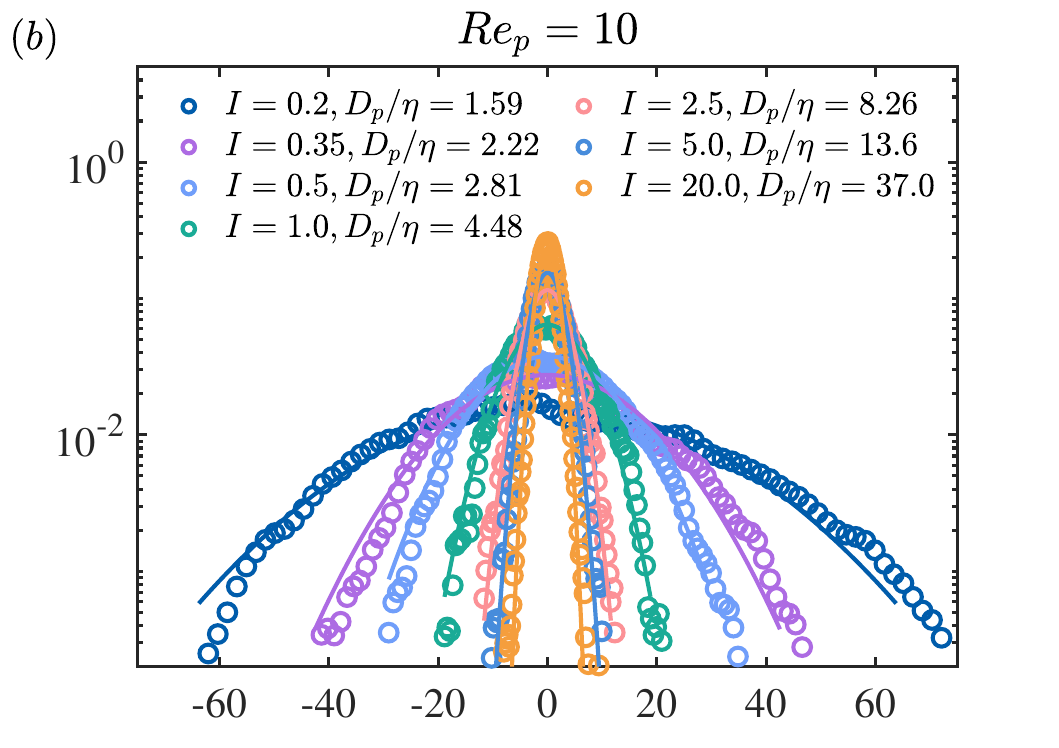}}
\subfigure{\includegraphics[width=0.45\textwidth]{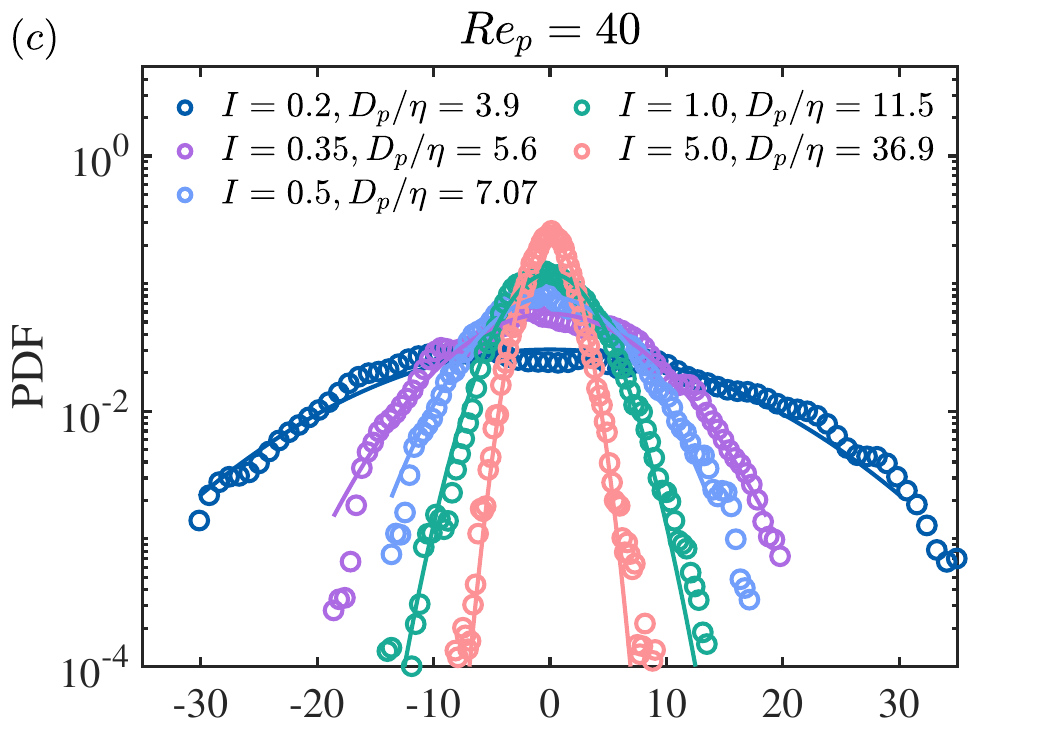}}
\subfigure{\includegraphics[width=0.45\textwidth]{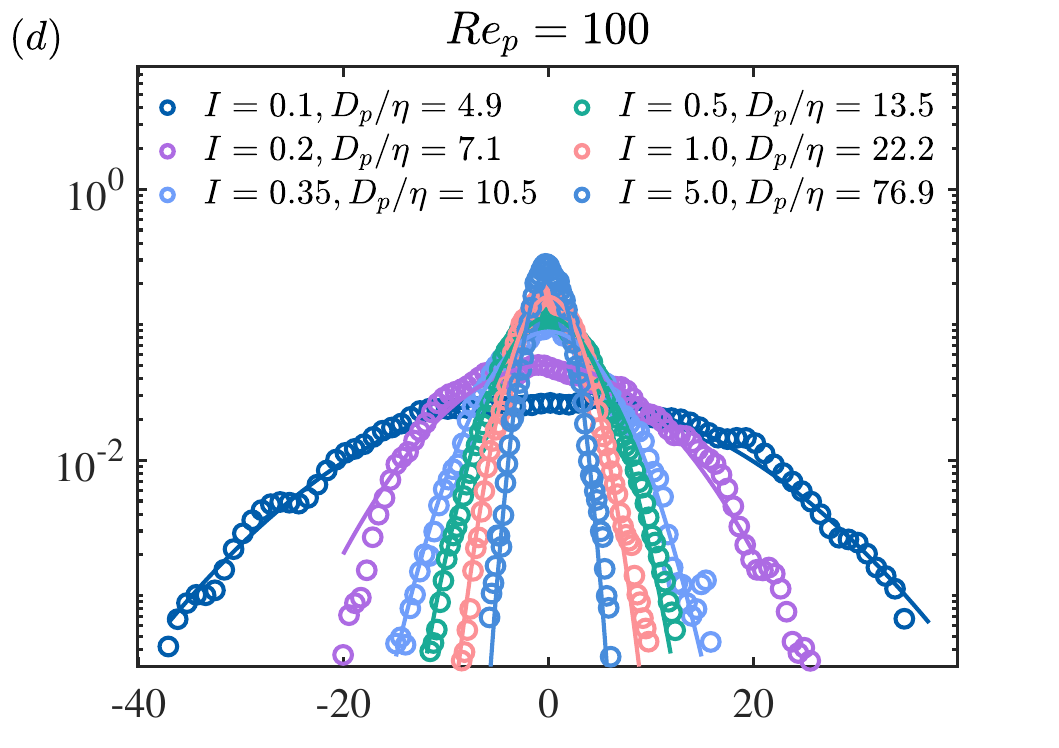}}
\subfigure{\includegraphics[width=0.45\textwidth]{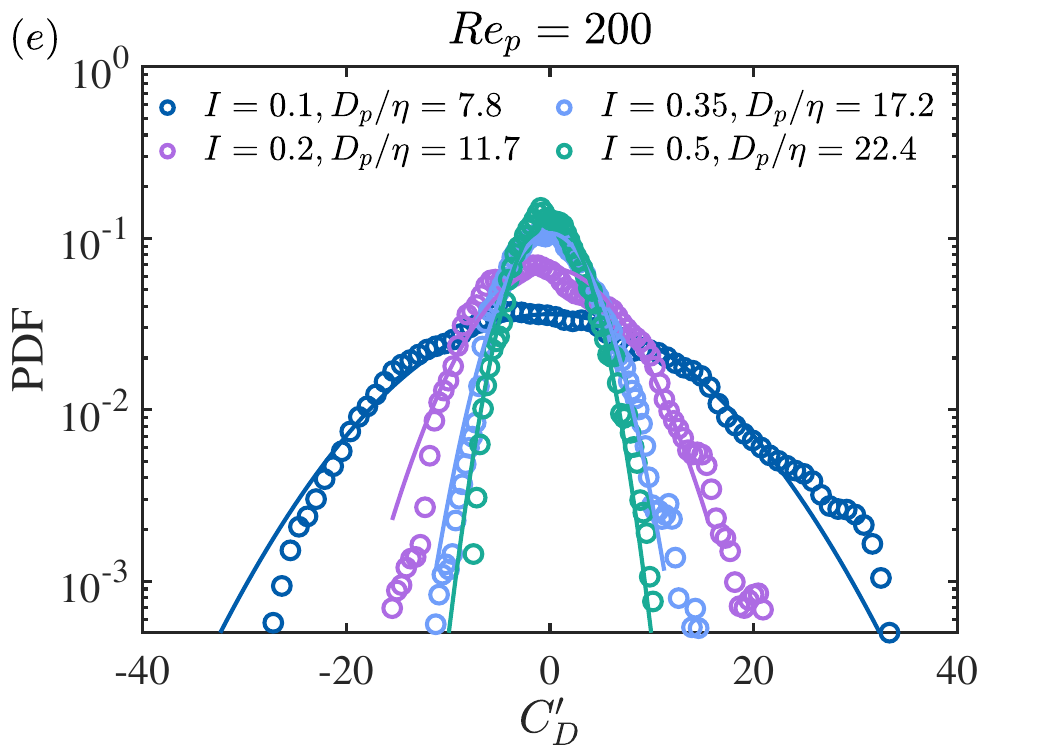}}
\subfigure{\includegraphics[width=0.45\textwidth]{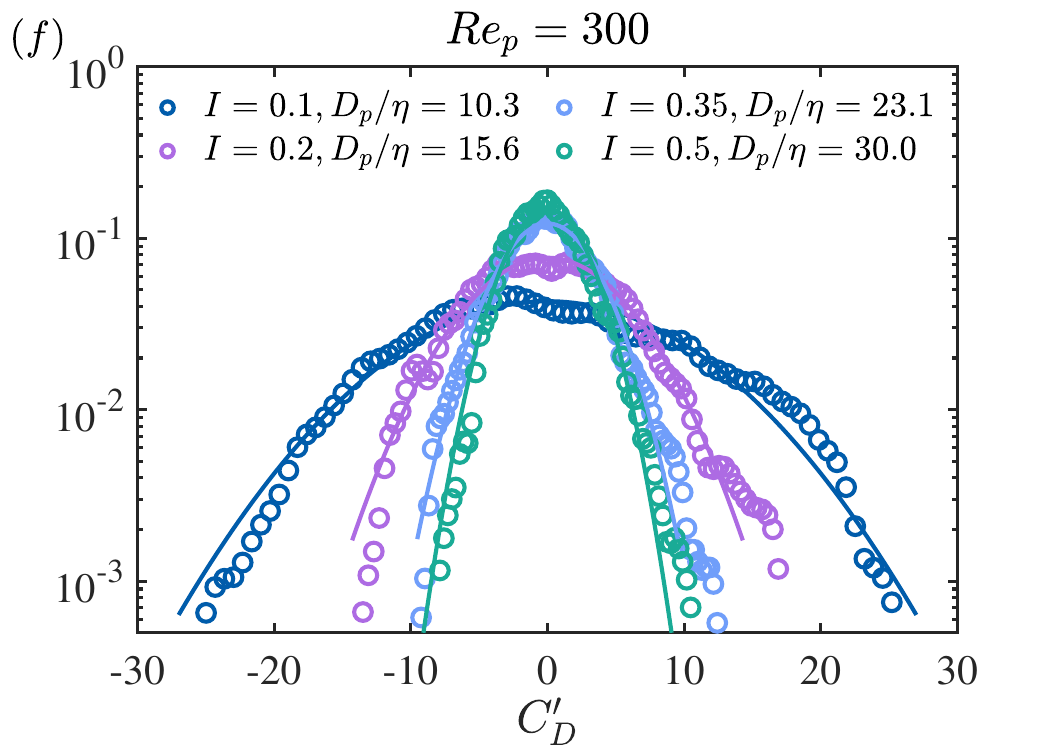}}
\caption{The probability distribution functions of the fluctuating particle drag coefficient. The symbols are the PRDNS data, and the solid lines are Gaussian distributions.}
\label{cd_distribution}
\end{figure}

Figure~\ref{cd_distribution} shows the probability distribution function (PDF) of the fluctuating particle drag coefficient. The {$R^2(C_D^\prime)$ in table~\ref{tab:parameters} of the Appendix \ref{Appendix A} presents the coefficient of determination between the PDF of the fluctuating drag coefficient and the Gaussian distribution, where the mean and variance of the Gaussian distribution are derived from the PRDNS data.} It can be seen that the fluctuating particle drag coefficient follows the Gaussian distribution very well under different conditions. A stationary Gaussian process is completely characterized by its mean, variance, and autocorrelation function. Originally proposed as a stochastic model for the velocity of a microscopic particle undergoing Brownian motion, the Langevin equation \citep{Langevin1908} can yield a Gaussian distribution of the particle velocity \citep{Pope_2000}. Therefore, the Langevin equation can be reasonably adopted to model the temporal variation of the fluctuating particle drag coefficient as follows:
\begin{equation}
 \displaystyle
 {\rm d}C_D^\prime(t) = -\frac{C_D^\prime(t)}{T_D}{\rm d}t+C_{D,rms}^\prime\sqrt{\frac{2}{T_D}}{\rm d}W,
\label{cdprime_equ}
\end{equation}
where $C_{D,rms}^\prime$ and $T_D$ represent the RMS value and integral time scale of the fluctuating particle drag coefficient, respectively, which require modeling, ${\rm d}W$ is the discrete Wiener process, and ${\rm d}t$ is the time step. We note that the Langevin equation is not the only option to model a Gaussian process, but we adopt it because of its simplicity and wide applications. 

\subsubsection{Modeling particle drag fluctuation intensity}

First, assuming that the instantaneous particle drag force follows the S-N drag law and that the slip velocity follows a Gaussian distribution, we employ a similar derivation for the correlation (\ref{deltacdp1}) to model $C_{D,rms}^\prime$ as
\begin{eqnarray}
  C_{D,rms}^\prime &=& \text{RMS}(C_D^\prime(t)) = \text{RMS}(C_D(t)) \nonumber\\
  & = &
  \frac{24\nu}{u_{rms}D_p} \text{RMS} \left[ \frac{u_s(t)\left( 1+0.15\tilde{Re}_p^{0.687}(t) \right)}{u_{rms}}\right] \nonumber\\
  & = & 
  \frac{24\nu}{u_{rms}D_p} \left[ 1+ \text{RMS} \left(\frac{0.15\tilde{Re}_p^{0.687}(t) u_s(t)}{u_{rms}}\right) \right] \nonumber\\
  & = & 
  \frac{24\nu}{u_{rms}D_p} \left[ 1+ 0.15\left( \frac{u_{rms}D_p}{\nu} \right)^{0.687} \text{RMS} \left(\frac{\left| \boldsymbol{U}_s(t) \right| ^{0.687} u_s(t)} {u_{rms}^{1.687}} \right) \right] \nonumber\\
  & = & 
  \frac{24}{Re_p^\prime} \left( 1+ 0.15Re_p^{\prime 0.687} f_2 \right),
  \label{cdrms}
\end{eqnarray}
where $Re_p^\prime=u_{rms}D_p/\nu$ is the turbulence-velocity-based particle Reynolds number, $\text{RMS}(\cdot)$ represents the computation of RMS and the function we need to close (\ref{cdrms}) is
\begin{equation}
    f_2=\text{RMS} \left(\frac{\left| \boldsymbol{U}_s(t) \right| ^{0.687} u_s(t)} {u_{rms}^{1.687}} \right).
    \label{f_2}
\end{equation}
Similar to $f_1$ in (\ref{f_1}), it is very difficult to obtain an explicit expression for $f_2$. Therefore, the numerical method used for $f_1$ is also employed to determine $f_2$.
\begin{equation}
    C_{D,rms}^\prime=\left\{
    \begin{array}{ll}
    \displaystyle
     \frac{24}{Re_p^\prime} \left(1+ 0.253Re_p^{0.687} \right), & I \leq 0.5, \\[15pt]
    \displaystyle
     \frac{24}{Re_p^\prime} \left[1+ 0.15Re_p^{\prime0.687}  \left( 1.702+0.347I^{-1.687} \right) \right], & I > 0.5. \\[4pt]
    \end{array} \right.
    \label{cdrmsp1}
\end{equation}

\begin{figure}
\centering
\subfigure{\includegraphics[width=0.45\textwidth]{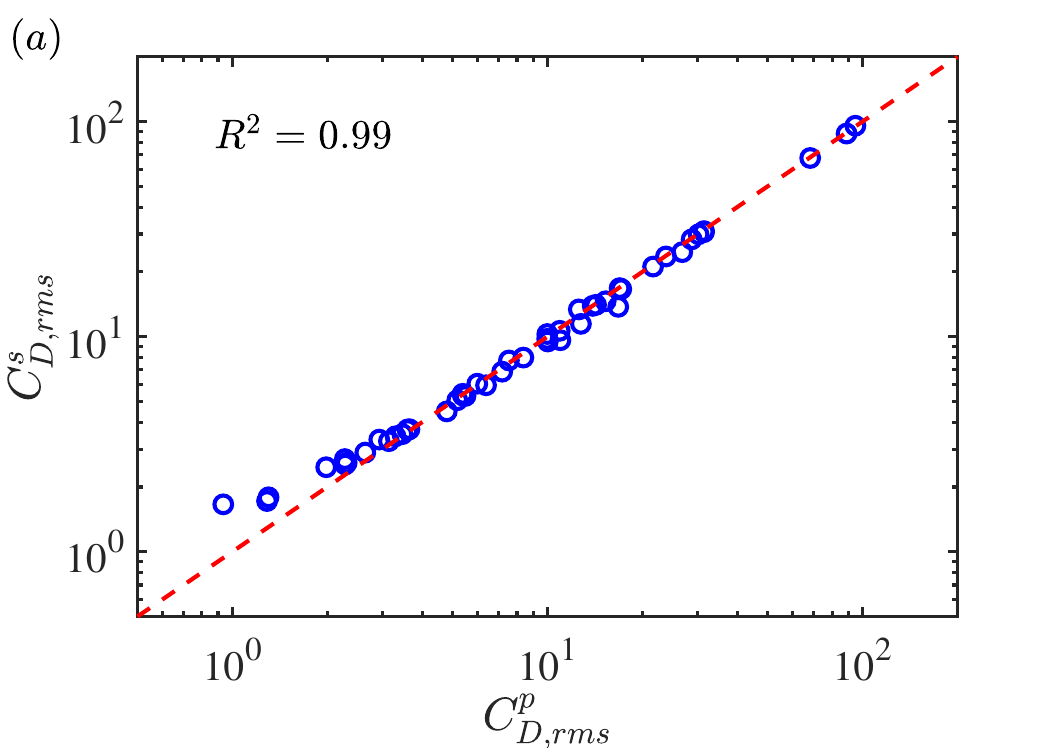}}
\subfigure{\includegraphics[width=0.45\textwidth]{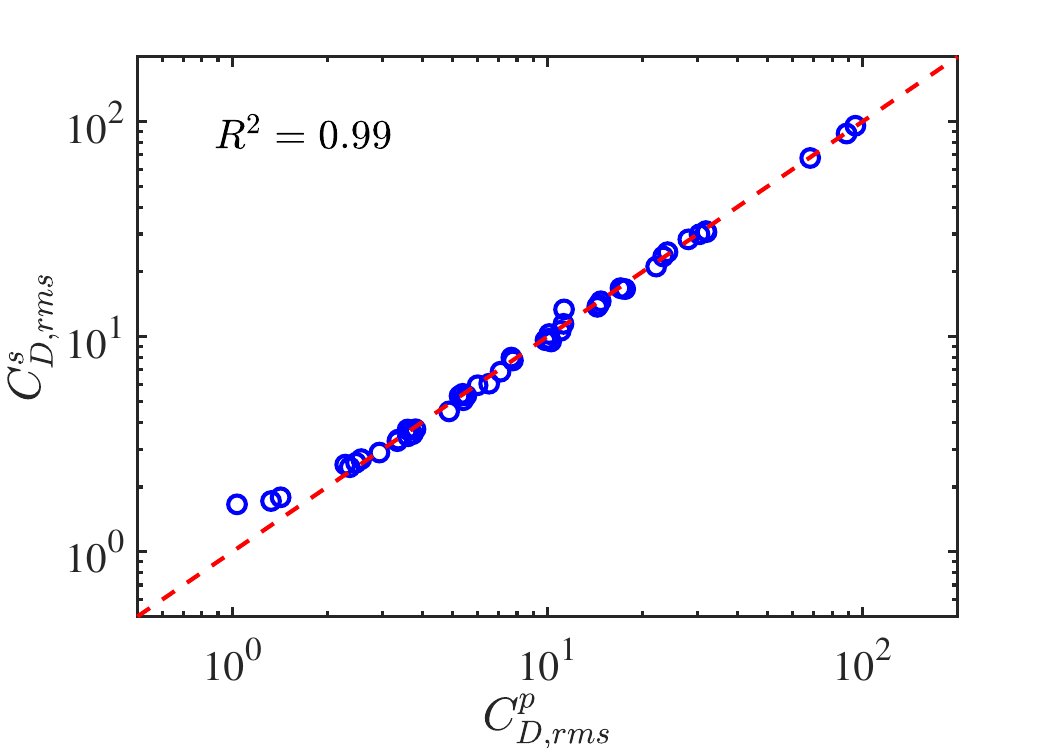}}
\caption{Comparison of the particle drag fluctuation intensities predicted by (a) the correlation (\ref{cdrmsp1}) and (b) the correlation (\ref{cdrmsp2}) with the DNS data.}
\label{cdrmsp_vs_cdrmss}
\end{figure}

On the other hand, only two of the three independent dimensionless parameters are included in the correlation~(\ref{cdrmsp1}). In turn, $f_2$ is determined by fitting the {PRDNS} and PPDNS data while accounting for an entire set of three dimensionless parameters for establishing the second correlation for particle drag fluctuation intensity, which is
\begin{equation}
     \displaystyle
     C_{D,rms}^\prime =
     \frac{24}{Re_p^\prime} \left[1+ 0.15Re_p^{\prime0.687} \left( 1.702+0.556I^{-0.891} \left( \frac{D_p}{\eta} \right)^{0.111} \right) \right].
    \label{cdrmsp2}
\end{equation}
Figure~\ref{cdrmsp_vs_cdrmss} compares the predicted particle drag fluctuation intensities calculated from correlations (\ref{cdrmsp1}) and (\ref{cdrmsp2}) against the {PRDNS} and PPDNS data.
Both correlations exhibit excellent agreement with the PRDNS data.
In addition, it is observed that $F_{D,rms}^\prime \rightarrow 0$ when $I \rightarrow 0$, thus, the singularity of $C_{D,rms}^\prime$ in the correlation (\ref{cdrmsp2}) can be avoided in terms of $F_{D,rms}^\prime$.

\subsubsection{Modeling particle drag fluctuation time scale}

The time scale $T_D$ involved in equation~(\ref{cdprime_equ}) is the integral time scale of the fluctuating particle drag coefficient, which is defined by
\begin{equation}    T_D=\int_{0}^{\infty}R_{C_D^\prime C_D^\prime}(\tau) d\tau,
\end{equation}
in which $R_{C_D^\prime C_D^\prime}(\tau)=\langle C_D^\prime(t) C_D^\prime(t+\tau)\rangle/\langle C_D^\prime(t)^2 \rangle$ is the autocorrelation of the fluctuating particle drag coefficient. 
If $T_D$ is normalized by $\tau_\eta$ ($\tau_\eta=(\nu/\varepsilon)^{1/2}$ is the Kolmogorov time scale.), it can be assumed to be fully determined by three independent dimensionless parameters, such as
\begin{equation}
    \frac{T_D}{\tau_\eta}=f_3\left( Re_p, I, {D_p}/{\eta} \right).
\end{equation}

It is usually natural to use power products to model $f_3$ \citep{Homann_Bec_Grauer_2013,Xia2022,Peng2023}, \emph{i.e.}
\begin{equation}
    \frac{T_D}{\tau_\eta}= a I^b Re_p^c \left({D_p}/\eta\right)^d.
    \label{Td_powerlaw}
\end{equation}
By taking the logarithm of both sides of equation~(\ref{Td_powerlaw}), the values of the unknown parameters $a,b,c$ and $d$ can be determined by fitting the {PRDNS} and PPDNS data via a multidimensional linear regression, which yields
\begin{equation}
    \frac{T_D}{\tau_\eta}= 1.175 I^{1.415} Re_p^{1.074} \left({D_p}/{\eta}\right)^{-1.132}.
    \label{Td_powerlaw1}
\end{equation}

\begin{figure}
\centering
\includegraphics[width=0.7\textwidth]{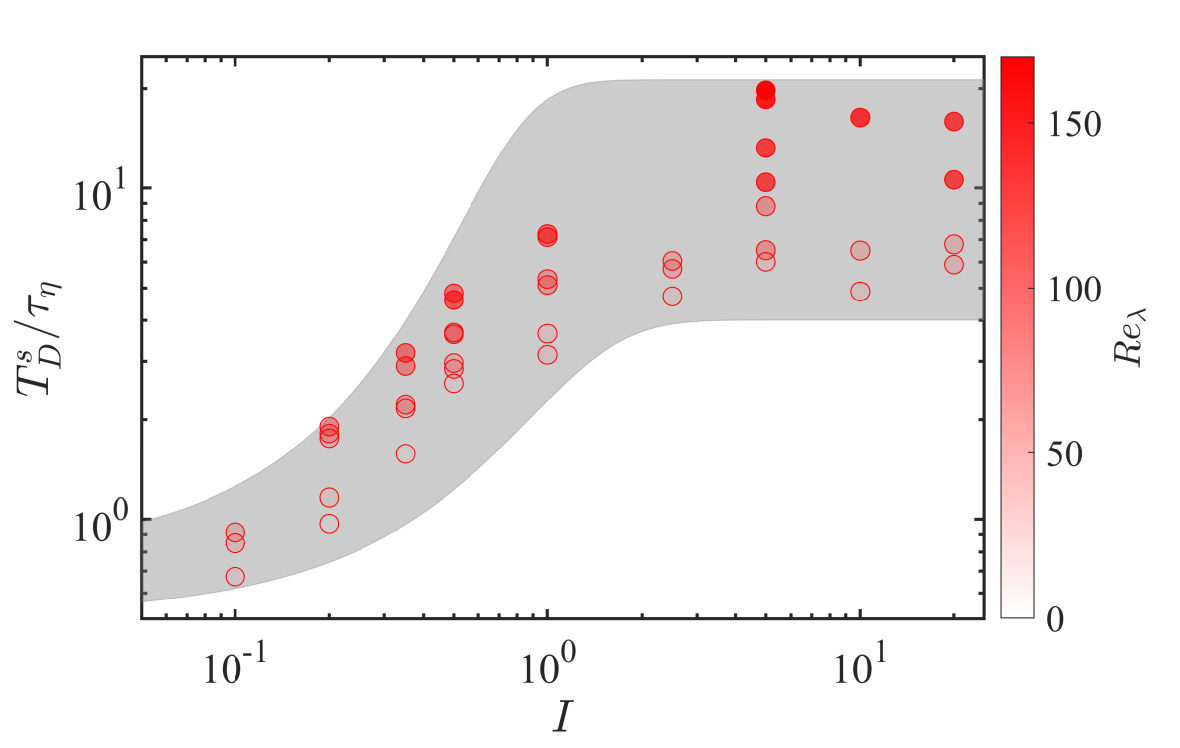}
\caption{Relationship between the particle drag time scale ratio and the turbulence intensity ratio.}
\label{Td_TauEta_I}
\end{figure}

\begin{figure}
\centering
\subfigure{\includegraphics[width=0.45\textwidth]{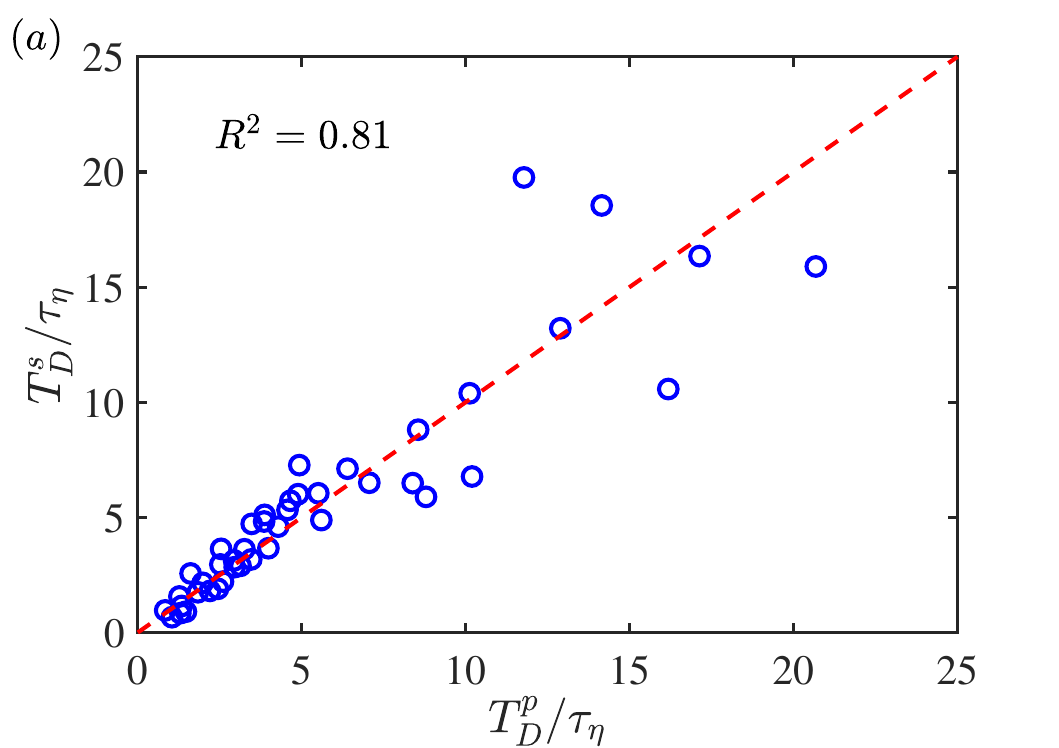}}
\subfigure{\includegraphics[width=0.45\textwidth]{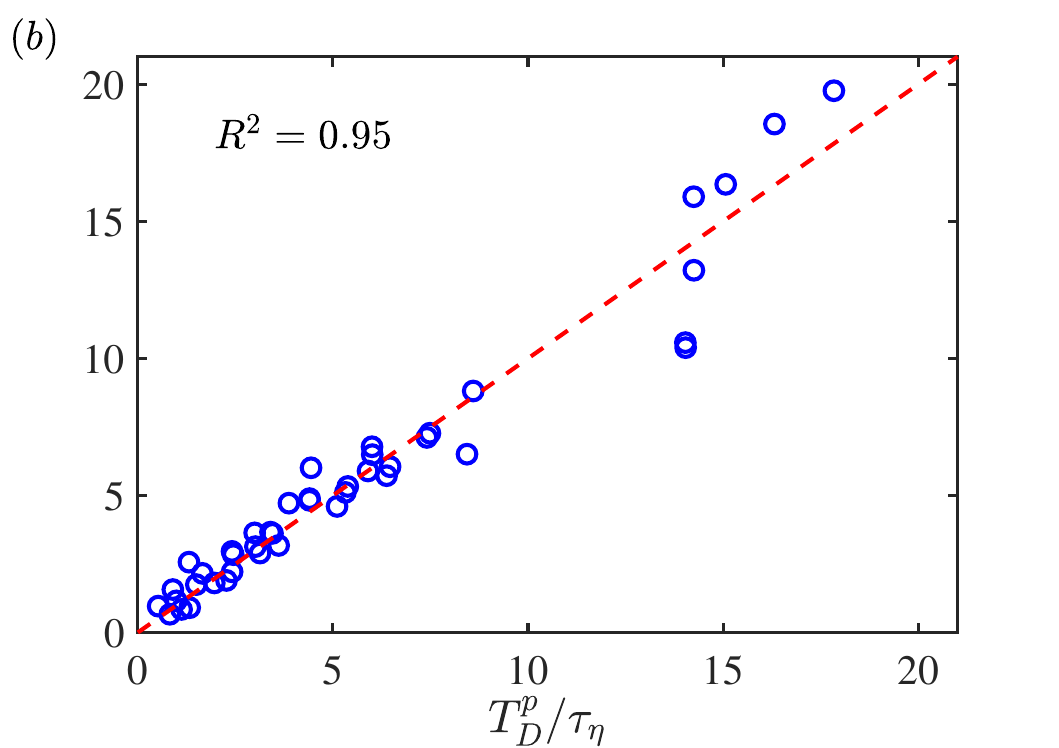}}
\caption{Comparison of the time scale ratio of the fluctuating particle drag coefficient predicted by (a) the correlation (\ref{Td_powerlaw1}) and (b) the correlation (\ref{EquTd}) with the DNS data.}
\label{TD}
\end{figure}

Upon a thorough examination of the PRDNS} and PPDNS data, we observed a robust correlation between the time scale ratio $T_D/\tau_\eta$, the Taylor-microscale-based Reynolds number $Re_\lambda$, and the turbulence intensity ratio $I$. This is reasonable since the fluctuation of the particle drag should probably correlate with the turbulence properties.
The time scale ratio $T_D/\tau_\eta$ is observed to increase with $Re_\lambda$.
Simultaneously, we find that the time scale ratios exhibit relative similarity at high turbulence intensities with comparable $Re_\lambda$, indicating a potential convergence towards an asymptotic value at high $I$, as shown in figure \ref{Td_TauEta_I}. 
Therefore, we propose another form for modeling the time scale ratio according to the above observations, as
\begin{equation}
    \frac{T_D}{\tau_\eta}=\frac{aRe_\lambda^b}{1+c e^{-dI}}.
    \label{EquTdModel}
\end{equation}
By fitting the PRDNS data, the second correlation for the time scale is obtained, which is
\begin{equation}
    \frac{T_D}{\tau_\eta}=\frac{0.634Re_\lambda^{0.654}}{1+4.614 e^{-2.282I}}.
    \label{EquTd}
\end{equation} 

Figure~\ref{TD} (a) compares the time scale of the fluctuation of particle drag between the {DNS} data and the correlation (\ref{Td_powerlaw1}). The prediction of the correlation (\ref{Td_powerlaw1}) is found to not agree well with the DNS data.
Figure~\ref{TD} (b) shows the comparison of the time scale ratio $T_D/\eta$ from the {DNS} data and the correlation (\ref{EquTd}). We can see that the prediction of the correlation (\ref{EquTd}) is much more accurate than the correlation (\ref{Td_powerlaw1}).

\subsection{Lateral force model}

\begin{figure}
\centering
\subfigure{\includegraphics[width=0.45\textwidth]{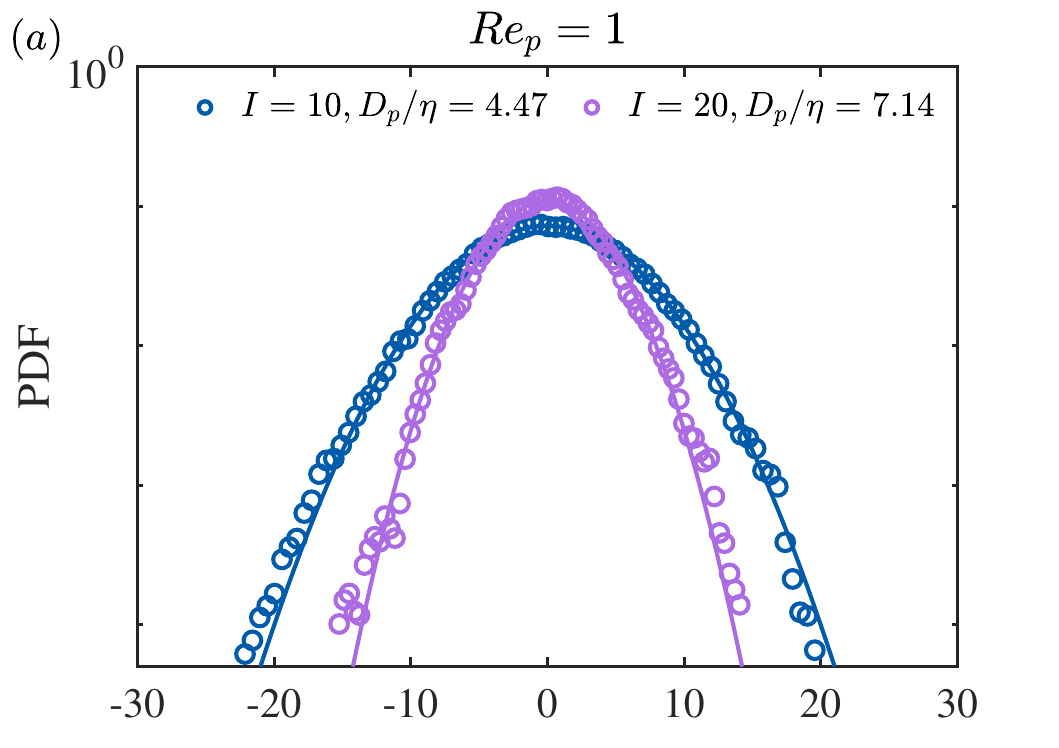}}
\subfigure{\includegraphics[width=0.45\textwidth]{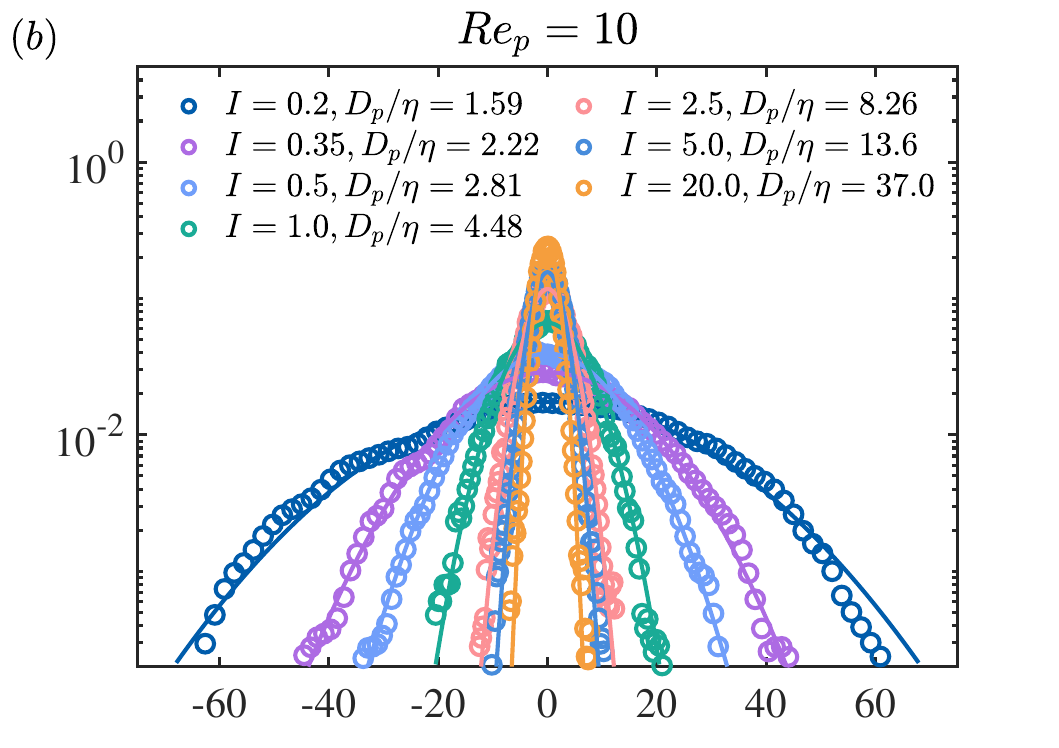}}
\subfigure{\includegraphics[width=0.45\textwidth]{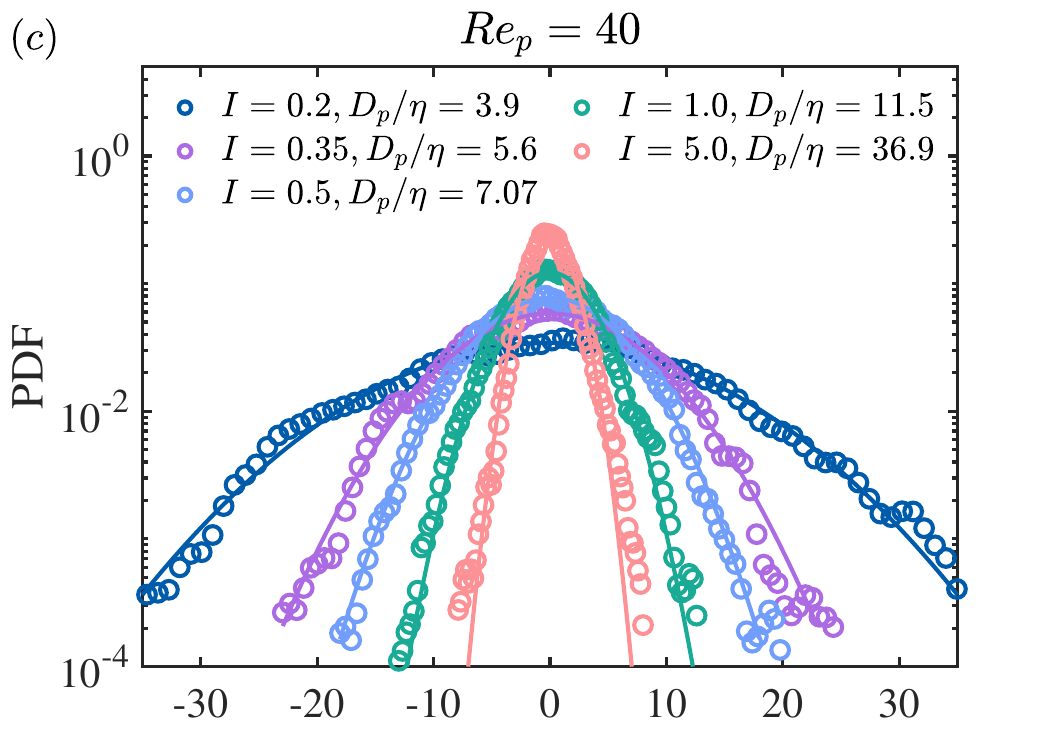}}
\subfigure{\includegraphics[width=0.45\textwidth]{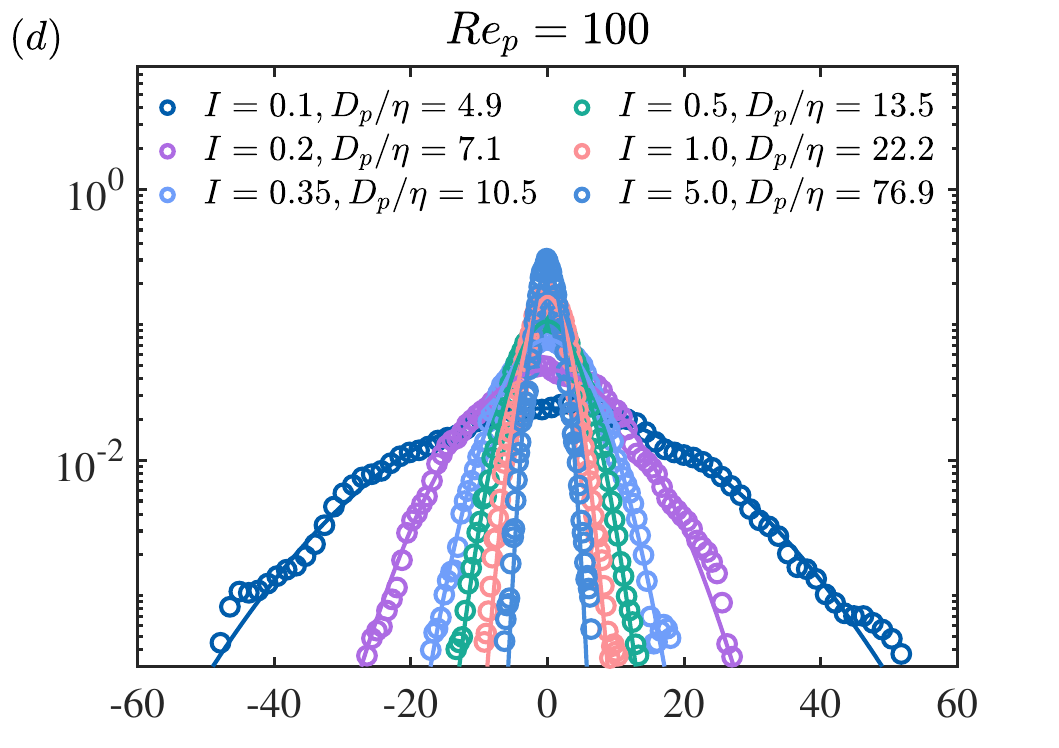}}
\subfigure{\includegraphics[width=0.45\textwidth]{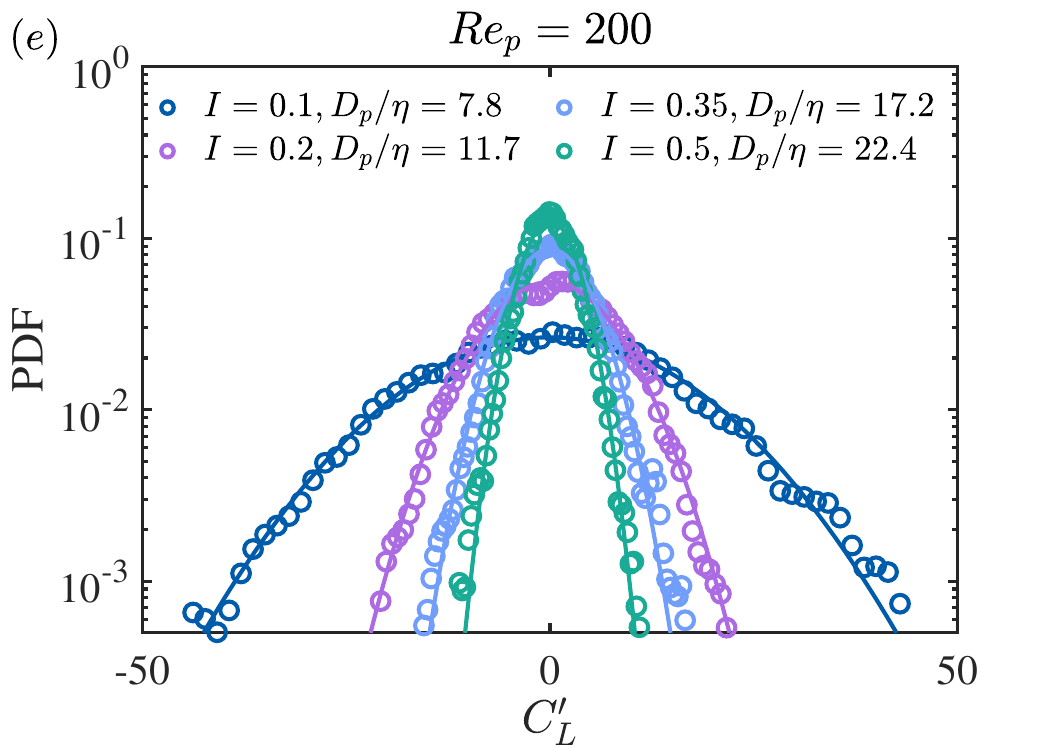}}
\subfigure{\includegraphics[width=0.45\textwidth]{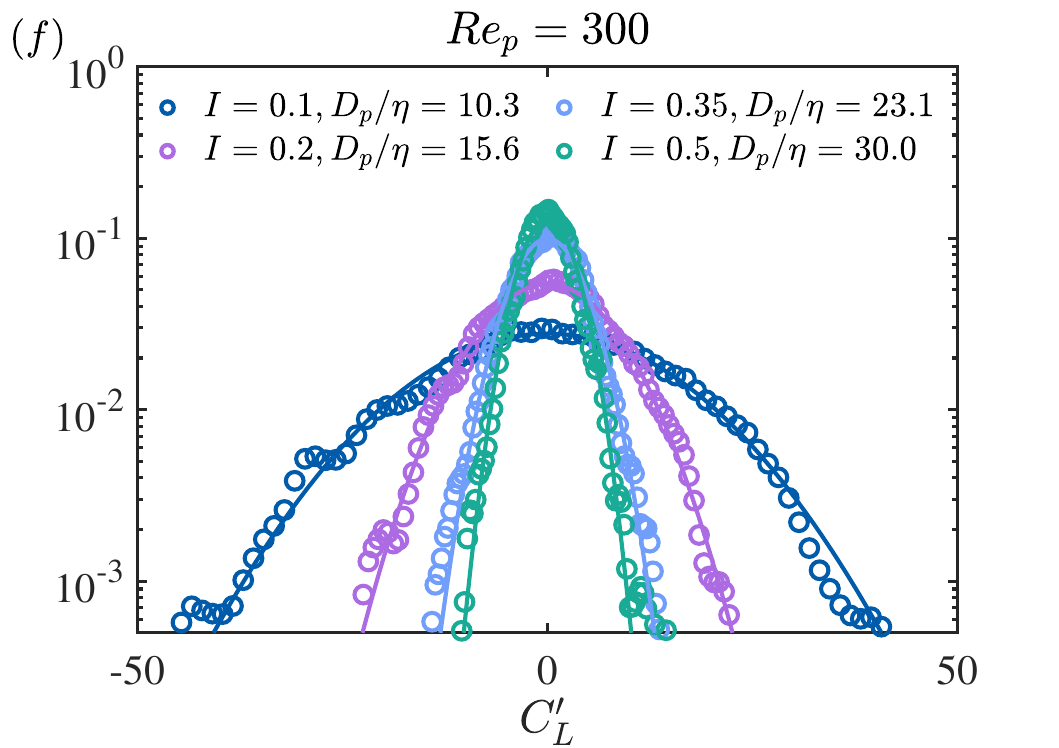}}
\caption{The probability distribution functions of the fluctuating particle lateral force coefficient. The symbols are the PRDNS data, and the solid lines are Gaussian distributions.}
\label{cl_distribution}
\end{figure}

In this section, we propose a stochastic model for the instantaneous lateral fluid force based on the Langevin equation, similar to the fluctuating drag force.
The three components of the fluid force that act on the particle include the lateral forces, which are the components perpendicular to the mean direction of the slip velocity. 
When the incoming flow is uniform at $Re_p<210$, the mean lift of the particle is zero due to the axisymmetric nature of the particle wake. In contrast, the mean lift is not zero due to the emergence of an asymmetric particle wake when $Re_p>210$; for example, the mean lift coefficient is around 0.06 when $Re_p=300$  \citep{Johnson1999,Kim2001,Ploumhans2002,kim2002,Constantinescu2003,Niazmand2003,wang2011,Liska2017}. 
Moreover, given uniform inflow conditions, the mean particle lift is approximately two orders of magnitude less than the mean particle drag. 
In turbulent inflow, we found that the mean lateral force on the particle at $Re_p=300$ is significantly less than that in uniform flow at the same particle Reynolds number. 
This may be due to turbulence disrupting the asymmetry of particle wakes, rendering them more symmetric in a statistical sense.
Therefore, we assume that the mean particle lateral force is roughly zero and focus solely on modeling the fluctuating component of the particle lateral force.

Figure~\ref{cl_distribution} shows the probability distribution functions of the fluctuating lateral force coefficient of the particle. The $R^2(C_L^\prime)$ in table~\ref{tab:parameters} of the Appendix \ref{Appendix A} represents the coefficient of determination between the PDF of the fluctuating lateral force coefficient and the Gaussian distribution, where the mean and variance of the Gaussian distribution are derived from the PRDNS data. It can be seen that it can be well approximated by the Gaussian distribution, similar to the fluctuating drag force. Consequently, the stochastic Langevin equation is also used to model the fluctuating particle lateral force as follows:
\begin{equation}
 \displaystyle
 {\rm d}C_L(t) = -\frac{C_L(t)}{T_L}{\rm d}t+C_{L,rms}\sqrt{\frac{2}{T_L}}{\rm d}W,
\label{clprime_equ}
\end{equation}
where $C_L(t)$ is the instantaneous particle lateral force coefficient with a zero mean, and $C_{L,rms}$ and $T_L$ are the RMS and integral time scale of the fluctuating lateral force coefficient, respectively.

\subsubsection{Modeling particle lateral force fluctuation intensity}

The lateral force is perpendicular to the mean slip velocity.
Therefore, the instantaneous particle lateral force can be written as
\begin{equation}
  {F}_L(t) = \frac{24}{\Tilde{Re}_p(t)} \left[ 1 + 0.15\Tilde{Re}_p^{0.687}(t) \right] \cdot \frac{1}{2}\rho_f |\boldsymbol{U}_s(t)| {v}_s(t)A,
\end{equation}
in which $V_s(t)$ is the instantaneous lateral slip velocity.
So, the instantaneous lateral force coefficient can be expressed as
\begin{eqnarray}
  C_L(t) &=& \frac{F_{L}(t)}{(1/2) \rho_f u_{rms}^2A} \nonumber\\
  & = & 
  \frac{24\nu}{u_{rms} D_p} \frac{v_s(t)}{u_{rms}} \left[ 1+0.15\tilde{Re}_p(t)^{0.687} \right].
\end{eqnarray}
As a result, $C_{L,rms}$ can be calculated by
\begin{eqnarray}
  C_{L,rms} &=& {\rm RMS} (C_L(t)) \nonumber\\
  & = &
  \frac{24\nu}{u_{rms}D_p} {\rm RMS} \left[ \frac{v_s(t)\left( 1+0.15\tilde{Re}_p(t)^{0.687} \right)}{u_{rms}}\right] \nonumber\\
  & = & 
  \frac{24\nu}{u_{rms}D_p} \left[ 1+ {\rm RMS} \left(\frac{0.15\tilde{Re}_p(t)^{0.687}v_s(t)}{u_{rms}}\right) \right] \nonumber\\
  & = & 
  \frac{24\nu}{u_{rms}D_p} \left[ 1+ 0.15\left( \frac{u_{rms}D_p}{\nu} \right)^{0.687} {\rm RMS} \left(\frac{\left| \boldsymbol{U}_s(t) \right| ^{0.687} v_s(t)} {u_{rms}^{1.687}} \right) \right] \nonumber\\
  & = & 
  \frac{24}{Re_p^\prime} \left( 1+ 0.15Re_p^{\prime 0.687} f_4 \right),
  \label{clrms}
\end{eqnarray}
where
\begin{equation}
    f_4={\rm RMS} \left(\frac{\left| \boldsymbol{U}_s(t) \right| ^{0.687} v_s(t)} {u_{rms}^{1.687}} \right).
\end{equation}
Similar to $f_1$ in (\ref{f_1}) and $f_2$ in (\ref{f_2}), the numerical method is used to determine $f_4$, which yields
\begin{equation}
    C_{L,rms}=\left\{
    \begin{array}{ll}
    \displaystyle
     \frac{24}{Re_p^\prime} \left(1+0.15Re_p^{0.687} + 0.162Re_p^{\prime0.687}I^{1.131} \right), & I \leq 0.5 \\[15pt]
    \displaystyle
     \frac{24}{Re_p^\prime} \left[1+ 0.15Re_p^{\prime0.687}  \left( 1.702+0.11I^{-1.875} \right) \right], & I > 0.5. \\[4pt]
    \end{array} \right.
    \label{clrmsp1}
\end{equation}

\begin{figure}
\centering
\subfigure{\includegraphics[width=0.45\textwidth]{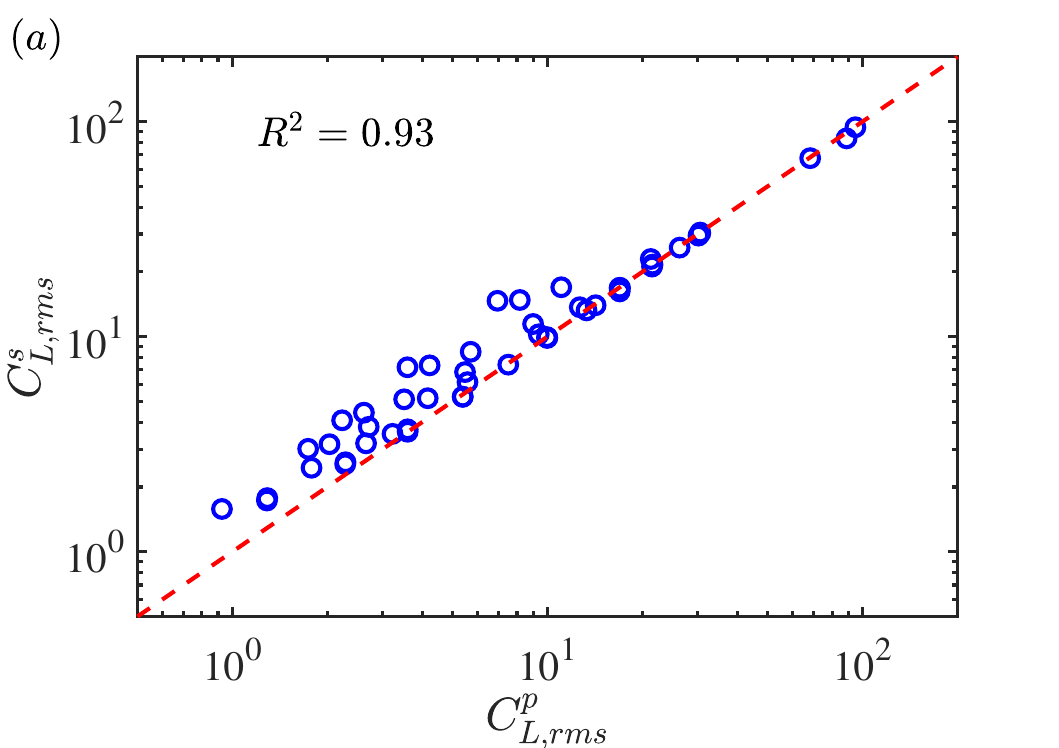}}
\subfigure{\includegraphics[width=0.45\textwidth]{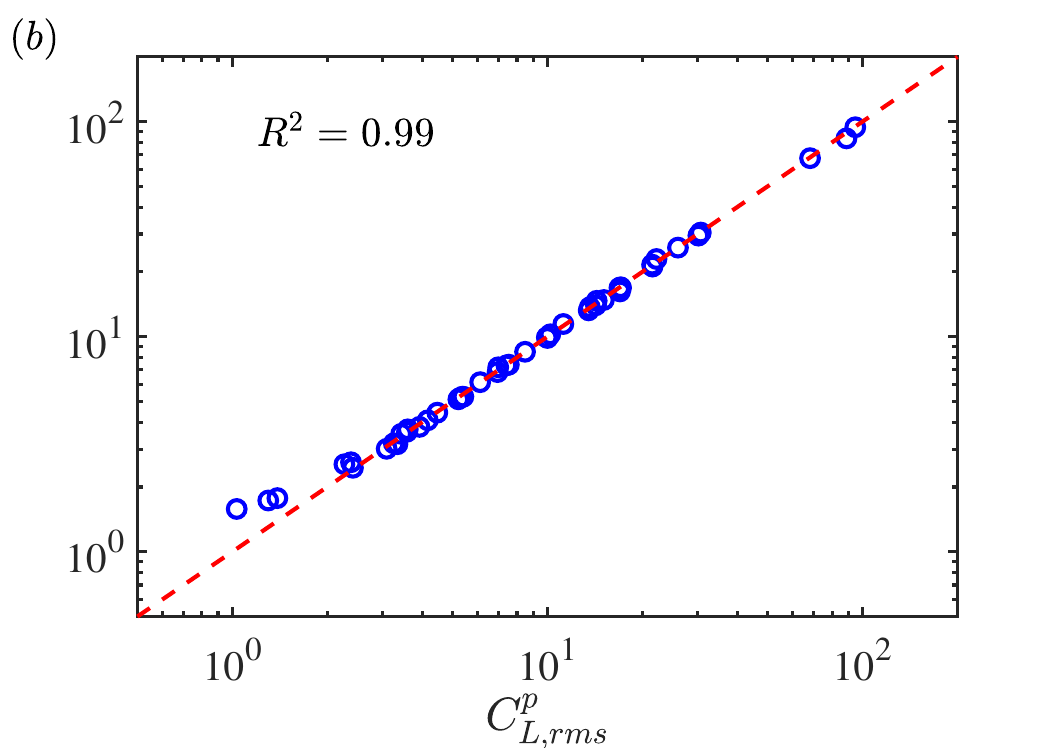}}
\caption{Comparison of the particle lateral force fluctuation intensities predicted by (a) the correlation (\ref{clrmsp1}) and (b) the correlation (\ref{clrmsp2}).}
\label{clrmsp_vs_clrmss}
\end{figure}

On the other hand, as in the correlation (\ref{cdrmsp2}), we introduce the complete three independent dimensionless parameters to determine $f_4$. The fitting expression is as follows:
\begin{equation}
     \displaystyle
     C_{L,rms}=
     \frac{24}{Re_p^\prime} \left[1+ 0.15Re_p^{\prime0.687}  \left( 1.689+0.226I^{-1.173} \left( \frac{D_p}{\eta} \right)^{0.432} \right) \right].
    \label{clrmsp2}
\end{equation}

Figure~\ref{clrmsp_vs_clrmss} compares the RMS of the fluctuating particle lateral force coefficient calculated by PRDNS and PPDNS with the correlations (\ref{clrmsp1}) and (\ref{clrmsp2}). 
The predictions obtained from the correlation (\ref{clrmsp2}) agree more closely with the {DNS} data.
Similarly to the drag fluctuation, it can be found that $F_{L,rms}^\prime \rightarrow 0$ when $I \rightarrow 0$, so the singularity of $C_{L,rms}^\prime$ in the correlation (\ref{cdrmsp2}) can be avoided in terms of $F_{L,rms}^\prime$.

\subsubsection{Modeling particle lateral force fluctuation time scale}
The time scale $T_L$ in equation (\ref{clprime_equ}) is  defined by
\begin{equation}    T_L=\int_{0}^{\infty}R_{C_L C_L}(\tau) d\tau,
\end{equation}
where $R_{C_L C_L}$ is the autocorrelation of the fluctuating particle lateral force coefficient. 

Similar to the model of $T_D$, we first assume that the time scale ratio $T_L/\tau_\eta$ is a power product of the particle Reynolds number $Re_p$, turbulence intensity ratio $I$ and the scale ratio $D_p/\eta$, \emph{i.e.}
\begin{equation}
    \frac{T_L}{\tau_\eta} = f_5(Re_p, I, D_p/\eta) = a I^b Re_p^c \left({D_p}/{\eta}\right)^d.
    \label{Tl_powerlaw}
\end{equation}
The values of the unknown parameters $a, b, c,$ and $d$ can be determined by fitting the {PRDNS} and PPDNS data by multidimensional linear regression after applying the logarithm to both sides of (\ref{Tl_powerlaw}), resulting in
\begin{equation}
    \frac{T_L}{\tau_\eta}= 0.953 I^{1.496} Re_p^{0.946} \left({D_p}/{\eta}\right)^{-1.048}.
    \label{TL_powerlaw1}
\end{equation}

\begin{figure}
\centering
\includegraphics[width=0.7\textwidth]{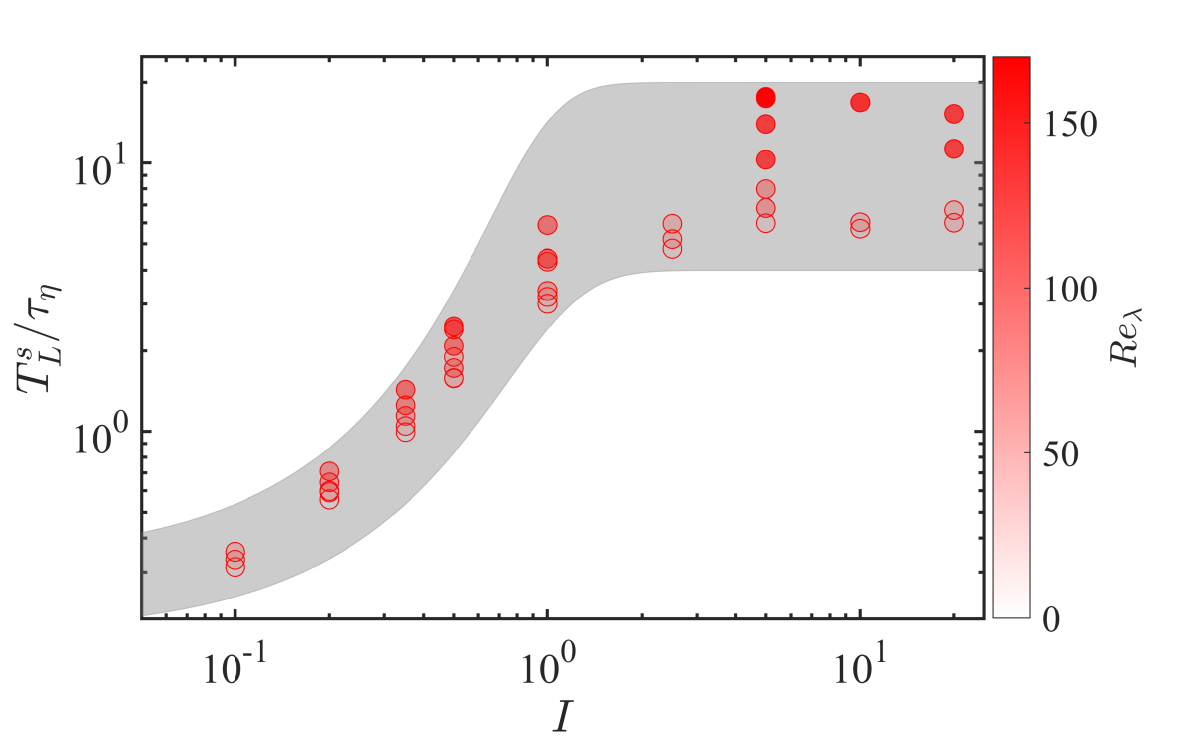}
\caption{Relationship between the particle lateral force time scale ratio and the turbulence intensity ratio.}
\label{TL_TauEta_I}
\end{figure}

\begin{figure}
\centering
\subfigure{\includegraphics[width=0.45\textwidth]{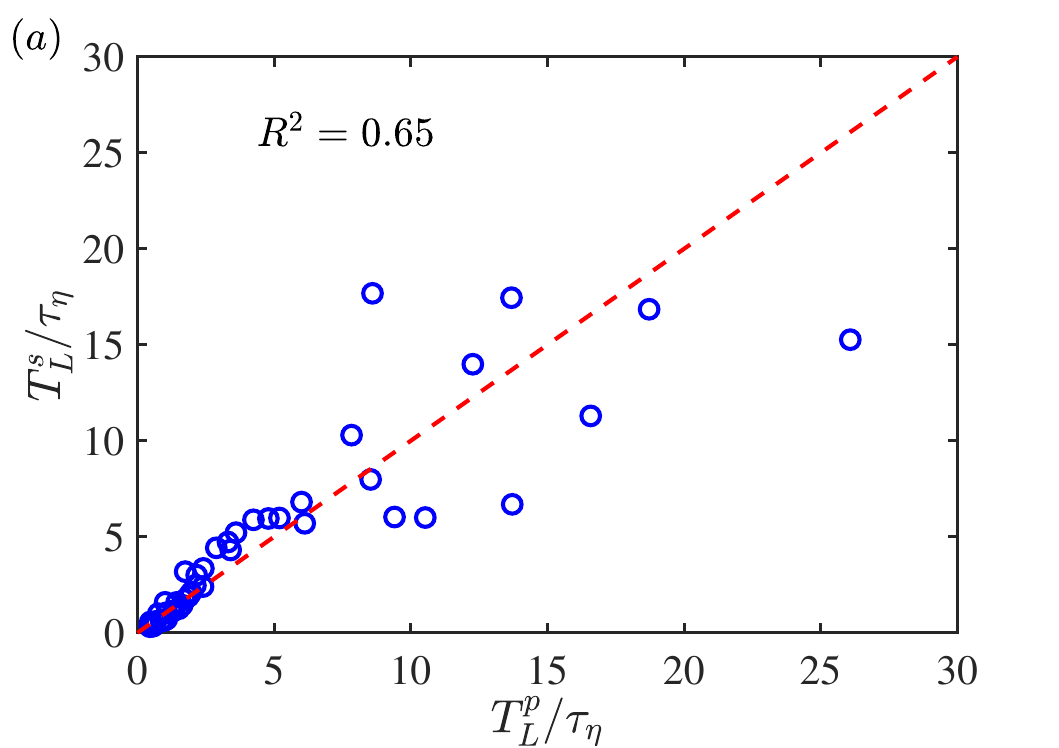}}
\subfigure{\includegraphics[width=0.45\textwidth]{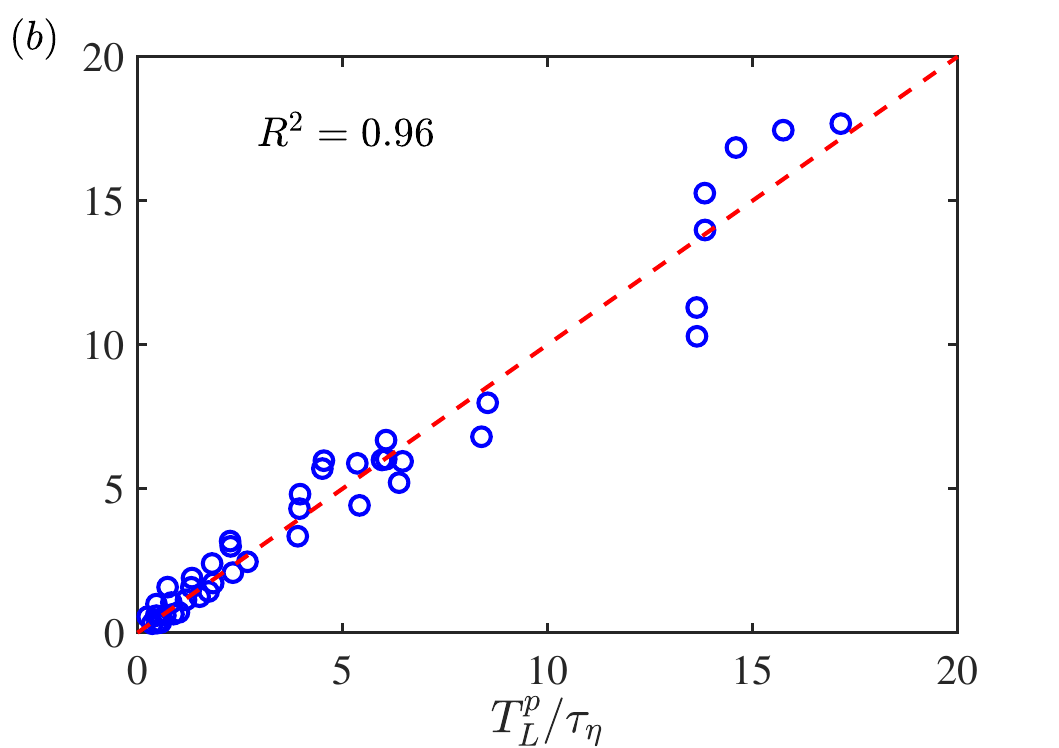}}
\caption{Comparison of the particle lateral force fluctuation time scale ratio predicted by (a) the correlation (\ref{TL_powerlaw1}) and (b) the correlation (\ref{EquTl}) with the DNS data.}
\label{TL}
\end{figure}

On the other hand, from the PRDNS data, it is found that the time scale ratio $T_L/\eta$ increases with the turbulence intensity ratio $I$ and reaches an asymptotic value at high $I$, as shown in figure \ref{TL_TauEta_I}. 
Therefore, another model with the same form as (\ref{EquTdModel}) is obtained as
\begin{equation}
    \frac{T_L}{\tau_\eta}=\frac{0.708Re_\lambda^{0.625}}{1+13.012 e^{-2.565I}}.
    \label{EquTl}
\end{equation} 
Figure~\ref{TL} compares the time scale ratio $T_L/\tau_\eta$ of the PRDNS and PPDNS data with the correlations (\ref{TL_powerlaw1}) and (\ref{EquTl}). The correlation (\ref{EquTl}) is seen to produce predictions far more accurate than the correlation (\ref{TL_powerlaw1}). 

\begin{table}
    \centering
	\def~{\hphantom{0}}
    {
	\caption{A summary of the present stochastic particle force model.} 
	\begin{tabular}{cc}
		\toprule
		Model   & Form  \\[5pt]
		\midrule
		Mean drag enhancement & $\displaystyle \Delta C_D=\frac{0.015Re_p^{0.687}}{1+0.15Re_p^{0.687}}I^{0.858}Re_\lambda^{0.474}$ \\[10pt]
            \midrule
		Stochastic drag force coefficient & $\displaystyle {\rm d}C_D^\prime(t) = -\frac{C_D^\prime(t)}{T_D} {\rm d}t+C_{D,rms}^\prime\sqrt{\frac{2}{T_D}}{\rm d}W,$ \\[10pt]
            & $\displaystyle C_{D,rms}^\prime = \frac{24}{Re_p^\prime} \left[1+ 0.15Re_p^{\prime0.687} \left( 1.702+0.556I^{-0.891} \left( \frac{D_p}{\eta} \right)^{0.111} \right) \right],$ \\[10pt]
            & $\displaystyle T_D=\frac{0.634Re_\lambda^{0.654}}{1+4.614 e^{-2.282I}}\tau_\eta.$ \\[10pt]
            \midrule
            Stochastic lateral force coefficient & $\displaystyle {\rm d}C_L(t) = -\frac{C_L(t)}{T_L}{\rm d}t+C_{L,rms}\sqrt{\frac{2}{T_L}}{\rm d}W,$ \\[10pt]
            & $\displaystyle C_{L,rms} = \frac{24}{Re_p^\prime} \left[1+ 0.15Re_p^{\prime0.687} \left( 1.689+0.226I^{-1.173} \left( \frac{D_p}{\eta} \right)^{0.432} \right) \right],$ \\[10pt]
            & $\displaystyle T_L=\frac{0.708Re_\lambda^{0.625}}{1+13.012 e^{-2.565I}}\tau_\eta.$ \\[10pt]
		\bottomrule
	\end{tabular}	    
    \label{tab:2}
    }
\end{table}

\begin{figure}
\centering
\subfigure{\includegraphics[width=0.45\textwidth]{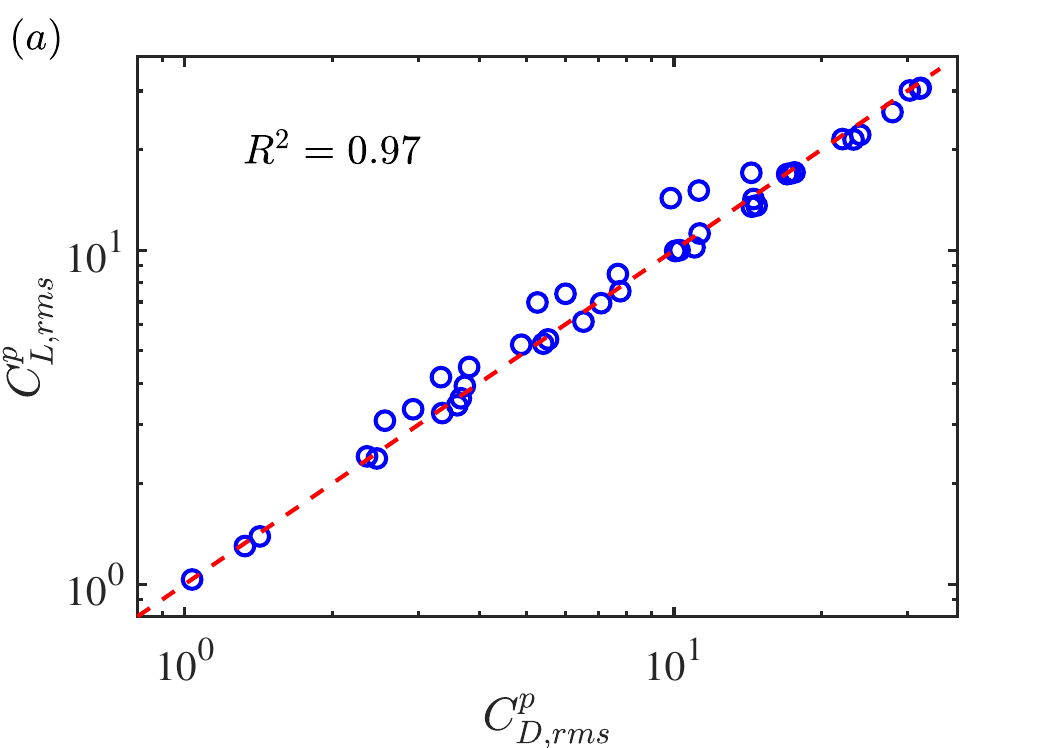}}
\subfigure{\includegraphics[width=0.45\textwidth]{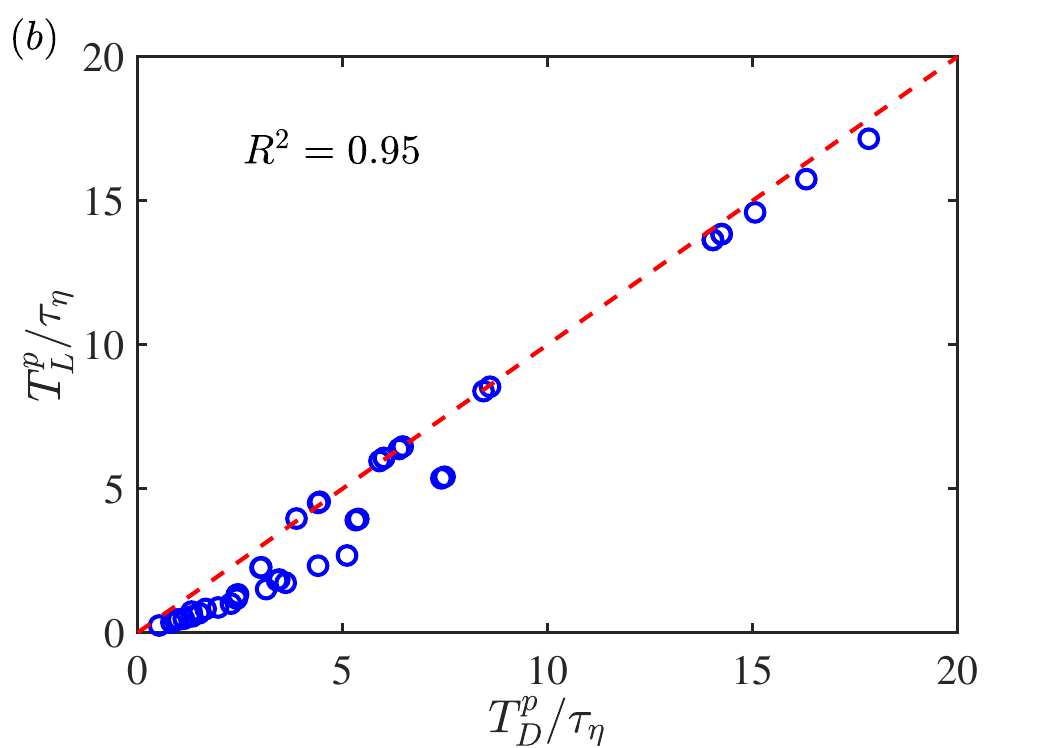}}
\caption{Similarities between the fluctuating drag force and lateral force models: (a) fluctuation intensity and (b) time scale.}
\label{cd_vs_cl}
\end{figure}

The similarities of the fluctuation intensities and time scales predicted by the fluctuating drag force and lateral force models are given in figure \ref{cd_vs_cl}. It can be seen that the predictions from the two models are quite similar. Therefore, the stochastic model for the fluctuating drag force can be adopted in all directions, which can significantly reduce the complexity of the model implementation. In addition, we note that the fluid force fluctuations in the three directions are statistically independent based on our PRDNS and PPDNS data; hence, they can be calculated separately.

{We summarize the proposed particle stochastic force model in table \ref{tab:2}.}

\section{Validation of the model}\label{sec:validation}

In this section, we will validate our model by simulating the motion of a finite-size particle in a turbulent flow and the dispersion of small particles in a turbulent channel flow to illustrate the prediction capabilities of the model for both finite-size and point particles. 
The inputs of the model are the mean flow and turbulence quantities, \emph{i.e.} mean flow velocity, TKE and TDR, similar to a point-particle Reynolds-averaged Navier–Stokes (PPRANS) simulation; thus, we will refer to it as "PPRANS+SFM", where SFM stands for the stochastic force model. 
In addition, a point-particle simulation with the S-N drag law is employed to test the turbulence effect, and we refer to it as "PPRANS" in the following. We also conduct DNS simulations to provide the corresponding references in the two problems.

\subsection{The motion of a finite-size particle in turbulence}

The first problem is with a finite-size particle released and freely moving in a turbulent flow. The background flow is homogeneous and isotropic turbulence superimposed by a streamwise uniform flow, similar to the PRDNS of a fixed particle.
The particle diameter is $D_p$, and the simulation domain size is $L_x \times L_y \times L_z=20D_p \times 10D_p \times 10D_p$. 
The periodic boundary condition is used for the flow field and the motion of the particles in the three directions. 
The initial position of the particle is $(x_p,y_p,z_p)=(5D_p,5D_p,5D_p)$, the same as in fixed particle cases, and the initial velocity of the particle is the same as the velocity of the fluid velocity at the particle position. Eight cases of different conditions are simulated using PRDNS for validation, as listed in table~\ref{tab:validate1}.

\begin{table}
  \begin{center}
\def~{\hphantom{0}}
\caption{Parameters of the motion of a finite-size particle in a turbulent flow, where $Re_{p0}=U_0D_p/\nu$ is the initial particle Reynolds number.}
  \begin{tabular}{cccc}
  \toprule
      ~~Case~~  & ~~$Re_{p0}$~~  & ~~$I$~~   & ~~$D_p/\eta$~~  \\[3pt]
      \midrule
         1      &   1            & 2.5         & 1.8      \\
         2      &   1            & 5.0         & 2.8      \\
         3      &   10           & 0.1         & 1.1      \\
         4      &   10           & 0.5         & 2.8      \\
         5      &   10           & 5.0         & 13.3     \\
         6      &   40           & 5.0         & 37.0     \\
         7      &   100          & 1.0         & 22.2     \\
         8      &   300          & 0.5         & 29.6     \\
         \bottomrule
  \end{tabular}
  \label{tab:validate1}
  \end{center}
\end{table}

\begin{figure}
\centering
\subfigure{\includegraphics[width=0.45\textwidth]{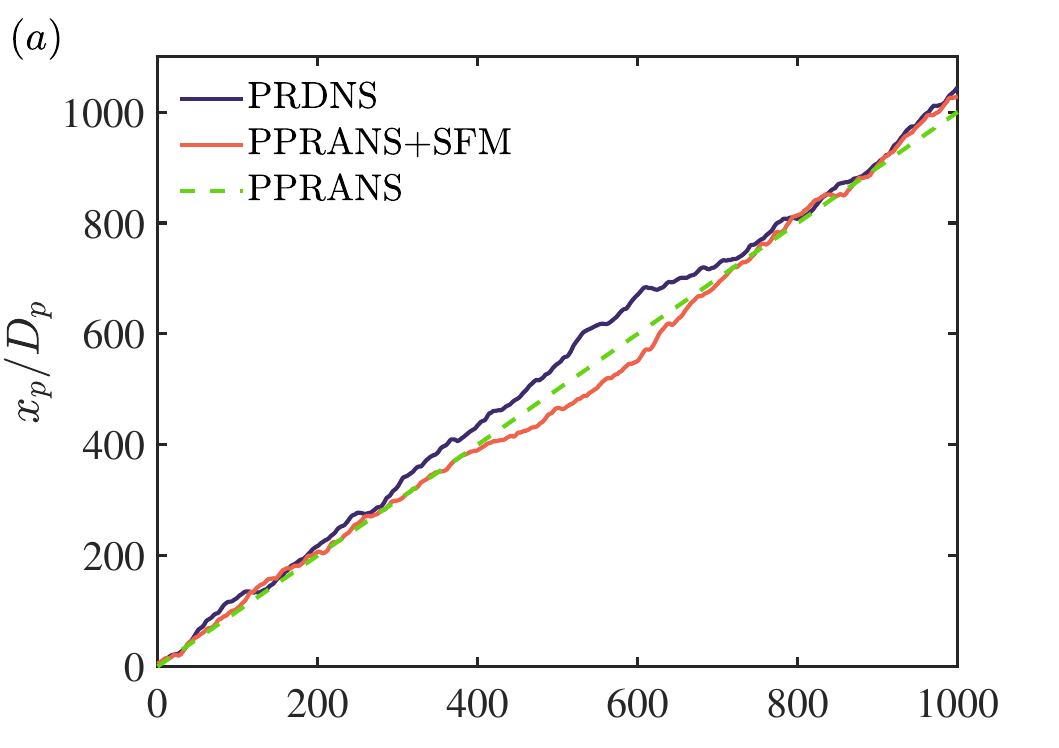}}
\subfigure{\includegraphics[width=0.45\textwidth]{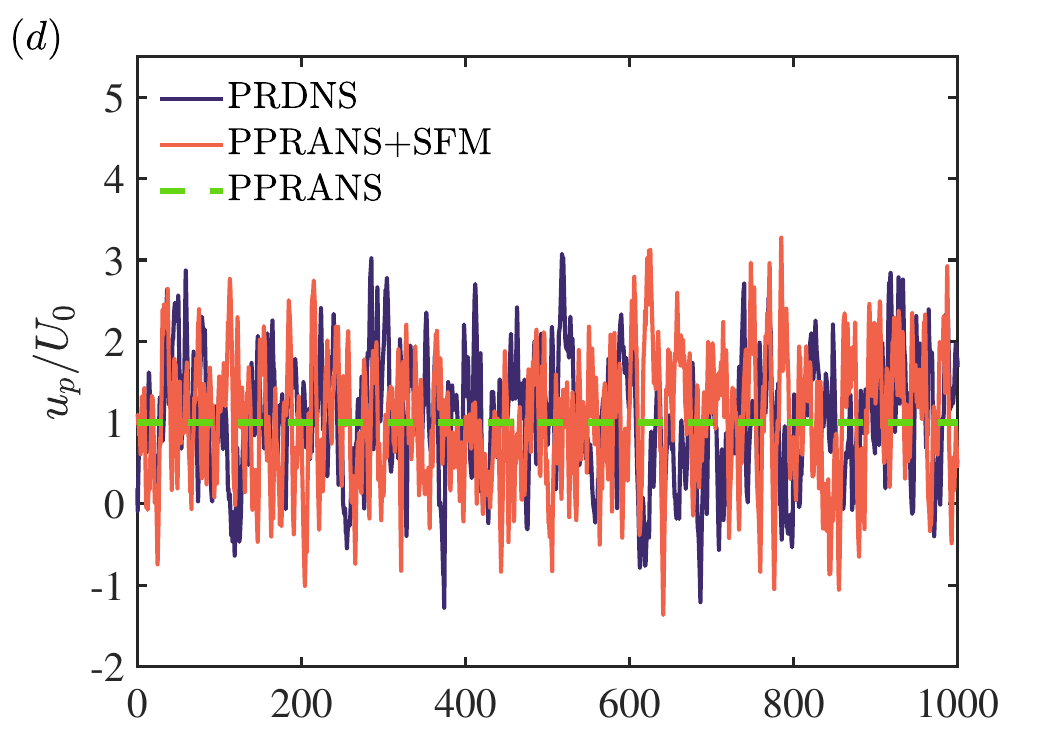}}
\subfigure{\includegraphics[width=0.45\textwidth]{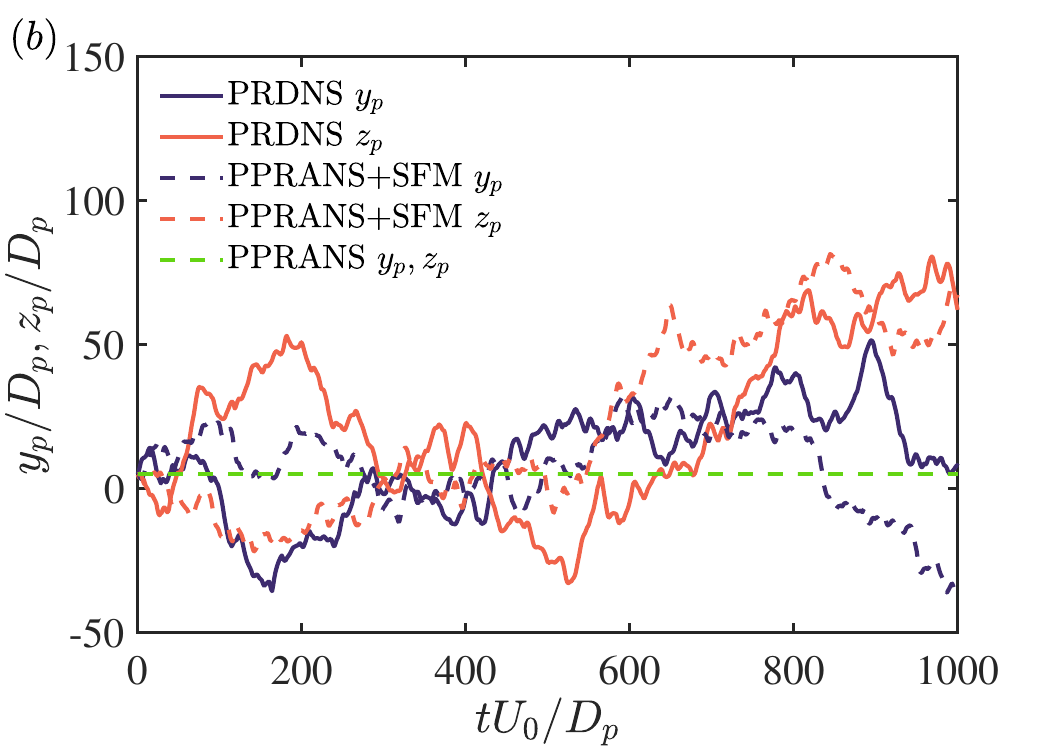}}
\subfigure{\includegraphics[width=0.45\textwidth]{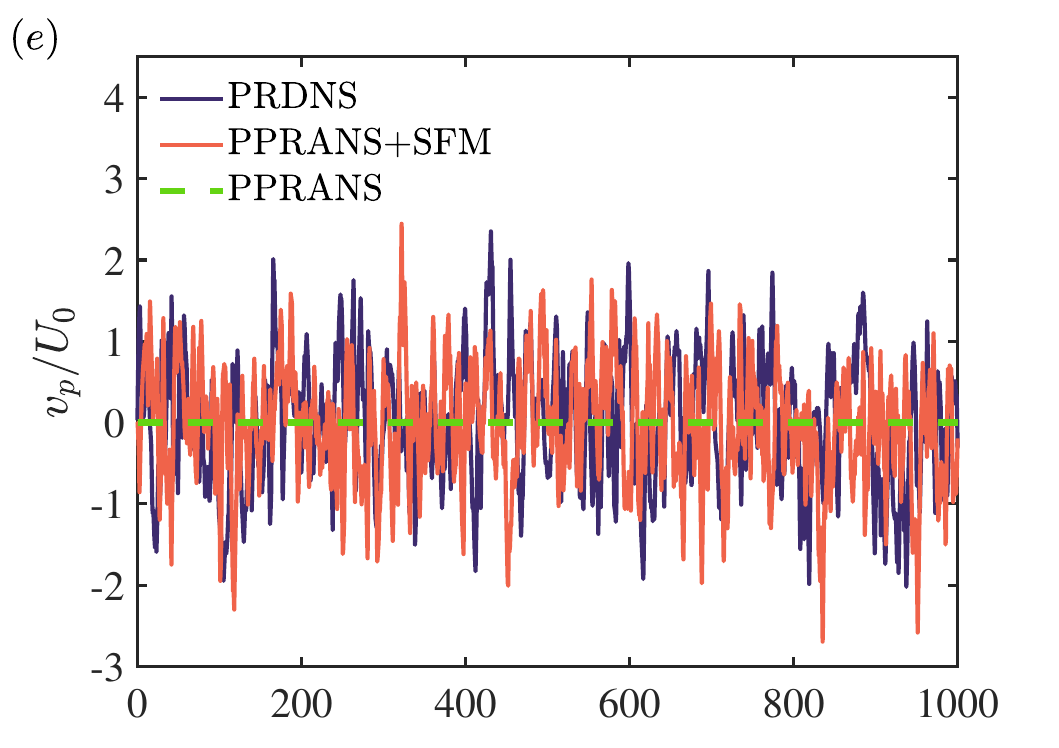}}
\subfigure{\includegraphics[width=0.45\textwidth]{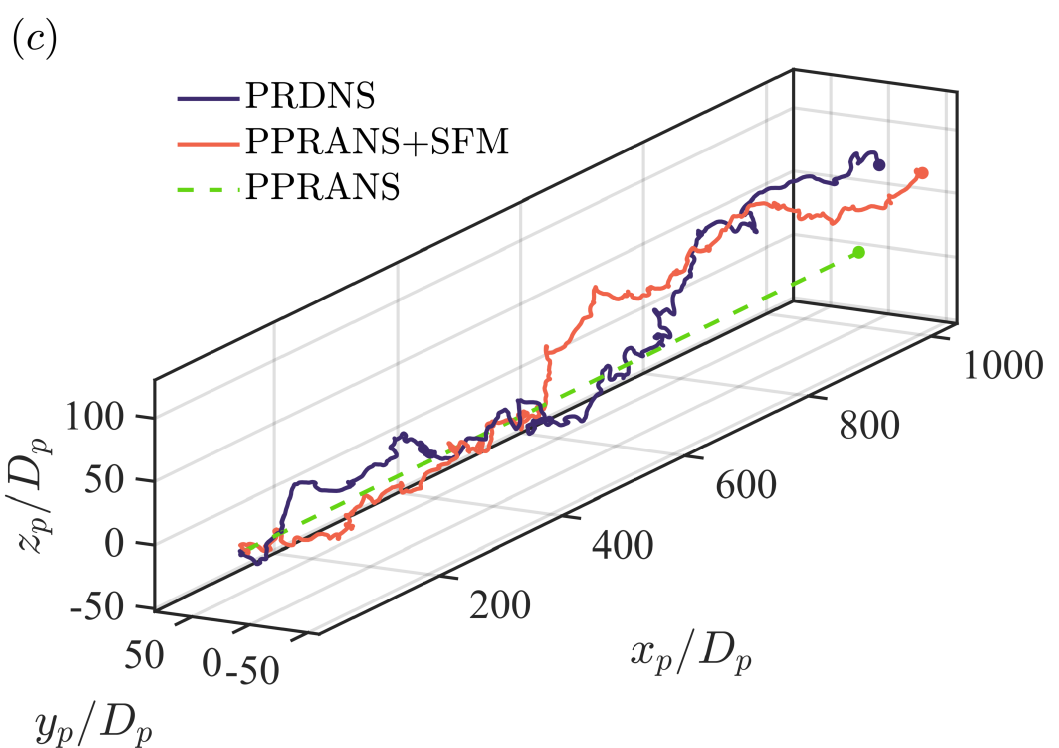}}
\subfigure{\includegraphics[width=0.45\textwidth]{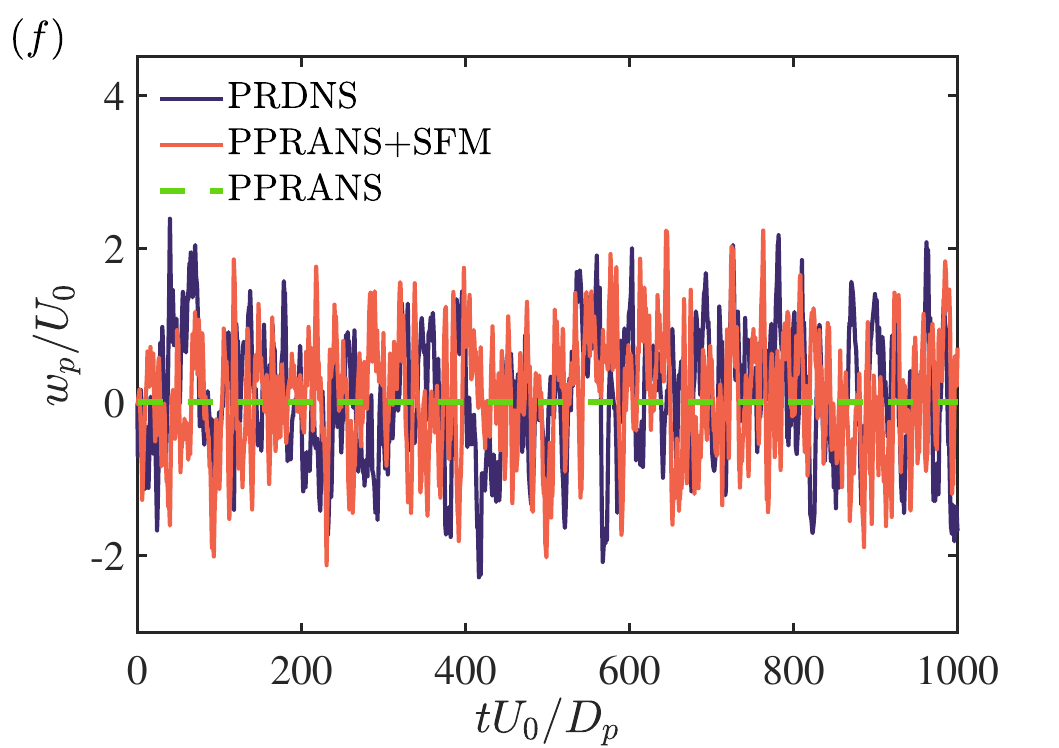}}
\caption{Time variations of (a,b,c) particle trajectories and (d,e,f) velocities with $Re_{p0}=100, I=1.0$ and $D_p/\eta=22.2$.}
\label{time_variation_of_particle}
\end{figure}

\begin{figure}
\centering
\subfigure{\includegraphics[width=0.45\textwidth]{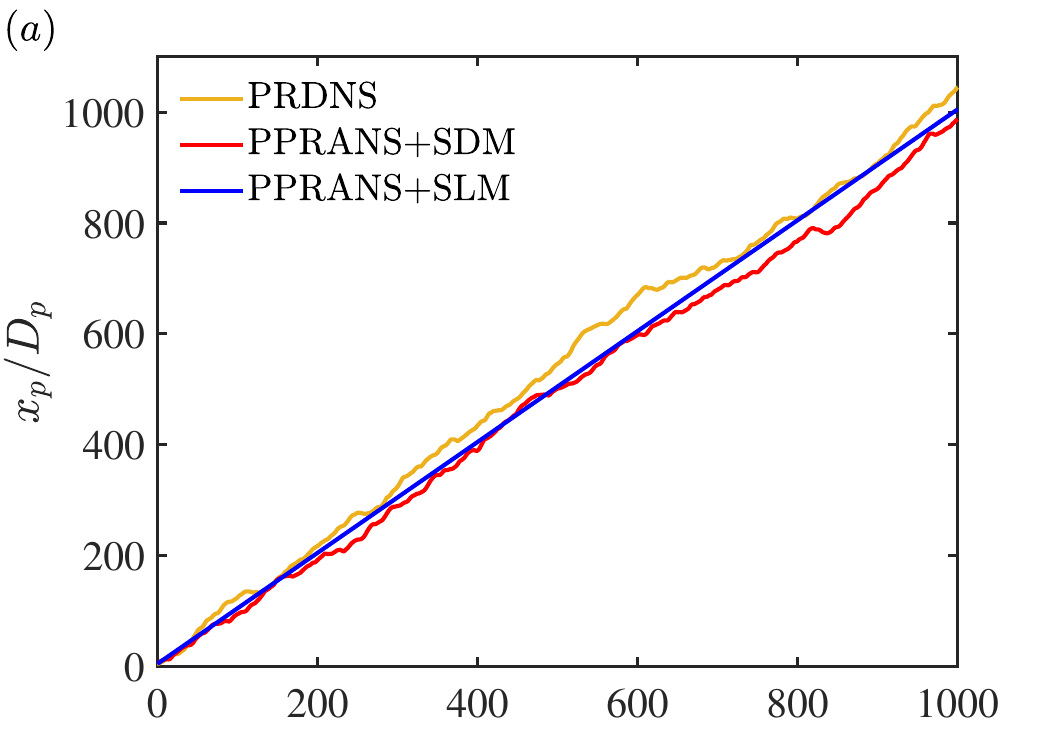}}
\subfigure{\includegraphics[width=0.45\textwidth]{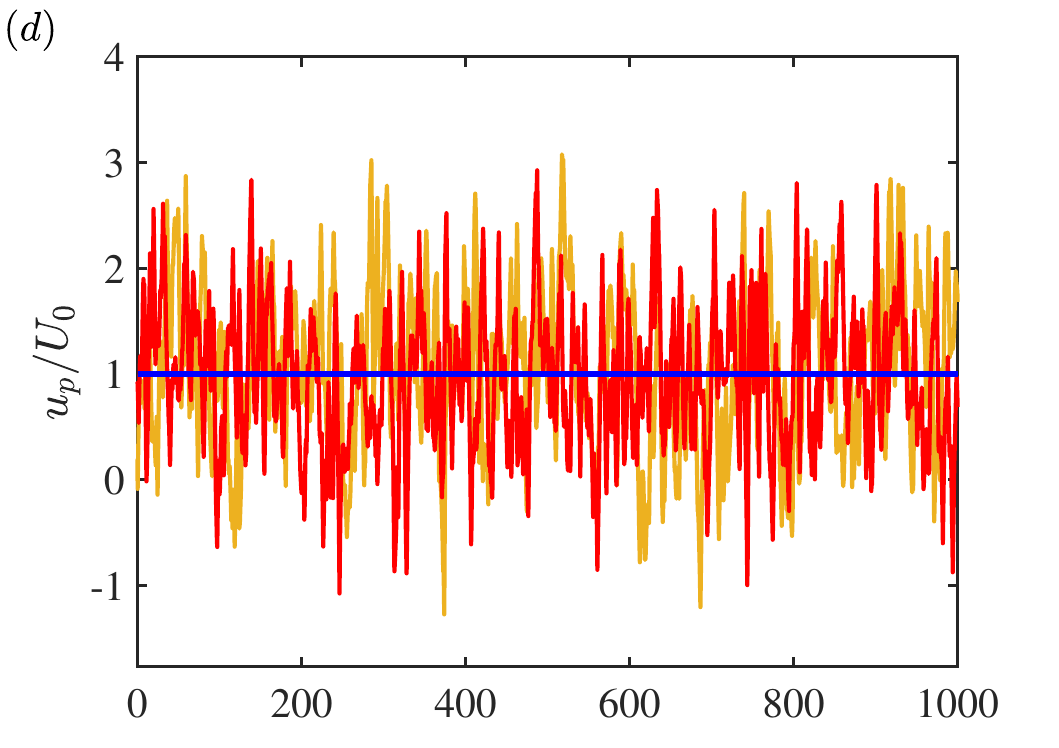}}
\subfigure{\includegraphics[width=0.45\textwidth]{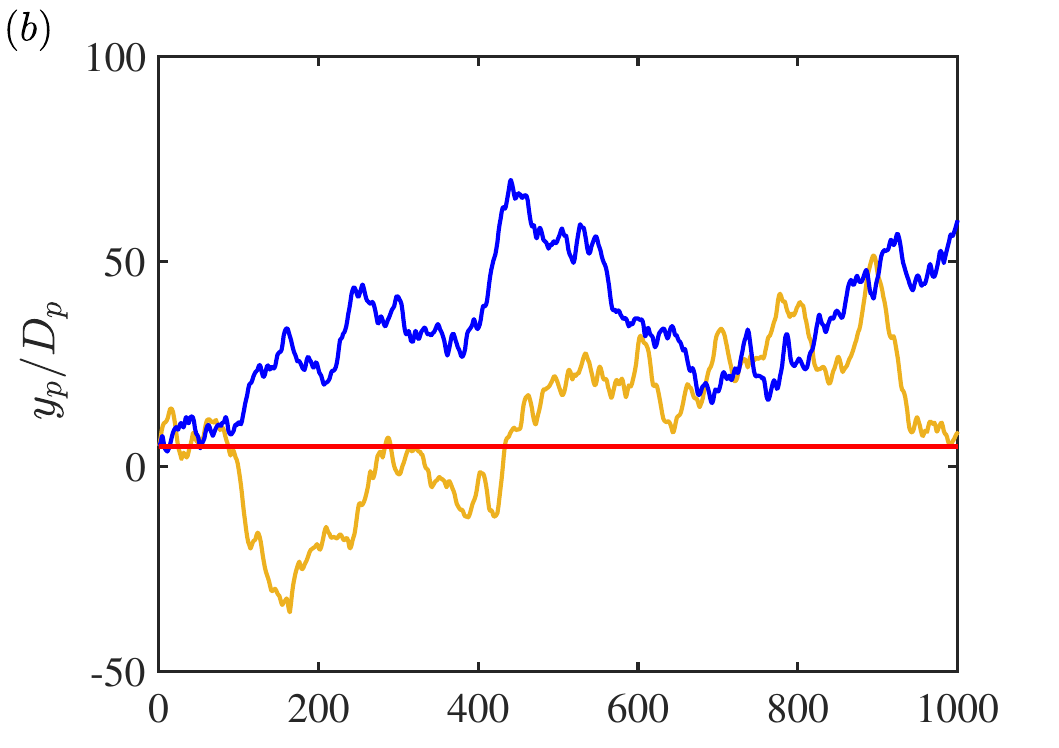}}
\subfigure{\includegraphics[width=0.45\textwidth]{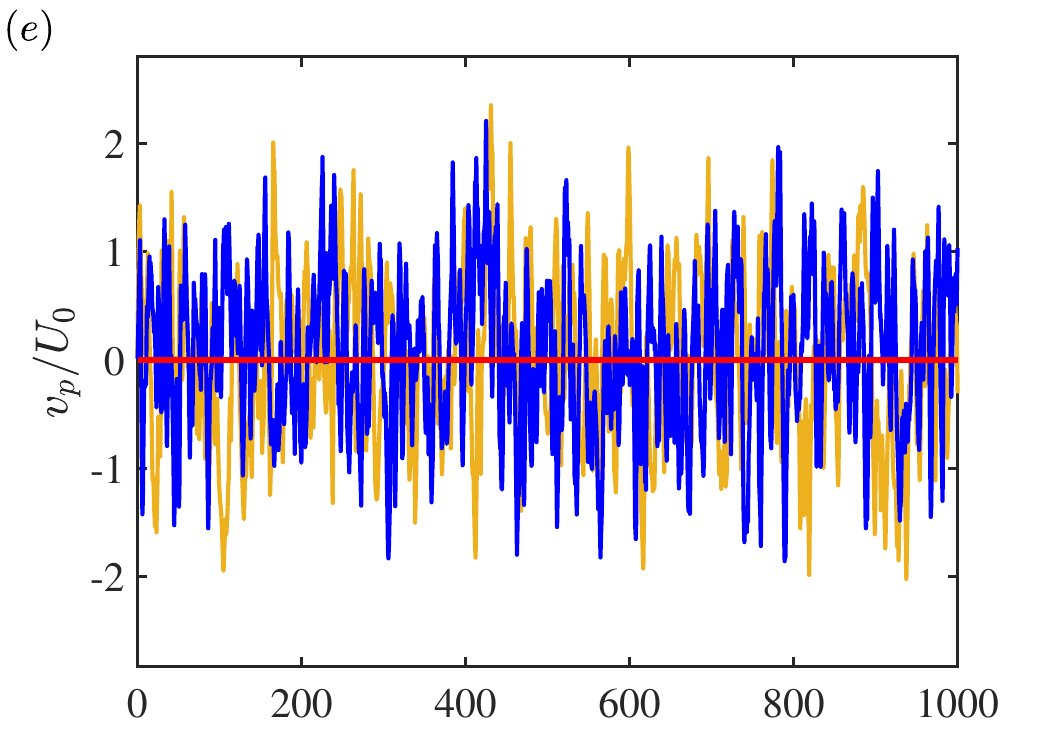}}
\subfigure{\includegraphics[width=0.45\textwidth]{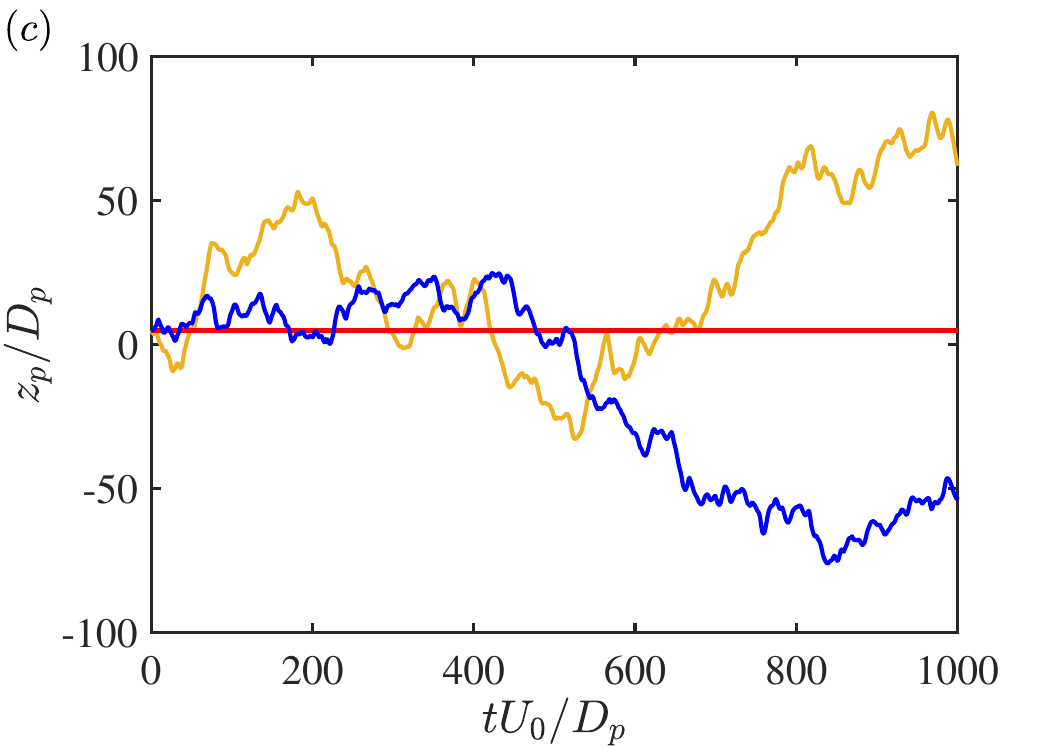}}
\subfigure{\includegraphics[width=0.45\textwidth]{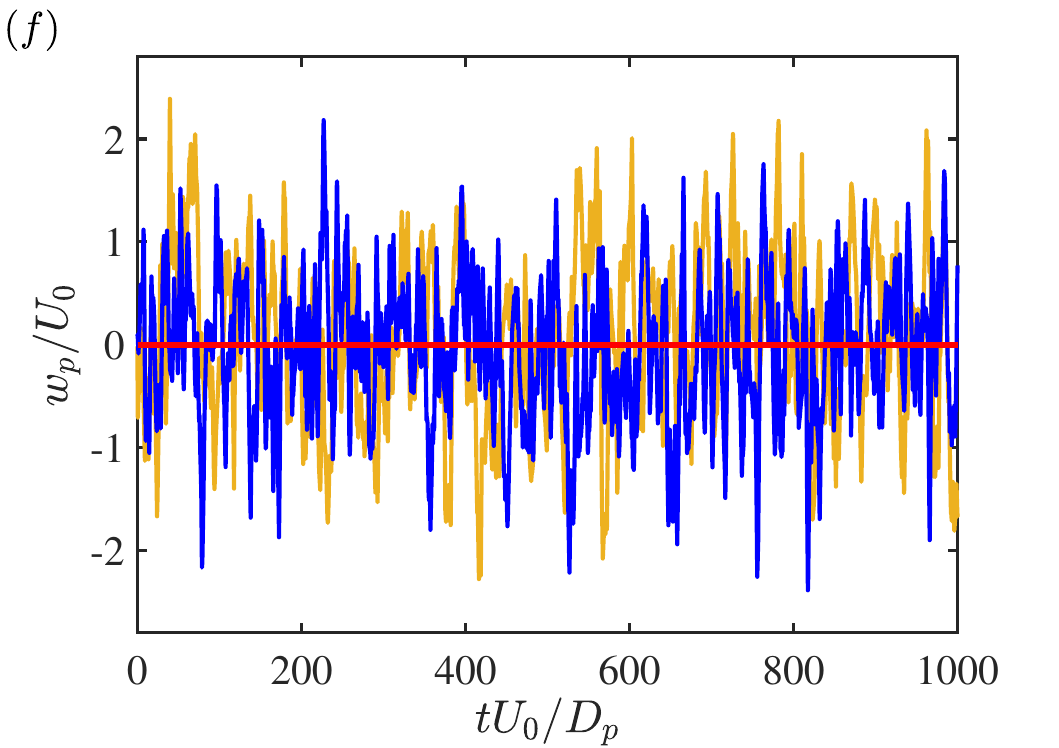}}
\caption{Time variations of (a,b,c) particle trajectories and (d,e,f) velocities with different versions of the stochastic force model. Here, PPRANS+SDM denotes that only the stochastic drag force model is incorporated, and PPRANS+SLM denotes that only the stochastic lateral force model is incorporated.}
\label{time_variation_of_particle_halfVersion}
\end{figure}

For PPRANS+SFM, the equations of particle motion are
\begin{equation}
    \frac{{\rm d} \boldsymbol{x}_p}{{\rm d}t}=\boldsymbol{v}_p,
    \label{particle_position}
\end{equation}
\begin{equation}
    \frac{{\rm d} \boldsymbol{v}_p}{{\rm d}t}=\left( 1+\Delta C_D \right) \left( 1 + 0.15Re_p^{0.687} \right) \frac{\boldsymbol{U} -\boldsymbol{v}_p}{\tau_p} + \frac{3\rho_f u_{rms}^2 }{4\rho_p D_p}\boldsymbol{C}^\prime,
    \label{particle_velocity}
\end{equation}
\begin{equation}
 \displaystyle
 {\rm d}\boldsymbol{C}^\prime(t) = -\frac{\boldsymbol{C}^\prime(t)}{T_D}{\rm d}t+C_{D,rms}^\prime\sqrt{\frac{2}{T_D}}{\rm d}\boldsymbol{W},
\end{equation}  
in which,
\begin{equation}
 \displaystyle
  \Delta C_D = \frac{0.015Re_p^{0.687}}{1+0.15Re_p^{0.687}}I^{0.858}Re_\lambda^{0.474},
\end{equation}
\begin{equation}
      \displaystyle
     C_{D,rms}^\prime =
     \frac{24}{Re_p^\prime} \left[1+ 0.15Re_p^{\prime0.687} \left( 1.702+0.556I^{-0.891} \left( \frac{D_p}{\eta} \right)^{0.111} \right) \right],
     \label{cdintensity}
\end{equation}
\begin{equation}    T_D=\frac{0.634Re_\lambda^{0.654}}{1+4.614 e^{-2.282I}}\tau_\eta.
    \label{timescale}
\end{equation}
{For PPRANS, the equations of particle motion are
\begin{equation}
    \frac{{\rm d} \boldsymbol{x}_p}{{\rm d}t}=\boldsymbol{v}_p,
    \label{particle_position2}
\end{equation}
\begin{equation}
    \frac{{\rm d} \boldsymbol{v}_p}{{\rm d}t}= \left( 1 + 0.15Re_p^{0.687} \right) \frac{\boldsymbol{U} -\boldsymbol{v}_p}{\tau_p}.
    \label{particle_velocity2}
\end{equation}}
Here $\boldsymbol{x}_p$ is the particle position, $\boldsymbol{v}_p$ the particle velocity, $\boldsymbol{U}$ the mean flow velocity, and $\tau_p=(\rho_pD_p^2)/(18\rho_f\nu)$ the particle relaxation time.
The stochastic model for the fluctuating drag force is adopted in all three directions. Moreover, the time advance of the particle motion is based on the second-order Adams-Bashforth scheme. 

\begin{table}[htbp]
  \begin{center}
\def~{\hphantom{0}}
\caption{The relative errors of the mean and RMS particle velocities of the PPRANS+SFM prediction (the data in the bracket represent the relative errors of the PPRANS prediction).}
  \begin{tabular}{ccccc}
  \toprule
      ~~Case~~  & ~~~$\overline{u}_p$~~~  & ~~~$u_{p,rms}$~~~   & ~~~$v_{p,rms}$~~~  & ~~~$w_{p,rms}$~~~ \\[3pt]
      \midrule
         1      & 2.0\% (3.1\%)           & 0.9\% (100\%)       & 6.5\% (100\%)      & 5.8\% (100\%) \\[3pt]
         2      & 2.1\% (2.3\%)           & 3.7\% (100\%)       & 5.3\% (100\%)      & 7.3\% (100\%) \\[3pt]
         3      & 0.3\% (0.3\%)           & 7.1\% (100\%)       & 4.5\% (100\%)      & 8.9\% (100\%) \\[3pt]
         4      & 1.0\% (1.1\%)           & 4.7\% (100\%)       & 3.7\% (100\%)      & 5.2\% (100\%) \\[3pt]
         5      & 4.2\% (6.2\%)           & 1.8\% (100\%)       & 0.8\% (100\%)      & 3.9\% (100\%) \\[3pt]
         6      & 1.9\% (6.2\%)           & 5.9\% (100\%)       & 8.9\% (100\%)      & 7.5\% (100\%) \\[3pt]
         7      & 0.4\% (6.2\%)           & 1.9\% (100\%)       & 0.7\% (100\%)      & 5.9\% (100\%) \\[3pt]
         8      & 3.5\% (3.8\%)           & 5.0\% (100\%)       & 9.5\% (100\%)      & 7.4\% (100\%) \\
         \bottomrule
  \end{tabular}
  \label{tab:validate1_error}
  \end{center}
\end{table}

Figure~\ref{time_variation_of_particle} shows the time variations in the position, velocity, and trajectory of the particles from the results of the PRDNS, PPRANS, and PPRANS+SFM simulations with $Re_{p0}=100, I=1.0$ and $D_p/\eta=22.2$ as an example. 
First, from the PPRANS result, it is clear that the particle moves along the streamwise direction steadily as the mean flow is in streamwise.
The trajectory of the particle is a straight line and the particle does not have transverse motion.
There are also no fluctuations in particle velocities because of the lack of turbulence effect in the PPRANS model.
Second, with the PPRANS+SFM model, the turbulence effect is reasonably incorporated, and the fluctuations in particle trajectory and velocities can be well predicted, which are in qualitatively good agreement with the PRDNS result.
To quantify the prediction errors of the stochastic model, table \ref{tab:validate1_error} displays the relative errors of the mean and RMS particle velocities predicted by the PPRANS+SFM as well as PPRANS to the PRDNS result. It can be seen that the overall relative errors of the mean and RMS particle velocities predicted by the PPRANS+SFM simulations are less than 10\% for all cases. This indicates that the proposed stochastic model can accurately predict the first- and second-order particle statistics in turbulence.

{Furthermore, to investigate the independent effects of the stochastic drag and lateral forces on particle motion, we present the results  in figure \ref{time_variation_of_particle_halfVersion}. It can be seen that if only the stochastic drag force model is incorporated, the streamwise particle velocity is basically in agreement with the PRDNS result. However, the spanwise and vertical particle velocities are zero, indicating that there is no displacement of the particle along the spanwise and vertical directions. 
If only the stochastic lateral force model is incorporated, the streamwise particle velocity is zero, whereas the spanwise and vertical particle velocities are consistent with the PRDNS results. 
Meanwhile, compared with the PPRANS + SFM result shown in figure \ref{time_variation_of_particle}, which incorporates all the stochastic force models, it is evident that including both the stochastic drag and the lateral force models can yield a prediction of particle motion that is closer to that of PRDNS.}

\subsection{Particle dispersion in a turbulent channel flow}

The second problem is the dispersion of small inertial particles in a turbulent channel flow. The friction Reynolds number is $Re_\tau=u_\tau H/\nu=180$, where $u_\tau=(\tau_w/\rho_f)^{1/2}$ is the friction velocity ($\tau_w$ is the mean wall-shear stress, $\rho_f$ is the fluid density), $H$ is the half-height of the channel, and $\nu$ is the fluid kinematic viscosity. The particle Stokes number $St=(\rho_pD_p^2u_\tau^2)/(18\rho_f\nu^2)$ is set to 25, 50, and 100, respectively, where the particle-to-fluid density ratio $\rho_p/\rho_f$ is 2650. The particle size is very small, \emph{i.e.} $D_p^+=0.41$, 0.58 and 0.82, therefore, PPDNS is performed for reference.
There are 30000 particles in each case, and the particles are initially evenly distributed in the channel, and only the S-N drag force is considered in PPDNS. In addition,  the particle concentration is very dilute, and the mean volume fractions are $\phi_v=1.2\times10^{-6}$, $3.4\times10^{-6}$, and $9.5\times10^{-6}$, respectively. Therefore, one-way coupling is assumed. 
The central difference scheme is utilized for the spatial discretization of the Navier-Stokes equations. 
The Adams-Bashforth scheme is employed for the convection term, while the Crank-Nicolson method is applied to the viscous term for time advancement. 
The size of the simulation domain is $L_x \times L_y \times L_z=4\pi H \times 2H \times 2\pi H$ and the grid number is $256 \times 256 \times 192$. The grid spacings in the streamwise and spanwise directions are uniform with $\Delta x^+=8.84$ and $\Delta z^+=5.89$. 
In the wall-normal direction, the grid is finer near the wall and coarser near the center of the channel, with $\Delta y^+=0.43\sim2.33$.

\begin{figure}
\centering
\subfigure{\includegraphics[width=0.49\textwidth]{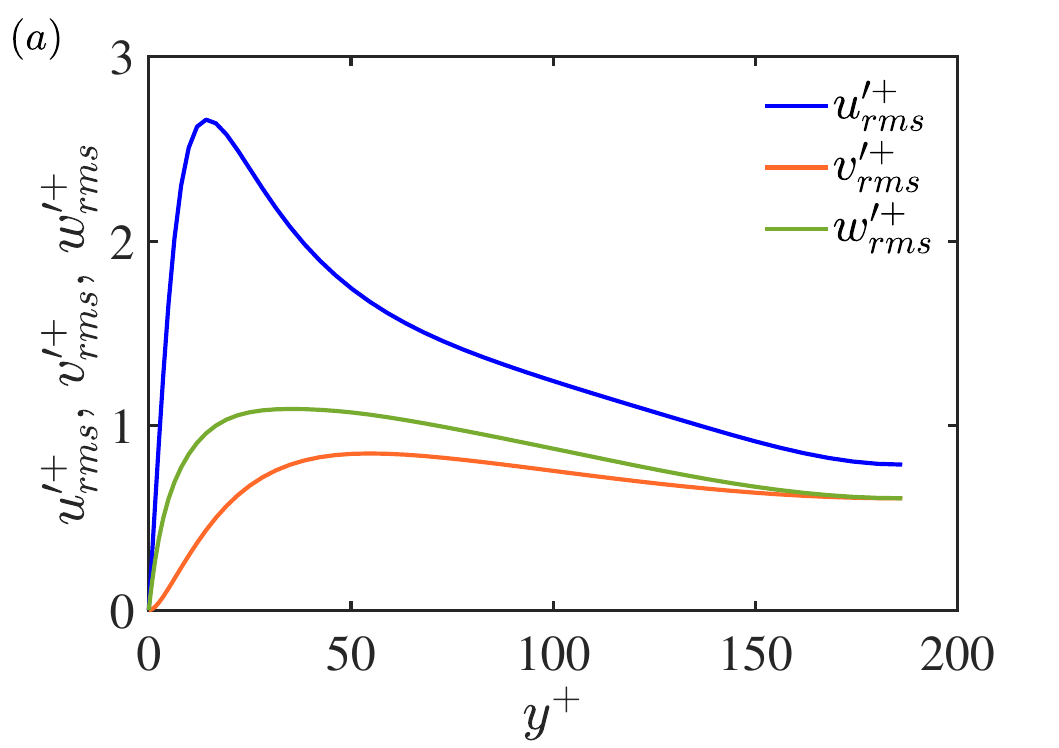}}
\subfigure{\includegraphics[width=0.49\textwidth]{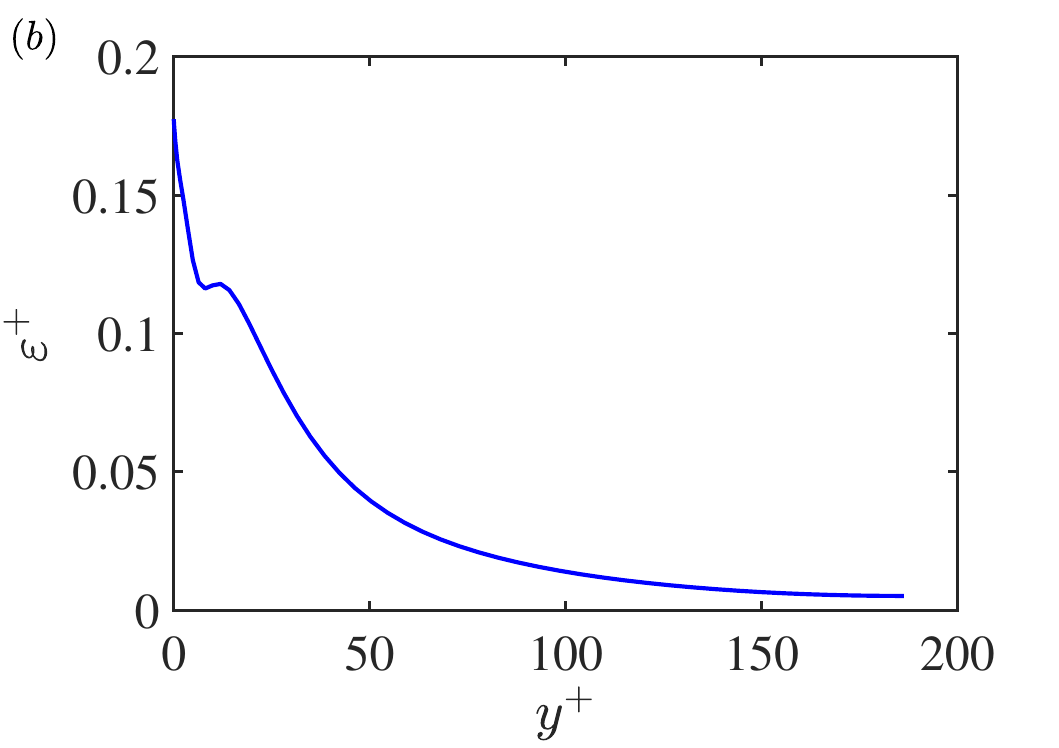}}
\caption{Profiles of (a) the turbulent velocity fluctuation intensities and (b) TDR along the wall-normal direction.}
\label{urms_epsi_yplus}
\end{figure}

\begin{figure}
\centering
\subfigure{\includegraphics[width=0.45\textwidth]{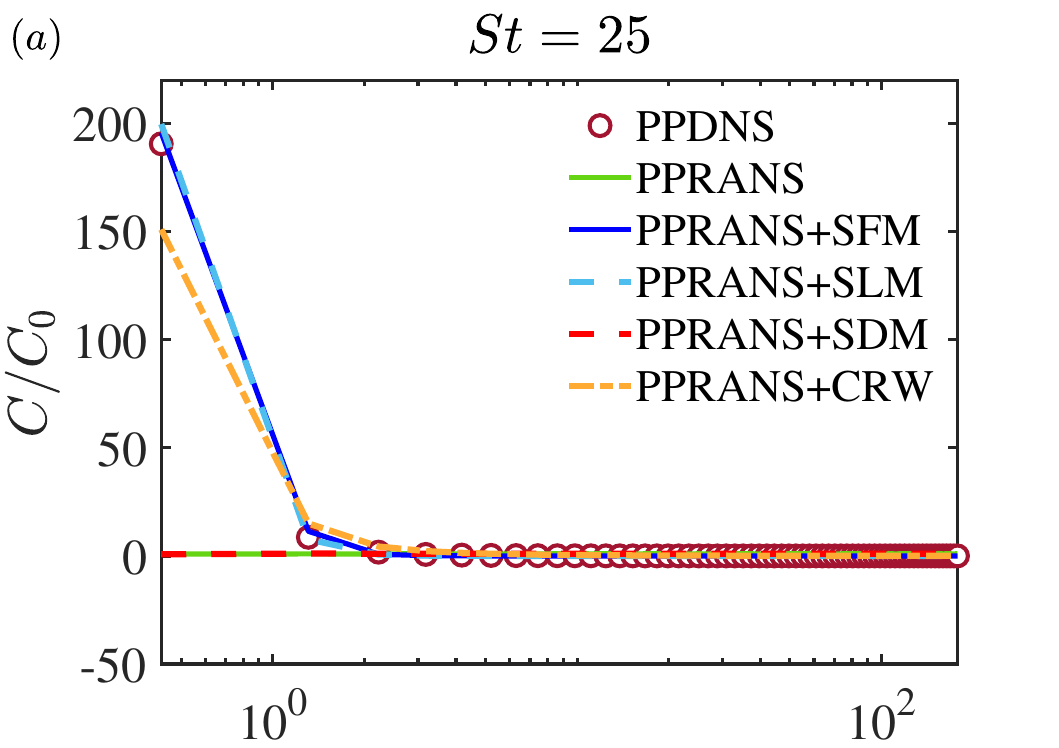}}
\subfigure{\includegraphics[width=0.45\textwidth]{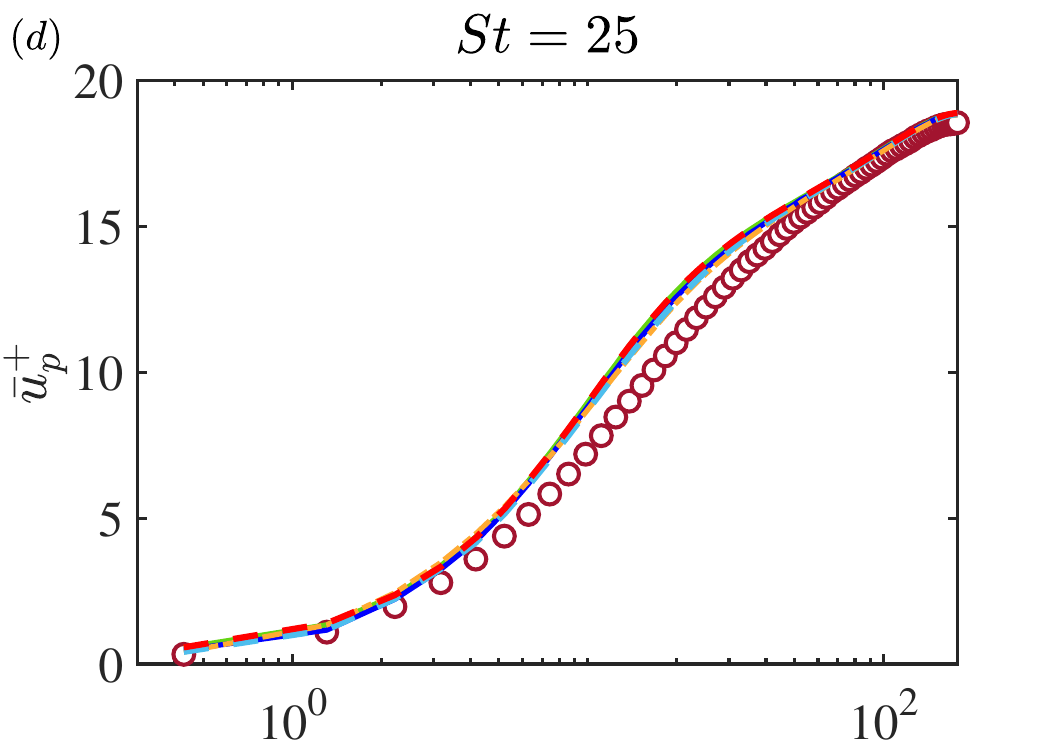}}
\subfigure{\includegraphics[width=0.45\textwidth]{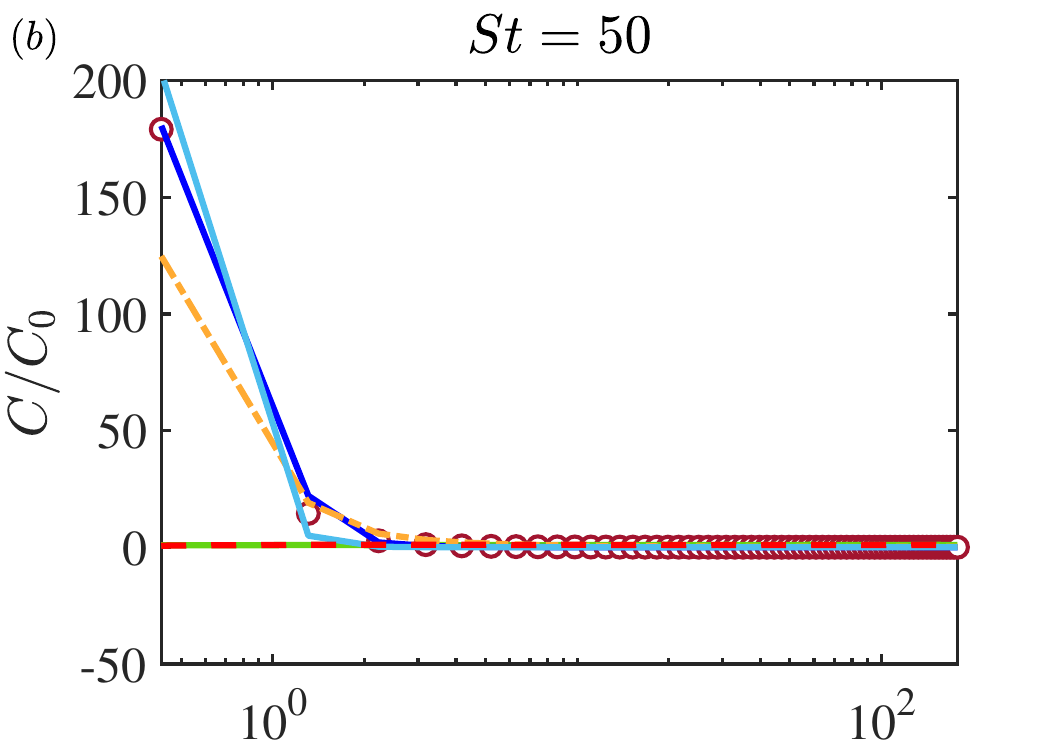}}
\subfigure{\includegraphics[width=0.45\textwidth]{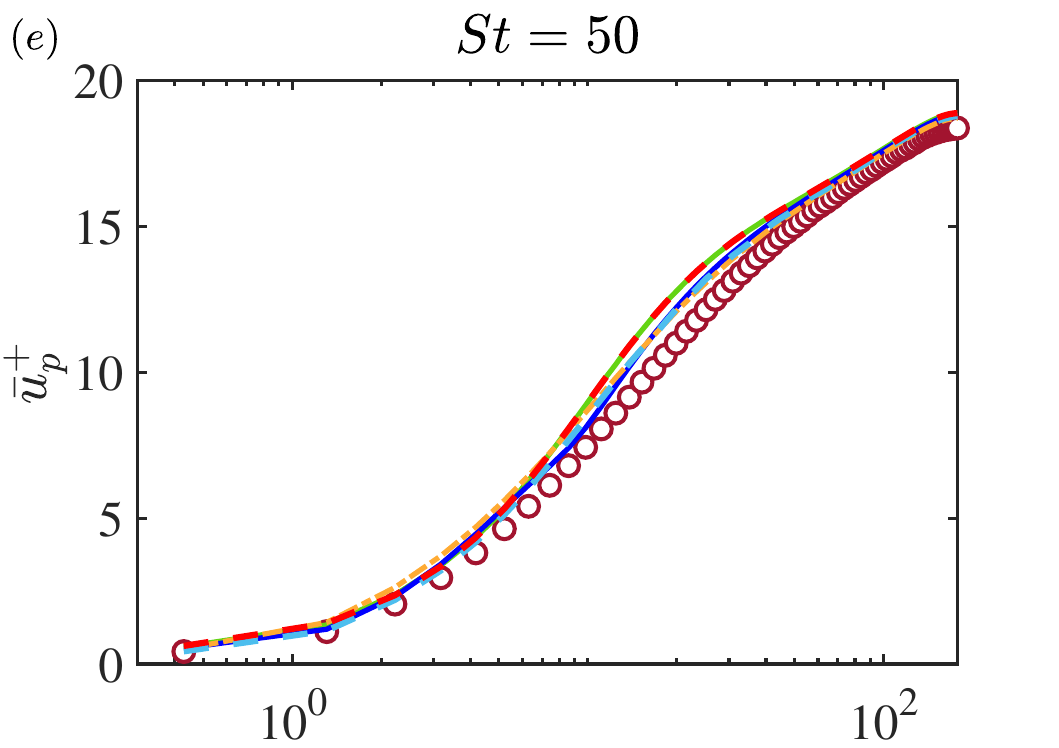}}
\subfigure{\includegraphics[width=0.45\textwidth]{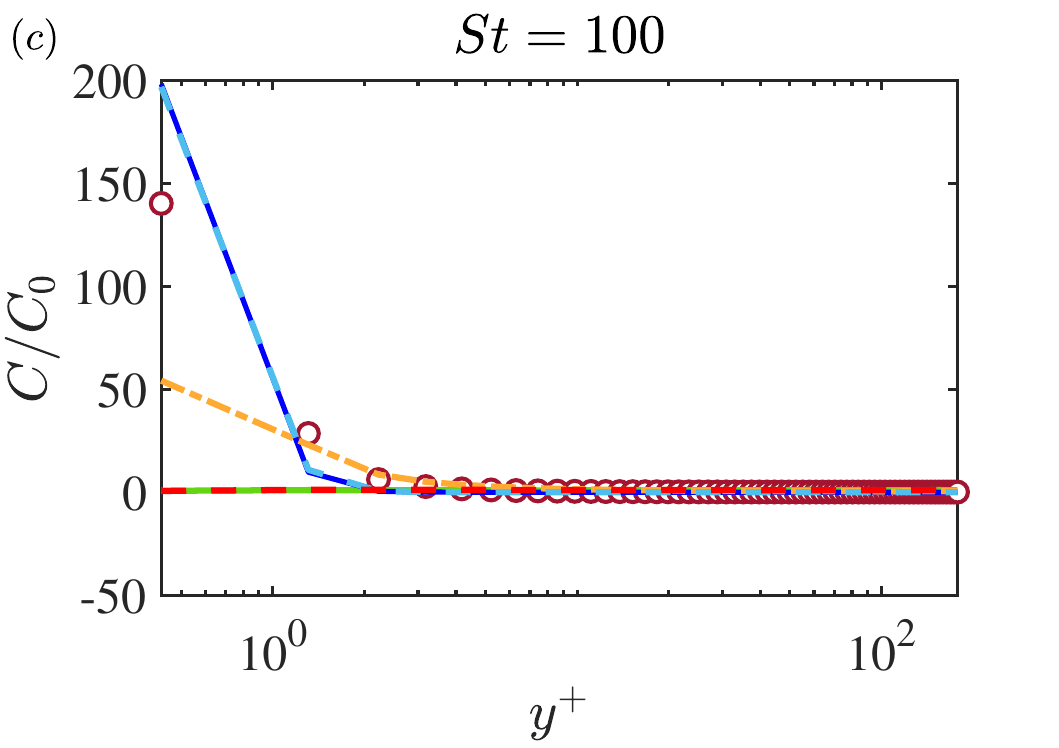}}
\subfigure{\includegraphics[width=0.45\textwidth]{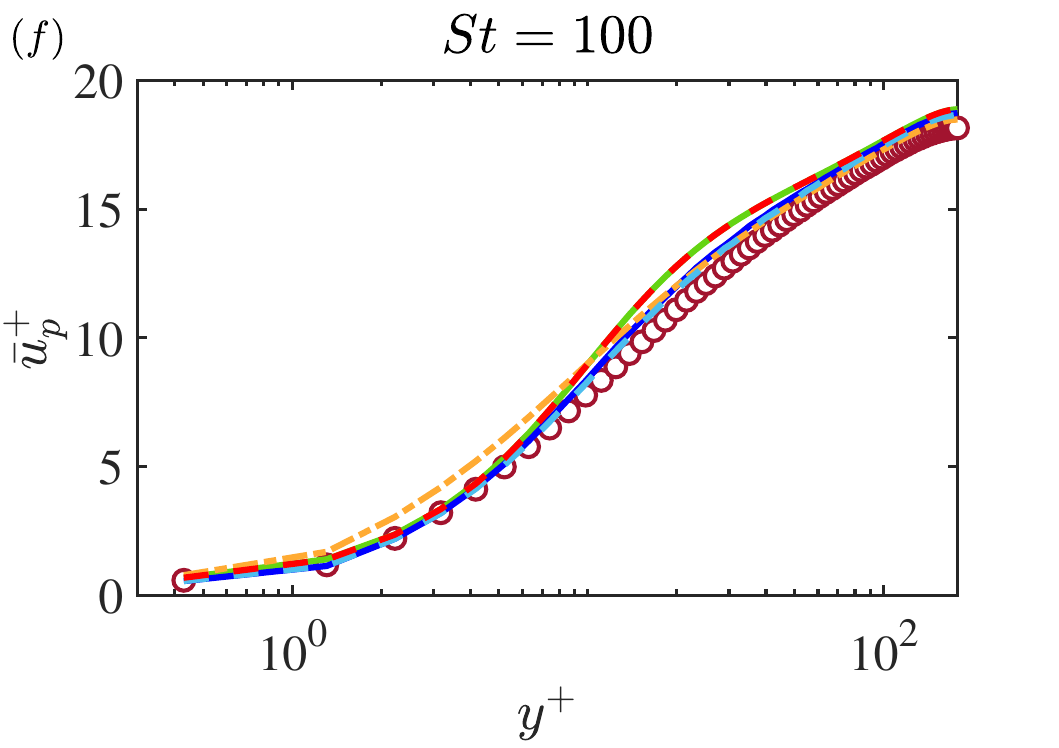}}
\caption{Profiles of (a,b,c) the normalized particle concentration and (d,e,f) the particle mean streamwise velocity along the wall-normal direction with different particle Stokes numbers. Here, PPRANS+SDM denotes that only the stochastic drag force model is incorporated, PPRANS+SLM denotes that only the stochastic lateral force model is incorporated, and PPRANS+CRW is the continuous random work model.}
\label{particle_concentration}
\end{figure}

Turbulent channel flow is highly anisotropic, which means that the intensities of turbulence fluctuation in different directions are quite different. To introduce the effect of turbulence anisotropy in our model, we use the turbulence intensities $u^\prime_{rms}$, $v^\prime_{rms}$ and $w^\prime_{rms}$ in the $x$, $y$ and $z$ directions to replace the isotropic turbulence intensity $u_{rms}$ in equation (\ref{particle_velocity}) in each direction. 
Furthermore, the turbulence-velocity-based particle Reynolds numbers and turbulence intensity ratios in the three directions involved in equations (\ref{cdintensity}) and (\ref{timescale}) are introduced by
\begin{equation}    Re_{p,x}^\prime=\frac{u^\prime_{rms} D_p}{\nu}, \quad  Re_{p,y}^\prime=\frac{v^\prime_{rms} D_p}{\nu}, \quad  Re_{p,z}^\prime=\frac{w^\prime_{rms} D_p}{\nu},
\end{equation}
and
\begin{equation}
    I_x=\frac{u^\prime_{rms}}{|\boldsymbol{U} -\boldsymbol{v}_p|}, \quad I_y=\frac{v^\prime_{rms}}{|\boldsymbol{U} -\boldsymbol{v}_p|}, \quad  I_z=\frac{w^\prime_{rms}}{|\boldsymbol{U} -\boldsymbol{v}_p|}.
\end{equation}
The profiles of the turbulent velocity fluctuations and TDR along the wall-normal direction are shown in figure \ref{urms_epsi_yplus}, where $u_{rms}^{\prime +}=u_{rms}^\prime /u_\tau$, $v_{rms}^{\prime +}=v_{rms}^\prime/u_\tau$, $w_{rms}^{\prime +}=w_{rms}^\prime/u_\tau$, $\varepsilon^+=\varepsilon \nu/u_\tau^4$ and $y^+=yu_\tau/\nu$.

In this section, in addition to PPDNS and PPRANS+SFM, we also investigate the dispersion of particles in the channel flow utilizing a continuous random walk (CRW) model \citep{Dehbi2008,Dehbi2010}. This model utilizes statistical flow characteristics obtained from RANS to generate the fluctuating velocities of the flow encountered by particles via a stochastic model in the Lagrangian framework. Therefore, we will refer to it as "PPRANS+CRW". 
For PPRANS+CRW, the equations of particle motion are 
\begin{equation}
    \frac{{\rm d} \boldsymbol{v}_p}{{\rm d}t}= \left( 1 + 0.15Re_p^{0.687} \right) \frac{\left( \boldsymbol{U}+\boldsymbol{u}^\prime \right) -\boldsymbol{v}_p}{\tau_p},
    \label{particle_velocity3}
\end{equation}
in which $\boldsymbol{u}^\prime$ is the flow fluctuating velocities encountered by particle and defined as  
\begin{equation}
    {\rm d} \left( \frac{u_1}{\sigma_1} \right)=-\left( \frac{u_1}{\sigma_1} \right) \cdot \frac{{\rm d}t}{\tau_L}+\sqrt{\frac{2}{\tau_L}} \cdot {\rm d}W_1+\frac{\partial \left( \frac{\overline{u_1 u_2}}{\sigma_1} \right)}{\partial x_2} \cdot \frac{{\rm d}t}{1+St_k},
\end{equation}
\begin{equation}
    {\rm d} \left( \frac{u_2}{\sigma_2} \right)=-\left( \frac{u_2}{\sigma_2} \right) \cdot \frac{{\rm d}t}{\tau_L}+\sqrt{\frac{2}{\tau_L}} \cdot {\rm d}W_2+\frac{\partial \sigma_2} {\partial x_2} \cdot \frac{{\rm d}t}{1+St_k},
\end{equation}
\begin{equation}
    {\rm d} \left( \frac{u_3}{\sigma_3} \right)=-\left( \frac{u_3}{\sigma_3} \right) \cdot \frac{{\rm d}t}{\tau_L}+\sqrt{\frac{2}{\tau_L}} \cdot {\rm d}W_3,
\end{equation}
where the subscripts 1, 2, and 3 represent the streamwise, vertical and spanwise directions, $u_i$ is the component of $\boldsymbol{u}^\prime$ in the $i$th direction, $\sigma _i$ is the RMS of the flow velocity fluctuation, ${\rm d}W_i$ is the discrete Wiener process, $\tau_L=\tau_L^+\nu/u_\tau$, and $\tau_L^+$ is approximated by the fit obtained by \cite{kallio1989} as
\begin{equation}
    \tau_L^+ = \left\{
    \begin{array}{ll}
    \displaystyle
     10, & y^+ \leq 5, \\[2pt]
    \displaystyle
      7.122 + 0.5731y^+ -0.00129{y^+}^2,        & 5 < y^+ < 200.
    \end{array} \right.
\end{equation}
And $St_k=\tau_p/\tau_L$.

\begin{figure}
\centering
\subfigure{\includegraphics[width=0.32\textwidth]{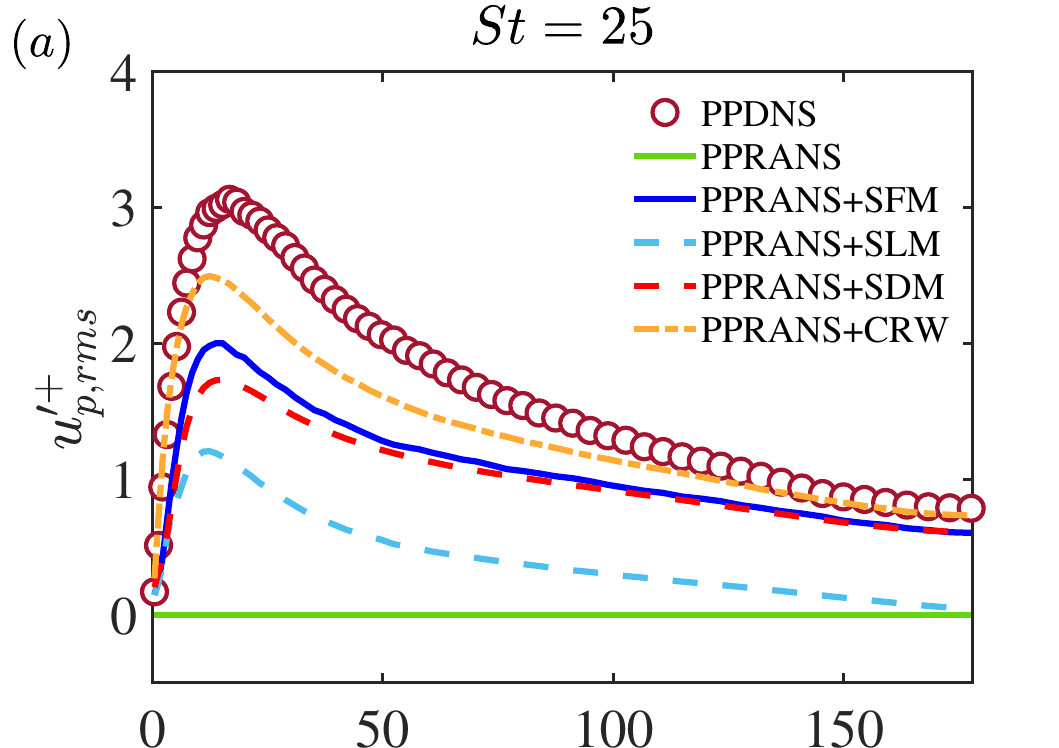}}
\subfigure{\includegraphics[width=0.32\textwidth]{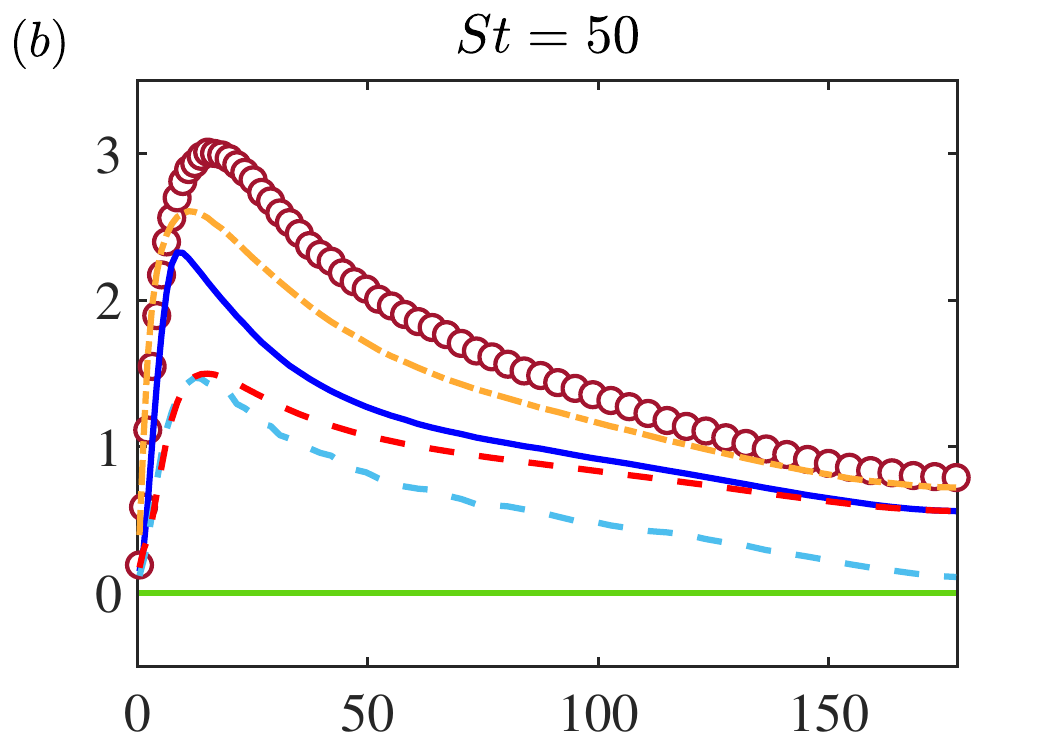}}
\subfigure{\includegraphics[width=0.32\textwidth]{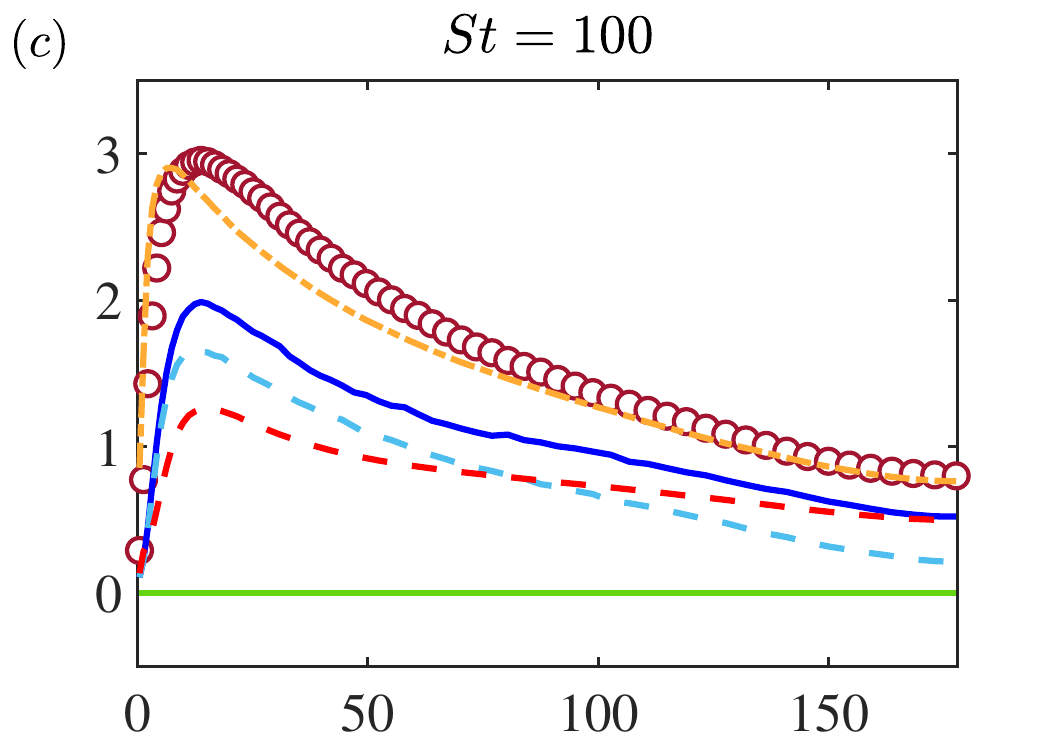}}
\subfigure{\includegraphics[width=0.32\textwidth]{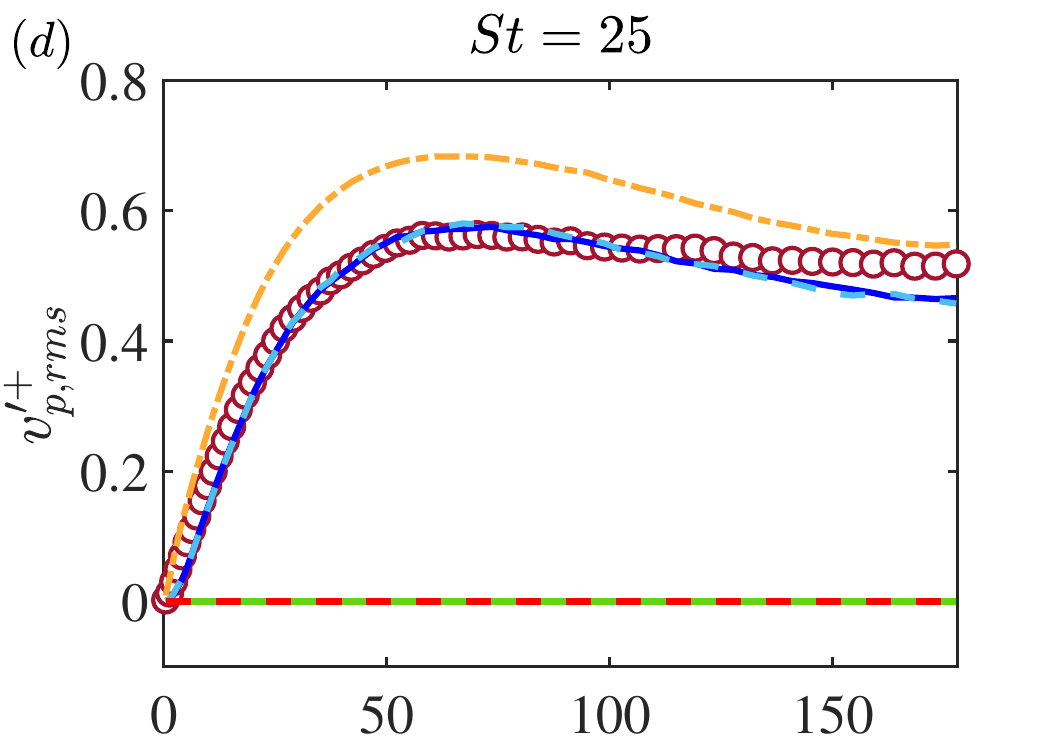}}
\subfigure{\includegraphics[width=0.32\textwidth]{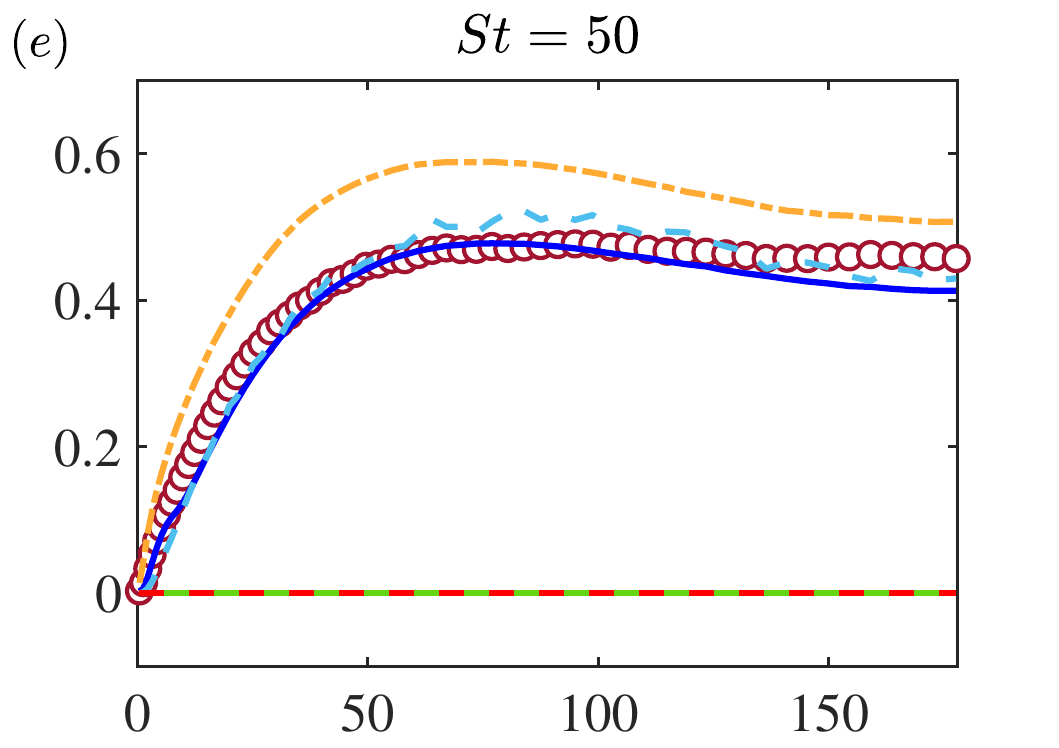}}
\subfigure{\includegraphics[width=0.32\textwidth]{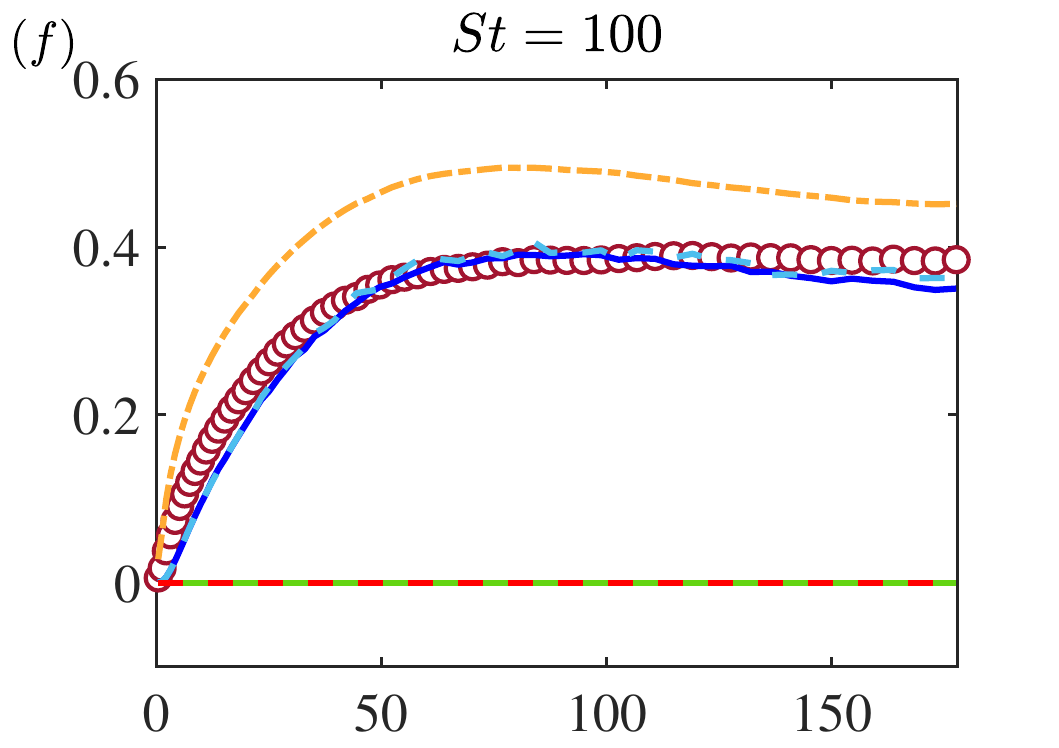}}
\subfigure{\includegraphics[width=0.32\textwidth]{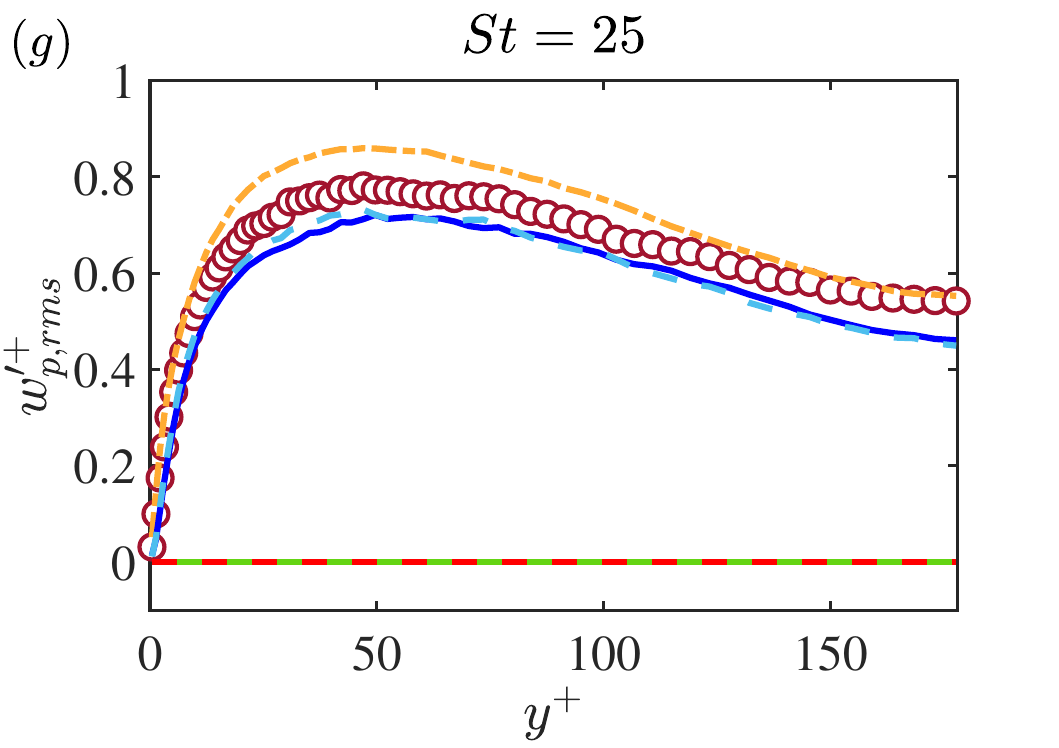}}
\subfigure{\includegraphics[width=0.32\textwidth]{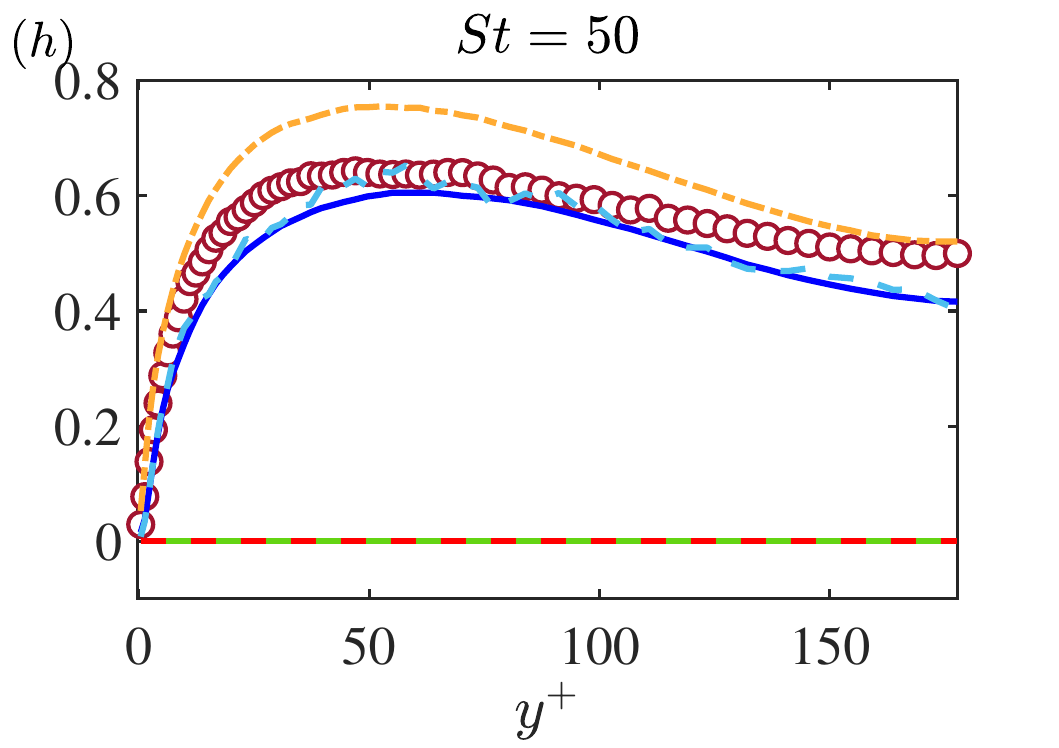}}
\subfigure{\includegraphics[width=0.32\textwidth]{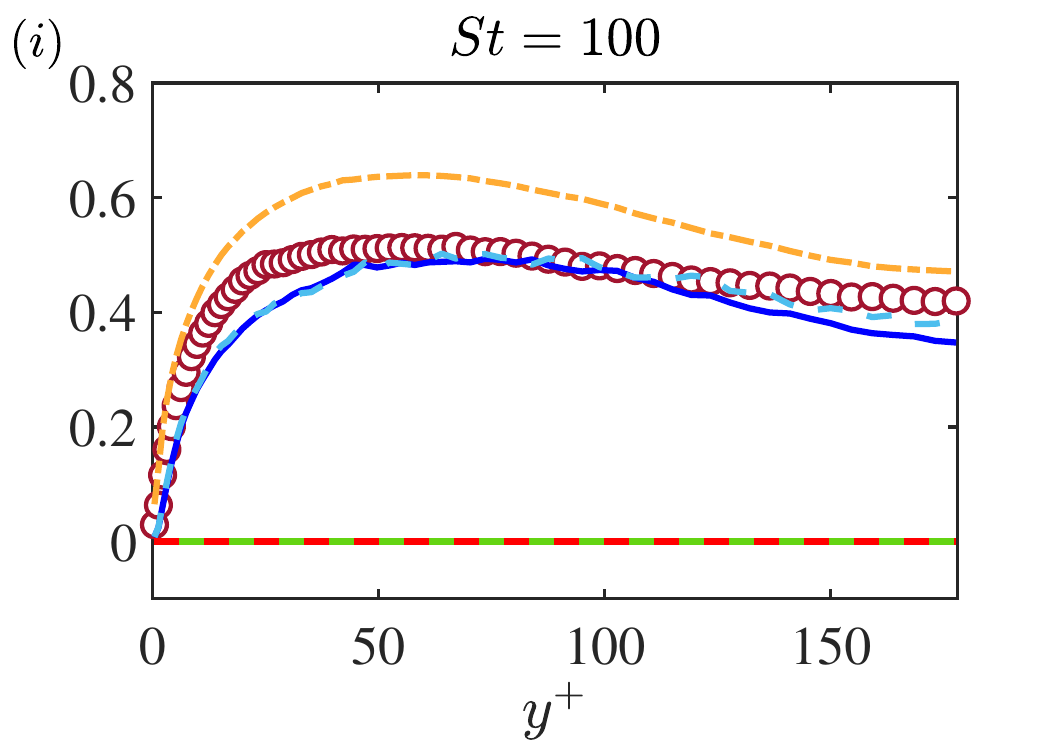}}
\caption{Profiles of the (a,b,c) streamwise, (d,e,f) wall-normal and (g,h,i) spanwise particle velocity fluctuation intensities along the wall-normal direction with different particle Stokes numbers. Here, PPRANS+SDM denotes that only the stochastic drag force model is incorporated, PPRANS+SLM denotes that only the stochastic lateral force model is incorporated, and PPRANS+CRW is the continuous random work model.}
\label{particle_rms_vel}
\end{figure}

Figure~\ref{particle_concentration} shows the profiles of the mean particle concentration and the mean streamwise particle velocity along the wall-normal direction with different particle Stokes numbers, where $C$ is the mean particle concentration and $C_0$ is the bulk concentration. 
It can be seen that if the stochastic force contributions due to turbulence are not considered (PPRANS), the particles are evenly distributed along the wall-normal direction, and there is no accumulation of particles near the wall caused by turbophoresis as observed in PPDNS. 
In contrast, incorporating both the stochastic drag and lateral force models based on the mean drag froce model(PPRANS+SFM) can predict the near-wall accumulation of particles, and the particle concentration distributions along the wall in the wall-normal direction coincide well with the PPDNS results. 
Moreover, in figure \ref{particle_concentration} (d,e,f), it can be seen that the mean particle streamwise velocity profiles predicted by PPRANS+SFM are also in good agreement with the PPDNS results.
Figure~\ref{particle_rms_vel} displays the profiles of the fluctuations in particle velocity along the wall-normal direction. 
It is seen that the predictions of PPRANS+SFM agree well with the results of PPDNS, especially for the wall-normal and spanwise particle velocity fluctuations. There are some discrepancies in the streamwise particle velocity fluctuations. This may be attributed to the fact that the particle is fixed in the flow field during model development using PRDNS, which implies that the particle has an infinite response time. Consequently, the proposed force model can only be reliably applied to highly inertial particles. In this validation case, despite the relatively high density ratio between the particle and fluid, it remains constrained. These limitations necessitate further investigation and enhancement in future studies. Moreover, due to the lack of turbulence effects, PPRANS cannot predict particle velocity fluctuations.

{Besides, we also present the results of incorporating only the stochastic drag force model or the stochastic lateral force model. As illustrated in figure \ref{particle_concentration}, when only the stochastic drag force model is taken into account, the particles are uniformly distributed throughout the channel. In contrast, if only the stochastic lateral force model is incorporated, the particle concentration distribution aligns closely with that of PPRANS+SFM. This is mainly because these statistics are calculated along the vertical direction, and the stochastic lateral force model can play an important role in the lateral motion. Moreover, it can be seen from figure \ref{particle_rms_vel} that if only the stochastic drag force model is considered, the streamwise particle velocity fluctuation is slightly lower than that of PPRANS+SFM, with the vertical and spanwise particle velocity fluctuations being zero. However, if the stochastic lateral force model is incorporated, the vertical and spanwise particle velocity fluctuations are consistent with those of PPRANS+SFM, while the streamwise particle velocity fluctuation is slightly lower.}

Furthermore, as illustrated in figure \ref{particle_concentration}, PPRANS+CRW also demonstrates good performance in predicting the mean concentration and mean velocity of the particles. However, it exhibits relatively inferior accuracy in predicting particle concentration near the wall compared to PPRANS+SFM. Additionally, figure \ref{particle_rms_vel} indicates that while PPRANS+CRW excels at predicting particle streamwise velocity fluctuations, PPRANS+SFM provides more accurate predictions of particle spanwise and wall-normal velocity fluctuations.

\subsection{Summary of the validation study}
The results of the two validation problems indicate that the proposed stochastic force models can accurately predict particle statistics in isotropic and anisotropic turbulence for both finite-size and point particles. 
To apply the stochastic force model, the mean flow velocity, TKE, and TDR must be known \emph{a priori}, which can be given from DNS data or RANS simulation using a specific turbulence closure, such as the $k-\varepsilon$ or $k-\omega$ model. 
If the effect of turbulence anisotropy needs to be included, more complicated Reynolds stress models can be adopted.

\section{Conclusions}\label{sec:conclusion}
This study conducts a series of direct numerical simulations to accurately compute the drag and lateral fluid forces on a fixed particle in homogeneous and isotropic turbulence superimposed with a uniform flow over a wide range of parameters: {the particle Reynolds number is 0.1$\sim$300, the scale ratio between the particle size and the Kolmogorov scale of turbulence is 0.22$\sim$76.9, and the turbulence intensity ratio is 0.1$\sim$20}.
The background turbulence is kept in a statistically steady state, and the particle wake effect is eliminated by a concurrent precursor technique.
Stochastic models based on the Langevin equation for the drag and lateral fluid forces were proposed with the high-fidelity PRDNS data. 
The proposed models were validated in two problems: the motion of a finite-sized particle in turbulence and the dispersion of small inertial particles in a turbulent channel flow. The predictions of the proposed stochastic models exhibit remarkable consistency with the DNS results in both cases. 
The inputs of the current model consist of the distributions of the mean flow velocity, the mean kinetic energy of the turbulence, and the mean turbulent dissipation rate, which can be effectively provided by various RANS turbulence closures. Consequently, it can be utilized for numerous particle-laden flow problems while maintaining an acceptable computational cost.

Several issues need further discussion. First, the particle is fixed in our PRDNS, indicating that the particle has an infinite response time $\tau_p$. Therefore, the proposed force model can only be reliably applied for highly inertial particles with $St \gg 1$ and $\rho_p/\rho_f \gg 1$ ($\rho_p$ and $\rho_f$ are the densities of particle and fluid, respectively). In other words, the model cannot be used for particles whose density is close to or smaller than that of the fluid, \emph{e.g.} tracer particles. It is also outside the scope of some specific theories for idealized inertialess particles, such as the Tchen-Hinze theory \citep{Hinze_1975}. 
Second, if we assume that there is a high particle-to-fluid density ratio, then the drag force is the most important, and other forces, such as the added mass and Basset history forces, can be neglected. To relieve the limits of the large density ratio assumption, one can simulate freely evolving particle suspensions in turbulent flows and develop new drag correlations \citep{tangDirectNumericalSimulations2016,tavanashadEffectDensityRatio2019,tavanashad2021}, which are left for future studies.
{Third, in the present work, only one particle is taken into account. Further research considering multiple particles \citep{vanderhoef2005,cate2006,beetstra2007,tenneti2011,tang2015} will be performed and the corresponding force models will be developed.}
Fourth, we only consider one-way coupling in the applications of the model, as the disperse phase is dilute. An extension to two-way coupling should be straightforward, and we will consider it in future work. The wake removal method is not responsible for the fluid-particle coupling, which should be essentially determined by the volume fraction or mass loading of the disperse phase.

\printcredits

\section*{Acknowledgement}{The authors acknowledge the financial support from the National Natural Science Foundation of China (Nos. 12388101 and 12472221) and the Fundamental Research Funds for the Central Universities (lzujbky-2024-oy10).}

\appendix

\section*{Appendix A. Simulation parameters}\label{Appendix A}

{Table~\ref{tab:parameters} lists the simulation parameters of a fixed particle in turbulence described in section \ref{Simulation_setup}.}
\begin{table}[H]
  \begin{center}
\def~{\hphantom{0}}
\caption{Simulation parameters: particle Reynolds number $Re_p=U_0D_p/\nu$; turbulent intensity $I=u_{rms}/U_0$; particle-to-turbulence scale ratio $D_p/\eta$; turbulence-velocity-based particle Reynolds number $Re_p^\prime=u_{rms}D_p/\nu$; Taylor-microscale-based Reynolds number $Re_\lambda=u_{rms}\lambda/\nu$; $R^2(C_D^\prime)$ and $R^2(C_L^\prime)$, the coefficient of determination between the PDF of the fluctuating drag and lateral force coefficients and a normal distribution, where the mean and variance of the normal distribution are derived from the statistical properties of these force coefficients.}
  \begin{tabular}{c|ccccccccc}
  \toprule
       Method & ~$Re_p$~ & ~$I$~ & ~$D_p/\eta$~ & ~$Re_p^\prime$~ & ~$Re_\lambda$~ & ~$R^2(C_D^\prime)$~ & ~$R^2(C_L^\prime)$~ & ~$N_x\times N_y\times N_z$~ & ~$h/\eta$~\\
      \midrule
       \multirow{28}*{PRDNS}
       ~ & 1     & 10       & 4.5  & 10  & 19  & 0.995 &0.996 & $400\times200\times200$  & 0.22 \\
       ~ & 1     & 20       & 7.1  & 20  & 30  & 0.998 &0.997 & $400\times200\times200$  & 0.36 \\
       ~ & 10    & 0.2      & 1.6  & 2   & 6   & 0.956 &0.984 & $400\times200\times200$  & 0.08 \\
       ~ & 10    & 0.35     & 2.2  & 3.5 & 10  & 0.978 &0.995 & $400\times200\times200$  & 0.11 \\
       ~ & 10    & 0.5      & 2.8  & 5   & 12  & 0.982 &0.995 & $400\times200\times200$  & 0.14 \\
       ~ & 10    & 1        & 4.5  & 10  & 19  & 0.983 &0.986 & $400\times200\times200$  & 0.22 \\
       ~ & 10    & 2.5      & 8.3  & 25  & 35  & 0.994 &0.995 & $400\times200\times200$  & 0.42 \\
       ~ & 10    &5         & 13.6 & 50  & 52  & 0.991 &0.996 & $400\times200\times200$  & 0.71 \\
       ~ & 10    & 20       & 37.0 & 200 & 114 & 0.990 &0.993 & $400\times200\times200$  & 1.85 \\
       ~ & 40    & 0.2      & 3.9  & 8   & 16  & 0.985 &0.997 & $400\times200\times200$  & 0.20 \\
       ~ & 40    & 0.35     & 5.6  & 14  & 24  & 0.992 &0.998 & $400\times200\times200$  & 0.28 \\
       ~ & 40    & 0.5      & 7.1  & 20  & 31  & 0.977 &0.986 & $400\times200\times200$  & 0.35 \\
       ~ & 40    & 1        & 11.5 & 40  & 47  & 0.987 &0.994 & $400\times200\times200$  & 0.58 \\
       ~ & 40    & 5        & 36.9 & 200 & 114 & 0.985 &0.990 & $400\times200\times200$  & 1.58 \\
       ~ & 100   & 0.1      & 4.9  & 10  & 16  & 0.979 &0.990 & $400\times200\times200$  & 0.25 \\
       ~ & 100   & 0.2      & 7.1  & 20  & 30  & 0.983 &0.994 & $400\times200\times200$  & 0.36 \\
       ~ & 100   & 0.35     & 10.5 & 35  & 43  & 0.982 &0.998 & $400\times200\times200$  & 0.52 \\
       ~ & 100   & 0.5      & 13.5 & 50  & 53  & 0.989 &0.998 & $400\times200\times200$  & 0.68 \\
       ~ & 100   & 1        & 22.2 & 100 & 79  & 0.994 &0.993 & $400\times200\times200$  & 1.11 \\
       ~ & 100   & 5        & 76.9 & 500 & 164 & 0.989 &0.979 & $800\times400\times400$  & 1.92 \\
       ~ & 200   & 0.1      & 7.8  & 20  & 26  & 0.989 &0.993 & $400\times200\times200$  & 0.19 \\
       ~ & 200   & 0.2      & 11.7 & 40  & 46  & 0.995 &0.994 & $400\times200\times200$  & 0.29 \\
       ~ & 200   & 0.35     & 17.2 & 70  & 64  & 0.985 &0.996 & $400\times200\times200$  & 0.43 \\
       ~ & 200   & 0.5      & 22.4 & 100 & 77  & 0.990 &0.990 & $400\times200\times200$  & 0.56 \\
       ~ & 300   & 0.1      & 10.3 & 30  & 33  & 0.988 &0.989 & $400\times200\times200$  & 0.26 \\
       ~ & 300   & 0.2      & 15.6 & 60  & 57  & 0.988 &0.993 & $400\times200\times200$  & 0.39 \\
       ~ & 300   & 0.35     & 23.1 & 105 & 80  & 0.967 &0.991 & $400\times200\times200$  & 0.58 \\
       ~ & 300   & 0.5      & 30.0 & 150 & 97  & 0.989 &0.987 & $400\times200\times200$  & 0.75 \\
       \midrule
       \multirow{15}*{PPDNS}
       ~ & 0.1   & 10       & 0.35 & 1    & 31  & 0.998 &0.999 & $400\times200\times200$  & 0.18  \\
       ~ & 0.1   & 20       & 0.71 & 2    & 31  & 0.995 &0.998 & $400\times200\times200$  & 0.35  \\
       ~ & 0.2   & 20       & 0.73 & 4    & 117 & 0.992 &0.995 & $400\times200\times200$  & 1.82 \\
       ~ & 0.4   & 5        & 0.89 & 2    & 20  & 0.994 &0.993 & $400\times200\times200$  & 0.11 \\
       ~ & 0.4   & 10       & 0.70 & 4    & 127 & 0.990 &0.997 & $800\times400\times400$  & 1.75 \\
       ~ & 0.5   & 5        & 0.67 & 2.5  & 54  & 0.997 &0.999 & $400\times200\times200$  & 0.34 \\
       ~ & 0.6   & 2.5      & 0.49 & 1.5  & 36  & 0.998 &0.998 & $400\times200\times200$  & 0.21 \\
       ~ & 0.8   & 2.5      & 0.97 & 2    & 16  & 0.987 &0.993 & $400\times200\times200$  & 0.03 \\
       ~ & 0.8   & 5        & 0.73 & 4    & 117 & 0.991 &0.989 & $400\times200\times200$  & 1.82 \\
       ~ & 1     & 1        & 0.44 & 1    & 20  & 0.987 &0.995 & $400\times200\times200$  & 0.22 \\
       ~ & 1     & 1        & 0.22 & 1    & 77  & 0.991 &0.999 & $400\times200\times200$  & 1.12 \\
       ~ & 1.2   & 5        & 0.99 & 6    & 143 & 0.983 &0.990 & $800\times400\times400$  & 2.05 \\
       ~ & 2.5   & 0.5      & 0.34 & 1.25 & 52  & 0.991 &0.999 & $400\times200\times200$  & 0.68 \\
       ~ & 2.8   & 1        & 0.80 & 2.8  & 48  & 0.982 &0.997 & $400\times200\times200$  & 0.57 \\
       ~ & 3.2   & 0.5      & 0.56 & 1.6  & 32  & 0.980 &0.999 & $400\times200\times200$  & 0.35 \\
       \bottomrule
  \end{tabular}
  \label{tab:parameters}
  \end{center}
\end{table}

\section*{Appendix B. Influence of the particle wake removing method and domain size}\label{Appendix B}

\begin{figure}[ht]
\centering
\includegraphics[width=0.8\textwidth]{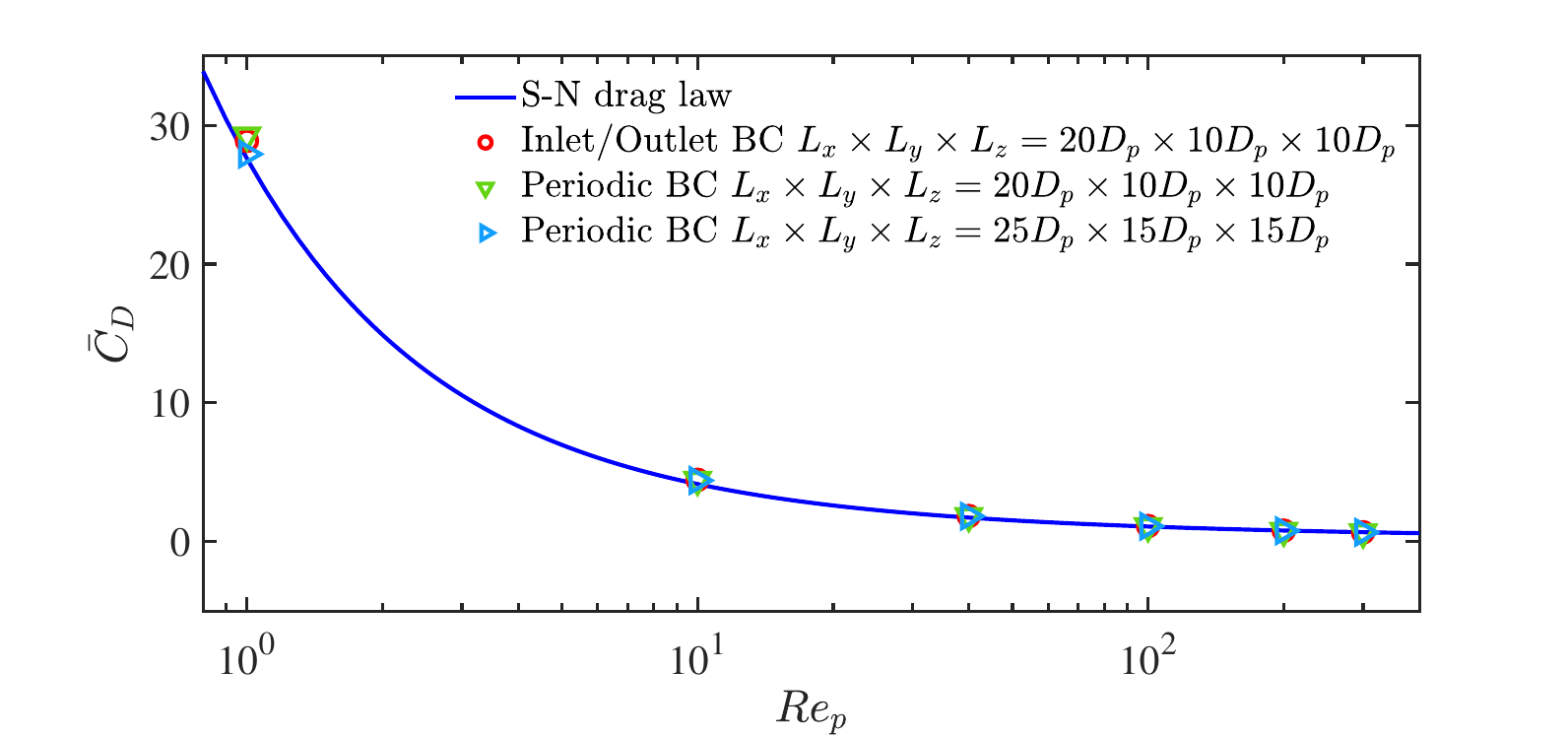}
\caption{Comparison of the mean drag coefficient of a particle in a uniform flow by PRDNS and the S–N drag law.}
\label{CD_Rep}
\end{figure}

Two types of boundary conditions are adopted to calculate the drag coefficient of the particles in uniform inflow. In the first type, the inlet and outlet boundary conditions are applied in the streamwise direction, while the free-slip boundary conditions were used in the two transverse directions. 
In the second type, periodic boundary conditions were applied in all three directions, and the concurrent precursor method is also employed to remove the particle wake. 
In both types, the computational domain size is $L_x \times L_y \times L_z = 20D_p\times10D_p\times10D_p$, the particle is fixed at $\boldsymbol{x}_p=(5D_p,5D_p,5D_p)$, and the grid size is $h=D_p/20$. The particle Reynolds numbers are $1, 10, 40, 100, 200$ and 300. 
In addition, another case using the periodic boundary condition with a larger domain size of $L_x \times L_y \times L_z = 25D_p\times15D_p\times15D_p$ (the particle is fixed at $\boldsymbol{x}_p=(5D_p,7.5D_p,7.5D_p)$) is carried out to check the domain independence. Figure~\ref{CD_Rep} shows the comparison of the mean drag coefficients of PRDNS with the two types of boundary conditions and the S-N drag law. It can be seen that the mean particle drag coefficients are in good agreement with the S-N drag law, and the maximum error is less than 10\% for both boundary conditions. This demonstrates the feasibility of the concurrent precursor method used in this study and confirms the independence of the domain size.

\bibliographystyle{elsarticle-harv}

\bibliography{cas-refs}


\end{document}